\documentstyle[psfig]{mn}

\newif\ifAMStwofonts

\def\Romain#1{\expandafter\uppercase\expandafter{\romannumeral #1}}
\def\ion#1#2{#1$\;${\small\rm\Romain#2}\relax}
\def\MBAB{M(B$_{\rmn{AB}}$\hspace*{-0.05cm})}

\ifoldfss
  \newcommand{\rmn}[1] {{\rm #1}}

  \ifCUPmtlplainloaded \else
    \NewTextAlphabet{textbfit} {cmbxti10} {}
    \NewTextAlphabet{textbfss} {cmssbx10} {}
    \NewMathAlphabet{mathbfit} {cmbxti10} {} 
    \NewMathAlphabet{mathbfss} {cmssbx10} {} 
  \fi
  \ifAMStwofonts
    \ifCUPmtlplainloaded \else
      \NewSymbolFont{upmath} {eurm10}
      \NewSymbolFont{AMSa} {msam10}
      \NewMathSymbol{\upi}     {0}{upmath}{19}
      \NewMathSymbol{\umu}     {0}{upmath}{16}
      \NewMathSymbol{\upartial}{0}{upmath}{40}
      \NewMathSymbol{\leqslant}{3}{AMSa}{36}
      \NewMathSymbol{\geqslant}{3}{AMSa}{3E}

       \let\le=\leqslant
       \let\ge=\geqslant
    \fi
  \fi
\fi 

\ifnfssone
  \newmathalphabet{\mathit}
  \addtoversion{normal}{\mathit}{cmr}{m}{it}
  \addtoversion{bold}{\mathit}{cmr}{bx}{it}
  \newcommand{\rmn}[1] {\mathrm{#1}}

  \newmathalphabet{\mathbfit} 
  \addtoversion{normal}{\mathbfit}{cmr}{bx}{it}
  \addtoversion{bold}{\mathbfit}{cmr}{bx}{it}
  \newmathalphabet{\mathbfss} 
  \addtoversion{normal}{\mathbfss}{cmss}{bx}{n}
  \addtoversion{bold}{\mathbfss}{cmss}{bx}{n}
  \ifAMStwofonts
    \ifCUPmtlplainloaded \else
      \UseAMStwoboldmath
      \makeatletter
      \new@mathgroup\upmath@group
      \define@mathgroup\mv@normal\upmath@group{eur}{m}{n}
      \define@mathgroup\mv@bold\upmath@group{eur}{b}{n}
      \edef\UPM{\hexnumber\upmath@group}
      \new@mathgroup\amsa@group
      \define@mathgroup\mv@normal\amsa@group{msa}{m}{n}
      \define@mathgroup\mv@bold\amsa@group{msa}{m}{n}
      \edef\AMSa{\hexnumber\amsa@group}
      \makeatother
      \mathchardef\upi="0\UPM19
      \mathchardef\umu="0\UPM16
      \mathchardef\upartial="0\UPM40
      \mathchardef\leqslant="3\AMSa36
      \mathchardef\geqslant="3\AMSa3E

       \let\le=\leqslant
       \let\ge=\geqslant
    \fi
  \fi
\fi 

\ifnfsstwo
  \newcommand{\rmn}[1] {\mathrm{#1}}

  \DeclareMathAlphabet{\mathbfit}{OT1}{cmr}{bx}{it}
  \SetMathAlphabet\mathbfit{bold}{OT1}{cmr}{bx}{it}
  \DeclareMathAlphabet{\mathbfss}{OT1}{cmss}{bx}{n}
  \SetMathAlphabet\mathbfss{bold}{OT1}{cmss}{bx}{n}
  \ifAMStwofonts
    \ifCUPmtlplainloaded \else
      \DeclareSymbolFont{UPM}{U}{eur}{m}{n}
      \SetSymbolFont{UPM}{bold}{U}{eur}{b}{n}
      \DeclareSymbolFont{AMSa}{U}{msa}{m}{n}
      \DeclareMathSymbol{\upi}{0}{UPM}{"19}
      \DeclareMathSymbol{\umu}{0}{UPM}{"16}
      \DeclareMathSymbol{\upartial}{0}{UPM}{"40}
      \DeclareMathSymbol{\leqslant}{3}{AMSa}{"36}
      \DeclareMathSymbol{\geqslant}{3}{AMSa}{"3E}

       \let\le=\leqslant
       \let\ge=\geqslant
    \fi
  \fi
\fi 

\ifCUPmtlplainloaded \else
  \ifAMStwofonts \else 
    \def\upi{\pi}
    \def\umu{\mu}
    \def\upartial{\partial}
  \fi
\fi

\title[The H$\alpha$ luminosity function and star formation rate up to $z\sim 1$]
{The H$\alpha$ luminosity function and star formation rate up to $z\sim 1$\thanks{Based on data obtained at the European Southern Observatory on Paranal, Chile.}}
\author[L.~Tresse, S.~J.~Maddox, O.~Le~F\`evre and J.-G.~Cuby]
{L.~Tresse,$^1$\thanks{Email: laurence.tresse@oamp.fr}  S.~J.~Maddox,$^2$ O.~Le~F\`evre$^1$ and J.-G.~Cuby$^3$ \\
$^1$Laboratoire d'Astrophysique de Marseille, Les Trois-Lucs, B.P.~8, 13376, Marseille Cedex 12, France \\
$^2$School of Physics and Astronomy, University of Nottingham, Nottingham, NG7 2RD \\
$^3$ESO, Alonso de Cordova 3107, Vitacura, Casilla 19001, Santiago 19, Chile
}

\date{Accepted -- -- --.
      Received -- -- --;
      in original form -- -- --}

\pagerange{\pageref{firstpage}--\pageref{lastpage}}
\pubyear{2001}

\begin{document}

\maketitle

\label{firstpage}

\begin{abstract}
  We describe ISAAC/ESO-VLT observations of the H$\alpha$
  $\lambda$6563 Balmer line of $33$ field galaxies from the
  Canada-France Redshift Survey (CFRS) with redshifts selected between
  0.5 and 1.1.  We detect H$\alpha$ in emission in 30 galaxies and
  compare the properties of this sample with the low-redshift sample
  of CFRS galaxies at $z\sim 0.2$ \cite{tre98}.  We find that the
  H$\alpha$ luminosity, L(H$\alpha$), is tightly correlated to \MBAB\ 
  in the same way for both the low- and high-redshift samples.
  L(H$\alpha$) is also correlated to L([\ion{O}{2}]$\lambda$3727), and
  again the relation appears to be similar at low and high redshifts.
  The ratio L([\ion{O}{2}])/L(H$\alpha$) decreases for brighter
  galaxies by as much as a factor~2 on average.  Derived from the 
  H$\alpha$ luminosity function, the comoving H$\alpha$ luminosity
  density increases by a factor~12 from $\langle z \rangle = 0.2$ to
  $\langle z \rangle = 1.3$. Our results confirm a strong rise of the
  star formation rate (SFR) at $z<1.3$, proportional to
  $(1+z)^{4.1\pm0.3}$ (with $H_0=50$ km~s$^{-1}$~Mpc$^{-1}$, $q_0=0.5$).
  We find an average SFR(2800\AA)/SFR(H$\alpha$) ratio of 3.2 using
  the Kennicutt (1998) SFR transformations. This corresponds to the
  dust correction that is required to make the near $UV$ data
  consistent with the reddening-corrected H$\alpha$ data within 
  the self-contained, $I$-selected CFRS sample.  
\end{abstract}


\section{Introduction}

Measuring the comoving space density of the star formation rate (SFR)
as a function of cosmic epoch has advanced rapidly over the last few
years, it has set important cons\-traints on our knowledge of galaxy
formation and evolution.  Since the major work of Madau et al.  (1996)
using the Canada-France Redshift Survey (CFRS) data at $z<1$
\cite{lilly96} and the Lyman-break galaxy population at $z>2$
\cite{steidel}, many observational studies and detailed models have
emerged to trace back the SFR history.

The SFR density is usually derived from the mean lumino\-sity density,
that is ${\cal L} = \int_{0}^{\infty} \phi(L)LdL = \phi^\ast L^\ast
\Gamma(\alpha+2)$, assuming galaxies follow a Schechter (1976)
luminosity function with a characteristic luminosity, $L^{\ast}$, a
faint-end slope, $\alpha$, and a normalization density parameter,
$\phi^{\ast}$.  The faint-end slope is found to be greater than $-2$,
implying a high-space density of low-luminosity galaxies, but although
they are very numerous, they contribute little to the mean luminosity
density.  However as the three Schechter parameters are highly
correlated, it is necessary to build the luminosity function over the
largest possible range of luminosities.  At first sight the SFR
density appears a simple and useful tool to trace back the evolution
of star formation and link it with the evolution of mass, but
discrepancies between different measurements have led to controversy.
Uncertainties in the conversion factors from luminosity to star
formation rate, coupled with the different survey selection criteria
mean that the SFR history of the Universe is still poorly determined,
and thus hotly debated.  The discrepancies are partly due to the
difficulty in relating what we observe to the intrinsic physical and
che\-mical properties of gala\-xies.  Each SFR indicator has pros and
cons in the sense that they are more or less dependent on the star
formation, and more or less affected by factors other than star
formation.  Another difficulty is the small number of galaxies in some
samples; this leads to large statistical uncertainties, because they
do not represent a fair sample of the galaxy population as a whole.

Large and deep galaxy surveys provide estimates of the continuum
luminosity density in different rest-frame wavelength ranges from the
far ultraviolet to the far infrared.  As well as continuum luminosity
density measurements, SFR indicators using line-emission measurements
have also been studied, further complicating the picture but
stimulating work for a better understanding of the physical star
formation processes within galaxies. Observationally speaking, it is
gene\-ral\-ly accepted that the SFR density rises from $z=0$ to
$z\sim1$, but at higher redshifts it is still unclear whether the SFR
density reaches a plateau, or decreases, or increases slowly.  There
has also been some controversy about the rise at $z<1$; Lilly et al.
(1996) found a steep rise of the 2800-\AA\ luminosity density
proportional to $(1+z)^{3.90\pm0.75}$, while Cowie, Songaila \& Barger
(1999) argue for a shallow rise of the 2500-\AA\ luminosity density
proportional to $(1+z)^{1.5}$.

In this present study, we investigate the H$\alpha$~$\lambda$6563
luminosity density evolution using data acquired at the ESO-VLT with
the ISAAC infrared spectro-imager (Cuby et al. 2000, Moorwood et al.
1999) mounted at the Nasmyth focus of the first of the four 8.2m VLT
Unit Telescopes.  We recall that H$\alpha$ luminosities are excellent
tracers of instantaneous star formation within galaxies since they are
directly proportional to the ionizing UV stellar spectra at
$\lambda<912$~\AA.  It suffers from stellar absorption and dust
attenuation, but dust affects H$\alpha$ at a lower level than the UV,
and can be mi\-tigated by reddening correction.  Moreover new models
(see e.g.  Charlot \& Longhetti 2001) derive SFR estimates in a
self-consistent way accounting for the depletion of heavy ele\-ments
onto dust grains, for the absorption of ionizing photons by dust in
\ion{H}{2} regions, and for contamination of Balmer emission by
stellar absorption. With these corrections and with the help of
several spectral lines, the SFR from H$\alpha$ becomes compatible with
the SFR using far-infrared estimators \cite{charlot02}.  Furthermore,
H$\alpha$ is a particularly useful SFR indicator at high redshift
since it is directly comparable to low redshift surveys, which makes
it straightforward to probe evolution.  We concentrate on H$\alpha$
measurements of CFRS galaxies between $0<z<1.1$ using the data from
Tresse \& Maddox (1998) at $z \le 0.3$, and new H$\alpha$ data at
$0.5<z<1.1$ taken with the ISAAC spectrograph.  These data allow us to
quantify for the first time the H$\alpha$ luminosity function and
density evolution up to $z\sim1$ within a single survey.  In Section~2
we present our data samples.  Section~3 details the ISAAC
spectroscopic acquisition, reduction and flux measurements. Section~4
discusses the reddening correction on our measurements. Section~5
presents the [\ion{O}{2}]/H$\alpha$ ratio and other properties of the
samples. We estimate the SFR density for our samples in Section~6, and
conclude with a discussion of our results in Section~7.  Throughout
this paper, we use $H_0=50$~km~s$^{-1}$ Mpc$^{-1}$ and $q_0=0.5$
mainly for ease of comparison with published results.

\section{Galaxy selection}

Our sample of galaxies is taken from the $I$-band-selected CFRS, of
which 591 galaxies have reliable redshifts $0 < z < 1.3$ and apparent
magnitudes $17.5 \le I_{AB} \le 22.5$, in five regions of the sky (see
Le F\`evre et al. 1995). As described below, our sample is made of the
low-$z$ sample of H$\alpha$ emitters at $z \le 0.3$ observed in the
five CFRS fields and of the high-$z$ sample of H$\alpha$ emitters at
$0.50 < z < 1.05$ observed in three CFRS fields. This offers the
opportunity of direct measurements of evolution within a single survey
with a well-controlled selection. Moreover this study does not suffer
from uncertainties in line identification as encounted by narrow-band
filter surveys.

The CFRS spectral range is 4500--8500~\AA, thus the
[\ion{O}{2}]$\lambda$3727 line is observable for all galaxies with $z
\ga 0.2$, and the H$\alpha$ line is observable only for galaxies at $z
\la 0.3$.  Our low-$z$ sample consists of 110 out of the 138 CFRS
gala\-xies to $z=0.3$. They have H$\alpha$-emission fluxes measured
from the original CFRS spectra. This sample was analysed by Tresse \&
Maddox (1998), who describe the sample in more detail.  With new near
infrared capabilities we can now measure H$\alpha$ at higher
redshifts, and we have focused on CFRS galaxies at $z>0.5$.  For our
new observations, we used the ISAAC spectrograph in the
short-wavelength mode; this configuration enables us to span the
wavelength range 0.9--2.5~$\mu$m.  The $SZ$- and $J$-band filters with
an atmospheric transmission above 0.5 cover the wavelength range
0.98--1.35~$\mu$m.  Thus H$\alpha$ will be in this range for galaxies
within the redshift interval 0.50--1.05. The CFRS contains 323
galaxies in this interval.

For our high-$z$ sample, we selected our targets using the following
two well-defined criteria chosen so that we could measure
H$\alpha$-emission fluxes as accurately as possible.  The first
criterion was aimed at maximizing the pro\-ba\-bility of detecting
H$\alpha$ in emission. In nearby samples, the [\ion{O}{2}] and
H$\alpha$ emission lines are correlated, even though there is a large
dispersion (see e.g. Kennicutt 1992, Tresse et al. 1999).  Thus we
selected galaxies where the CFRS spectra show [\ion{O}{2}] emission
with observed equivalent width EW$\ga 10$~\AA.  There are 251 targets
out of 323 with EW([\ion{O}{2}]) $\ga 10$~\AA.  The second criterion
was to select galaxies with H$\alpha$ redshifted away from strong OH
night-sky emission lines.  For each of the EW([\ion{O}{2}]) $\ga
10$~\AA\ targets we cross-correlated the expected redshifted
wavelength of H$\alpha$ with a list of OH-sky lines, and rejected
galaxies where H$\alpha$ would be too close to a sky line.  For the
galaxy line-widths we assumed FWHM(H$\alpha$)~$= 13.1 \times
(1+z)$~\AA, corresponding to a rotation speed of 600~km~s$^{-1}$, and
for the OH-sky lines we assumed FWHM~$\sim 8$~\AA, corresponding to
the instrumental resolution for our observing set-up.  These FWHM are
large enough so we could be fairly confident that the observed
H$\alpha$ will be in between the OH bands, even with the uncertainty
from the redshift measurement.  Rejecting the galaxies at redshifts
where H$\alpha$ is likely to be affected by a sky line, leaves 84
targets.

\section{ISAAC spectroscopy}

\subsection{Observations}

Our spectroscopic observations were done in service mode during
Period-63 and Period-64 with the ISAAC spectrograph at the ESO-VLT.
Spectra were taken at Medium Re\-solution, and with the
Short-Wavelength channel equipped with a 1024$\times$1024 Hawaii
Rockwell array.  The pixel scale is 0.147 arcsec per pixel.  We opted
for long slits of 2 arcsec width, ensuring that most of the light of
the target was in the slit (see the 5x5 arcsec$^2$ postage stamps from
HST data shown in Fig.~1 of Schade et al. 1995).  This set-up gives a
re\-solution of $R \simeq 1500$, and covers a wavelength range of
0.16~$\mu$m in the $SZ$ filter, and 0.29~$\mu$m in the $J$ filter.  We
chose six central wavelengths which are away from the strongest OH
bands and cover the wavelength range for the expected positions of
H$\alpha$ for all the target galaxies. The central wavelengths are
1.02 and 1.06~$\mu$m for the $SZ$ filter, 1.11, 1.18, 1.26 and
1.33~$\mu$m for the $J$ filter.  For each observation, the appropriate
central wavelength was selected according to the redshift of the
target. Since the observations were carried out in service mode, the
choice of integration time had to be made before-hand. We set them by
comparing to the commissioning data which were available at that time,
and adjusting the time for each galaxy according to its [\ion{O}{2}]
flux and the spectral type.  Thus the individual integration times
(called DIT) were set to be either 600s, 900s or 1200s.  During the
observations the telescope was nodded between A and B positions, 20
arcsec apart along the slit.  The total integration time was
2$\times$DIT, consisting of one sequence~AB.  Dark frames, flat-fields
and (Xe+Ar) arc lamp spectra were taken with the same filter, central
wavelength and slit width for each of the targets observed during the
night. In total, 33 observation blocks (OB) were observed.
Table~\ref{tab1} lists the target name, the exposure time and the
filter used for each OB.
\begin{figure}
\centerline{\psfig{file=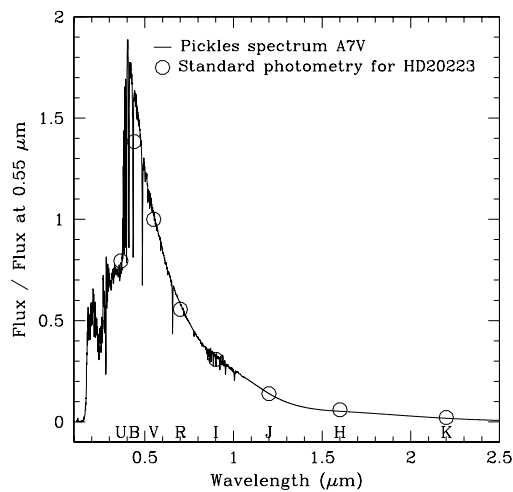,width=7cm}}
\caption{Broad-band  photometry (open points) and the inferred spectrum
  for the standard star HD20223. The spectrum is for an A7V star taken
  from the Pickles (1998) library, normalized to the observed
  photometry for HD20223. The agreement between the photometry and
  spectral flux is excellent in all bands, including J, H and K.
\label{fig1}
}
\end{figure}

\subsection{Reduction and calibration}

We used the {\scriptsize IRAF/CL} package for our data reduction. We
first subtracted the dark image from the A, B and flat ima\-ges. Next
we divided the A and B images by the flat res\-ponse.  The resulting A
and B images were then subtracted from each other to form (A$-$B) and
(B$-$A) images. At this stage, the images have been corrected for the
dark current, bias level and flat-field, and have been sky subtracted.
To combine A and B data, we first apply a spatial offset to the
(B$-$A) image, equal to the nod between the A and B positions so that
the galaxy signal appears at the same position in each image.  Then we
average the (A$-$B) image and the offset (B$-$A) image after rejecting
3$\sigma$ discrepant pixels. This procedure removes most cosmic rays,
sky residuals and bad pixels. Then a 1D spectrum was extracted at the
position of the emission line, and calibrated in wavelength using the
arc lines.

Finally the 1D spectra were spectrophotometrically ca\-li\-brated with
a standard star observed with the same setup as the target. For most
galaxies the standard star observations were taken during the same
night. For 2 galaxies, standard stars from neighbouring nights were
used.  At the time of our observations, carefully flux calibrated
standard star spectra covering the near infrared bands did not
e\-xist, and the only information we had for the standard stars was
the spectral type and the magnitude in several pass bands. Therefore
to determine the spectral response function we compared the observed
stellar spectra to spectra from the Pickles (1998) stellar library of
composite spectra. As a check on the reliability of the Pickles flux
calibration, we compared the spectra to the broad-band magnitudes. We
found that the agreement between the spectral shape and the UBVRIJHK
magnitudes is excellent as shown for an exam\-ple in Fig.~\ref{fig1}.
\begin{table*}
\phantom{.}
\vspace*{2cm}
\phantom{.}
\hspace*{-4cm}
\psfig{file=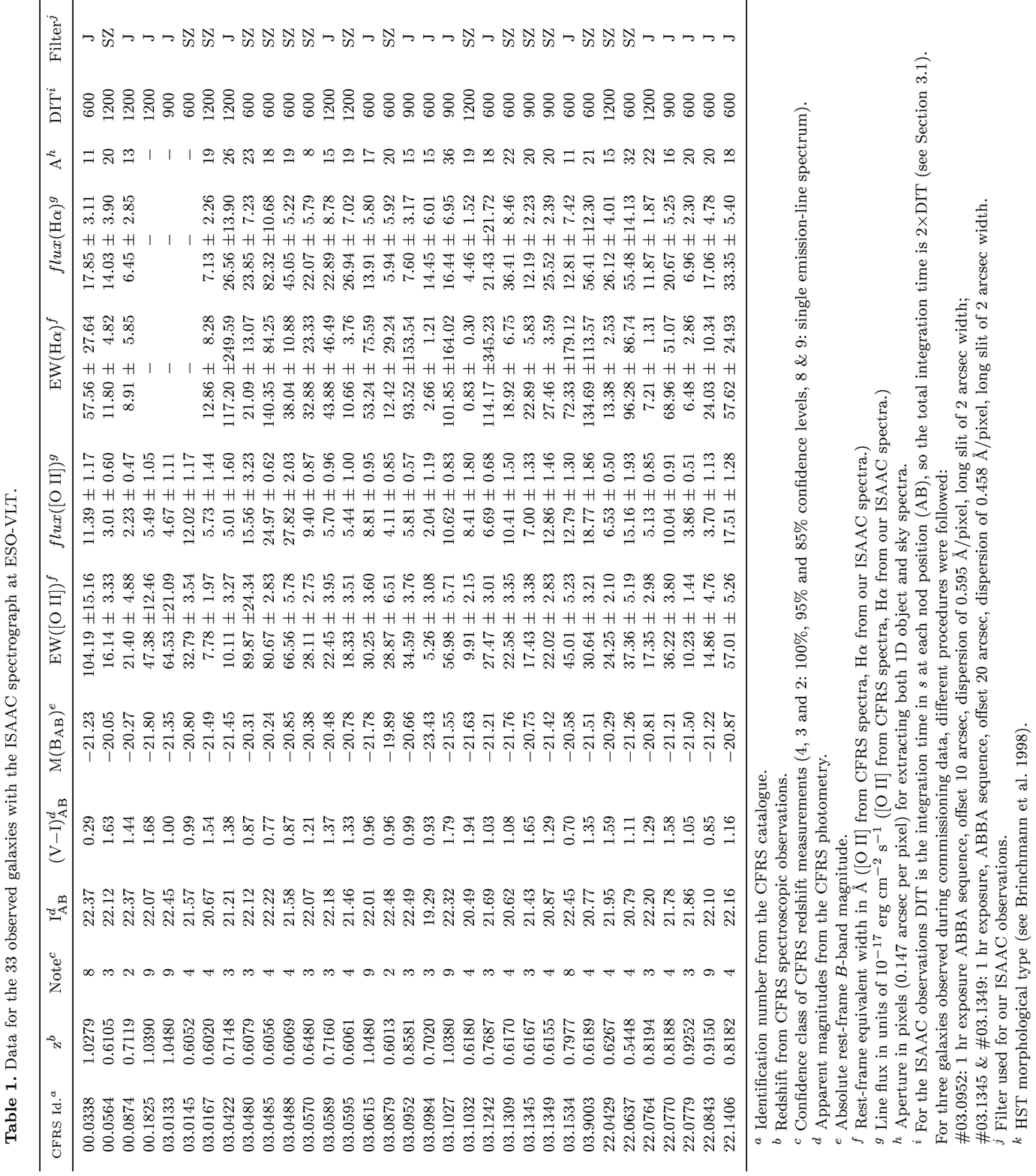,angle=90,width=300mm}
\label{tab1}
\end{table*}
\newcommand{\subpage}[1]{                                               
\begin{minipage}{55mm}{
\psfig{figure=#1,width=55mm,height=44mm}}
\end{minipage}}
\begin{figure*}   
\begin{minipage}{170mm}{                                                               
 \subpage{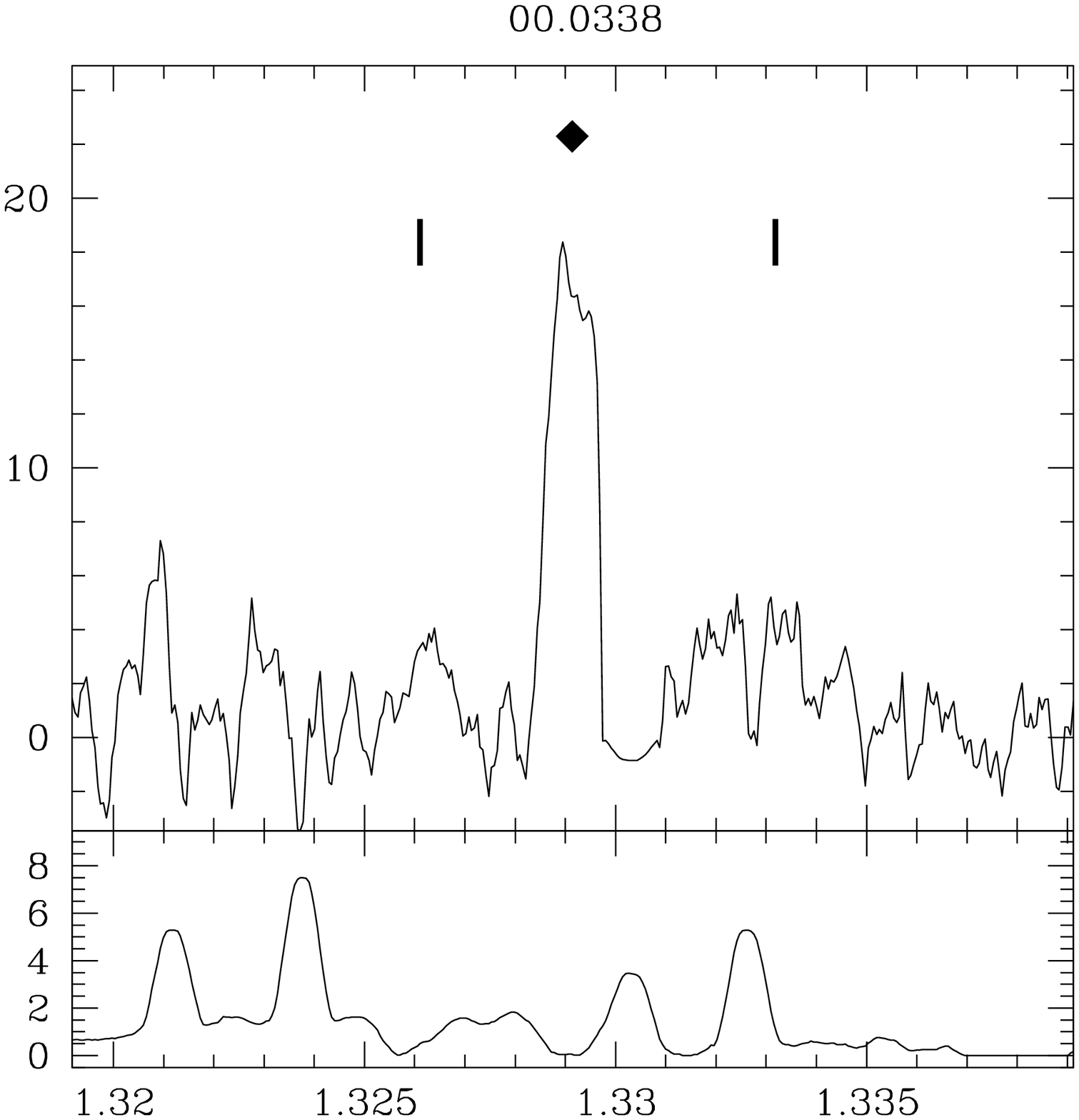}
 \subpage{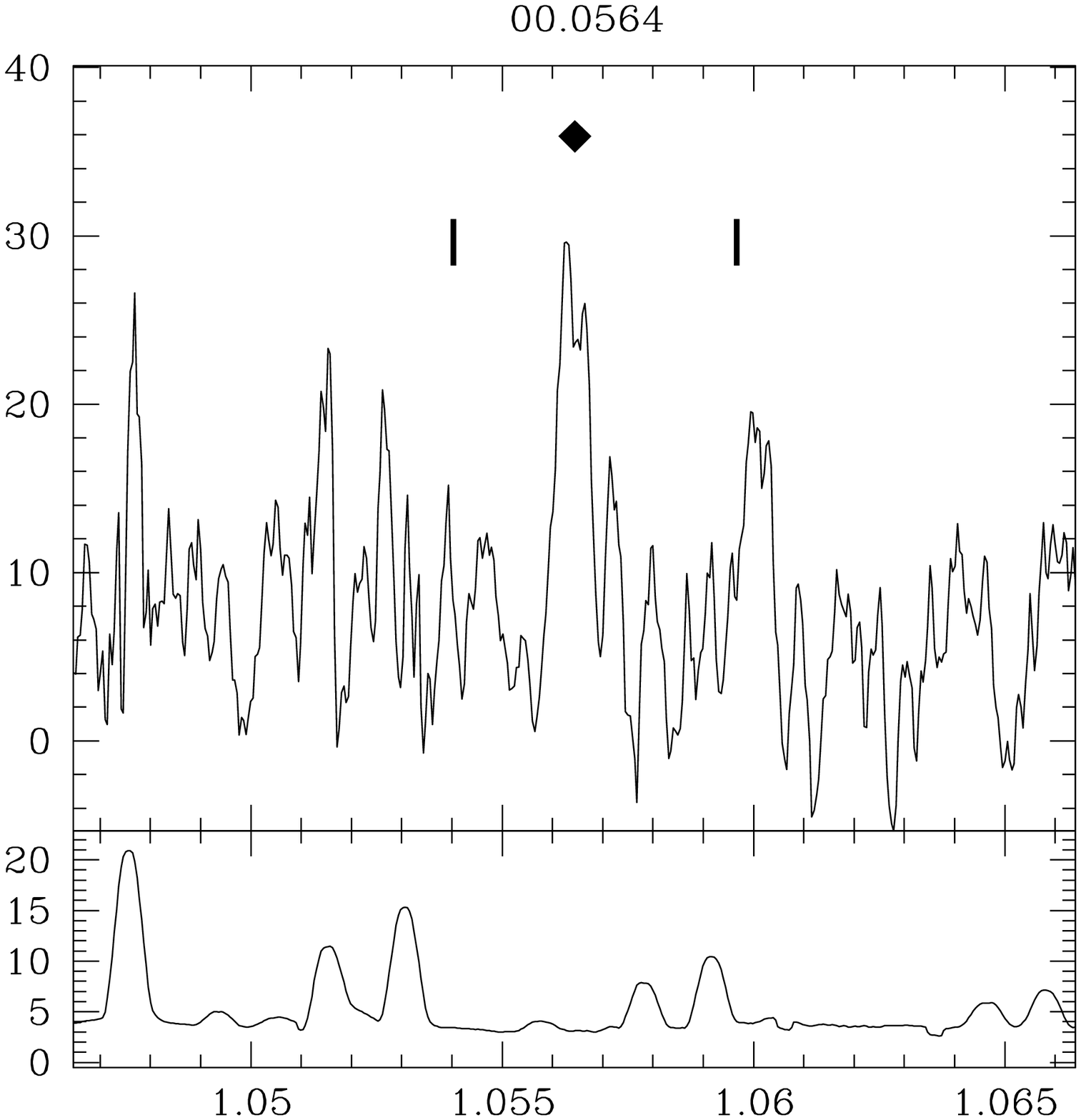}                                                              
 \subpage{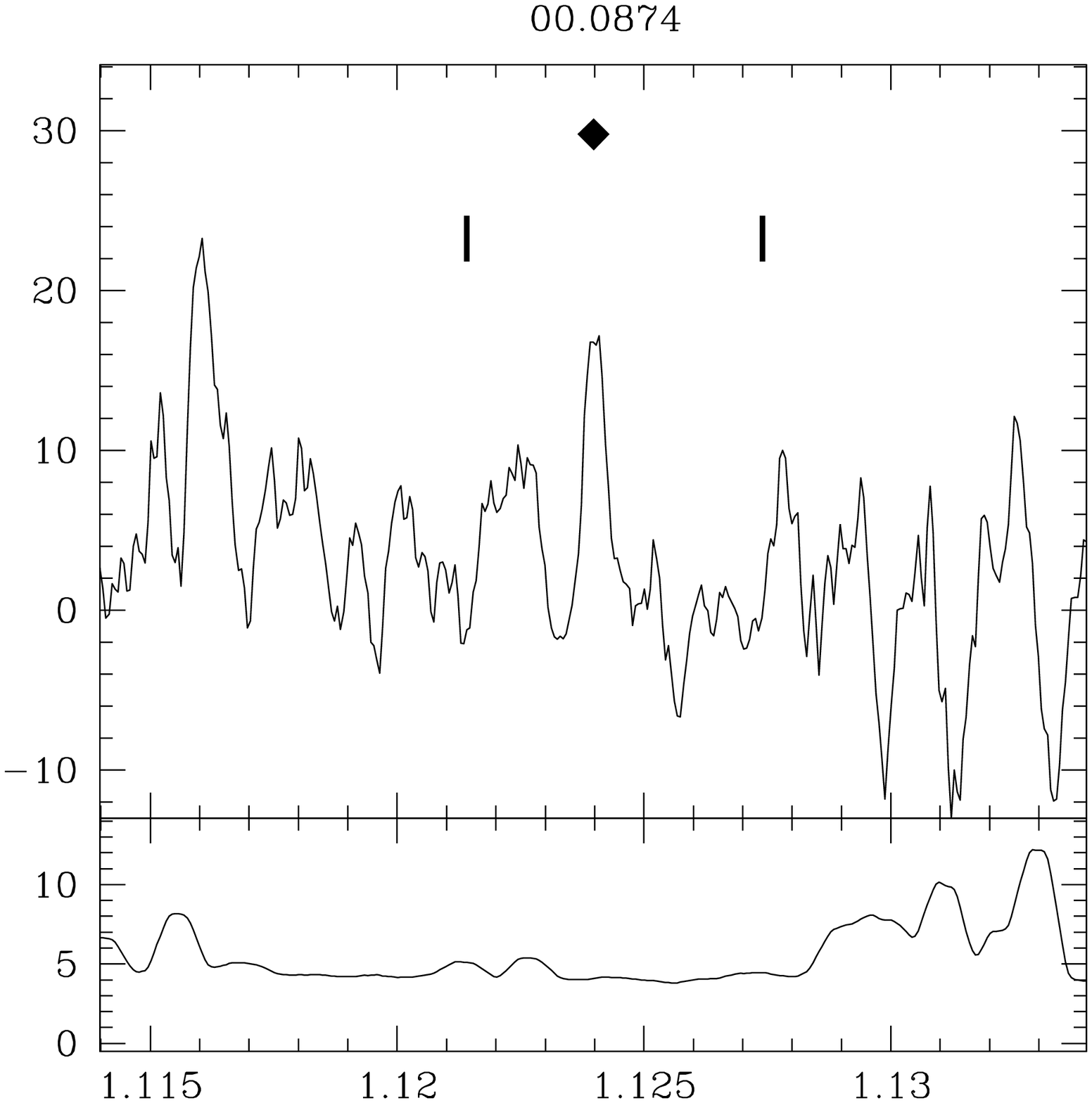} \\                                                           
 \subpage{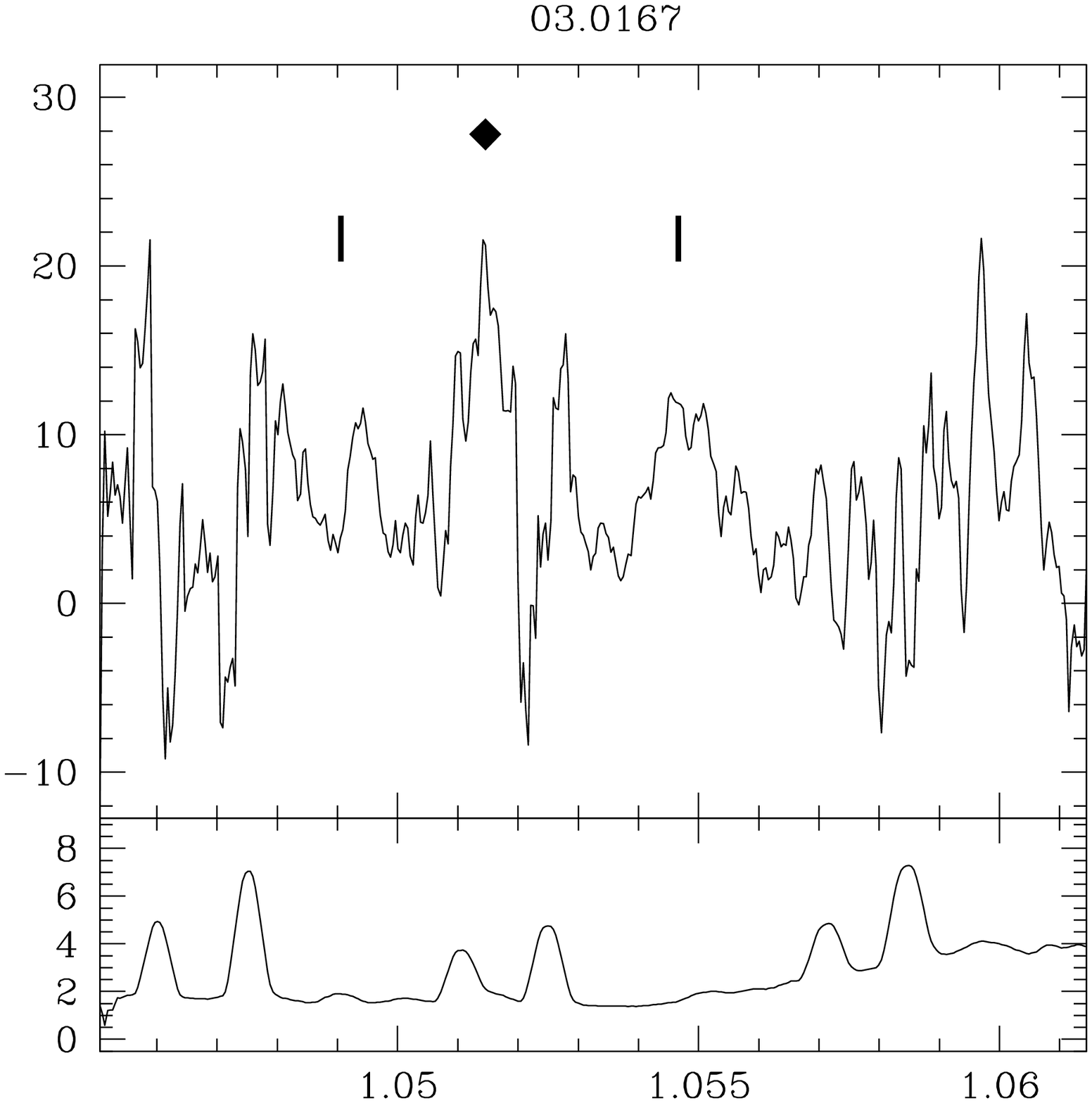}                                                              
 \subpage{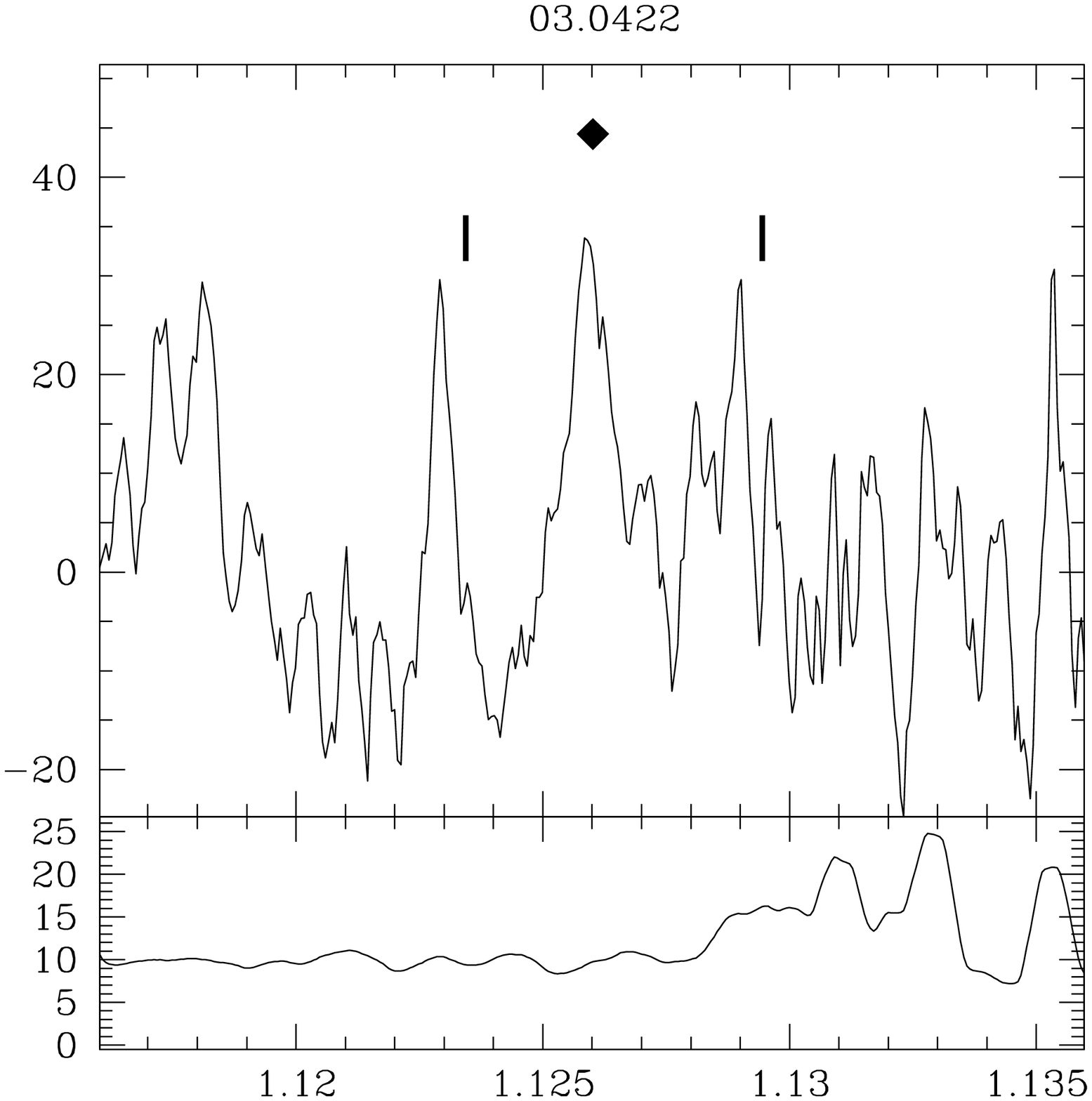}                                                              
 \subpage{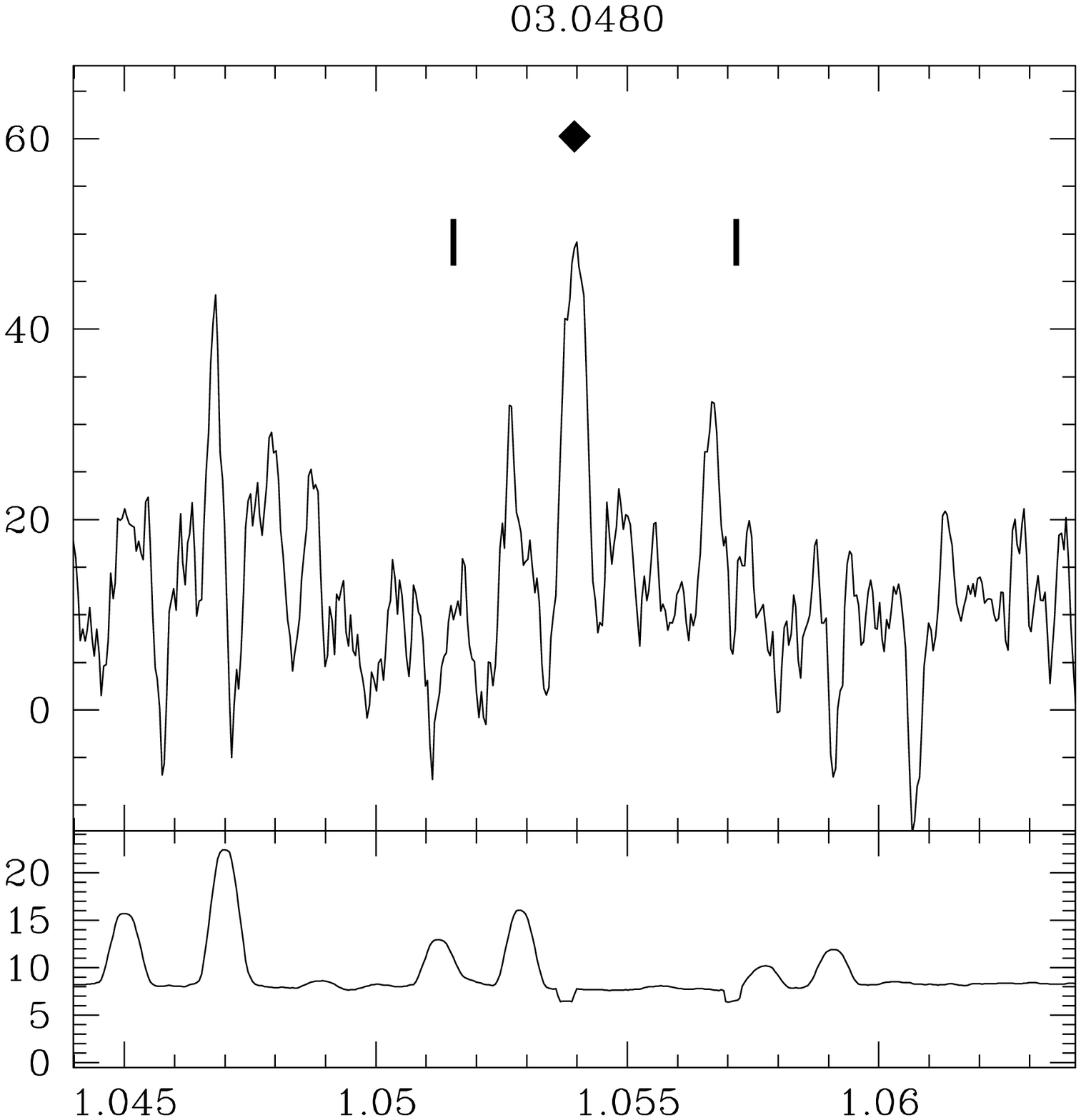} \\                                                           
 \subpage{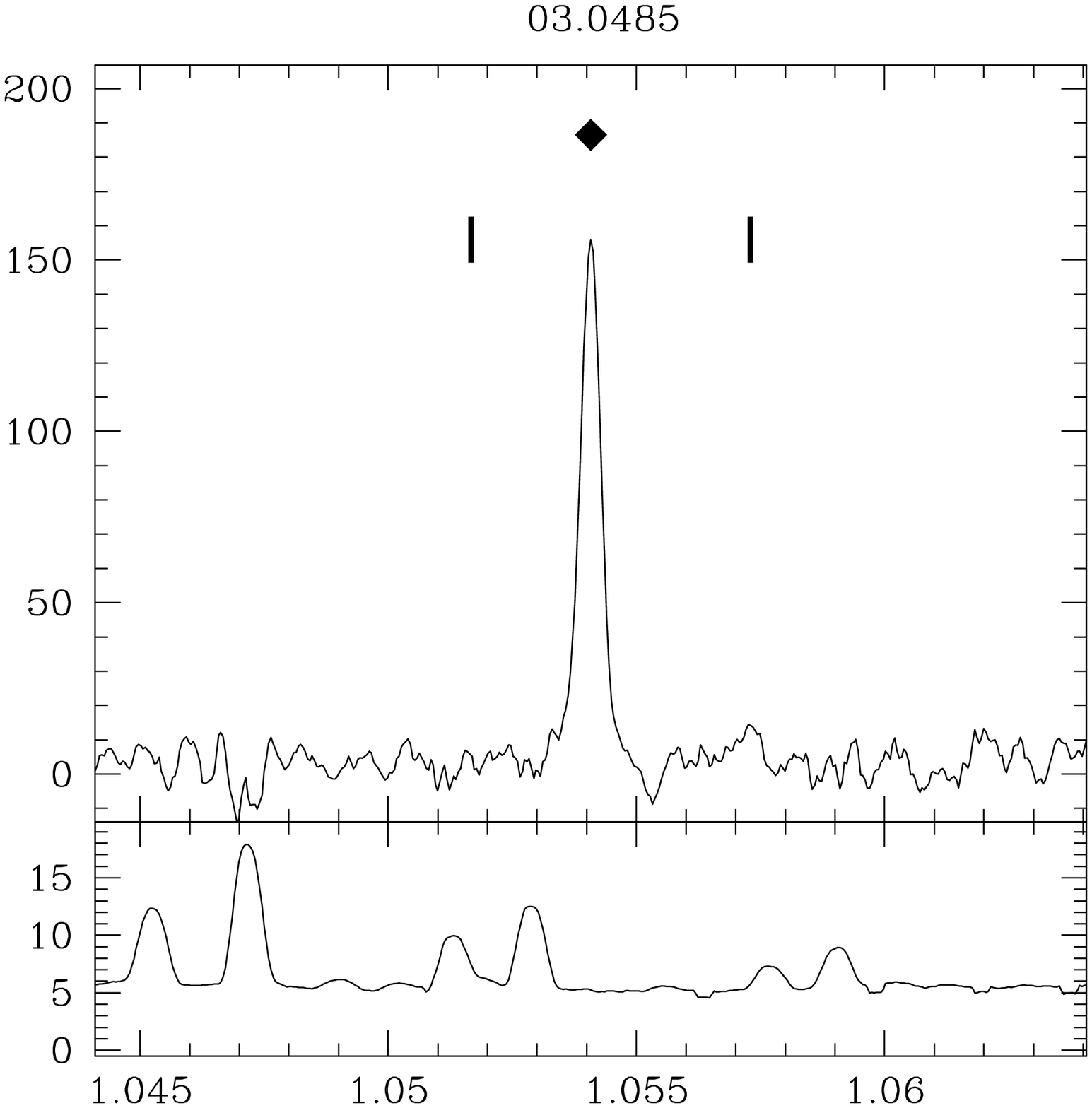}                                                              
 \subpage{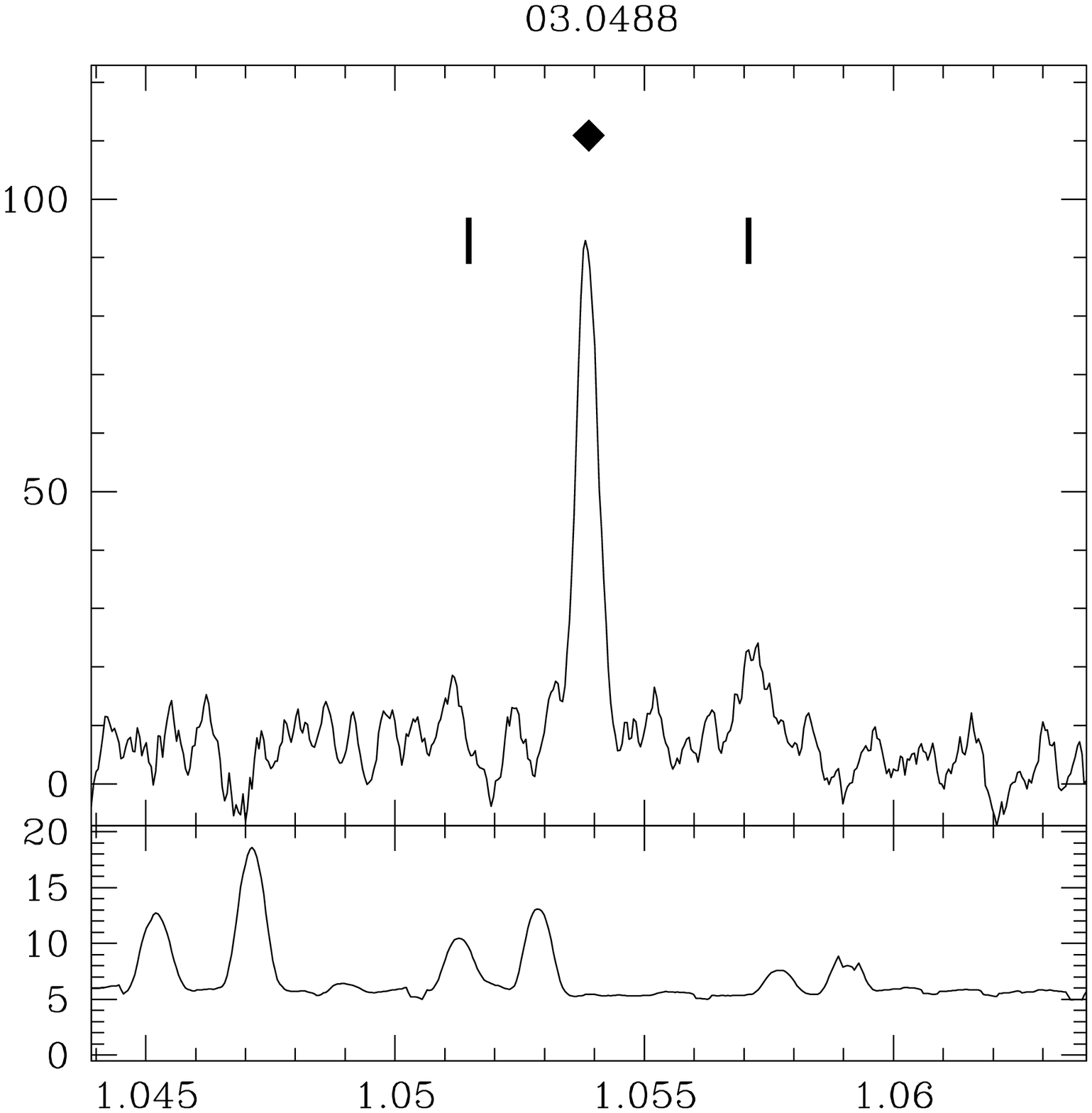}                                                              
 \subpage{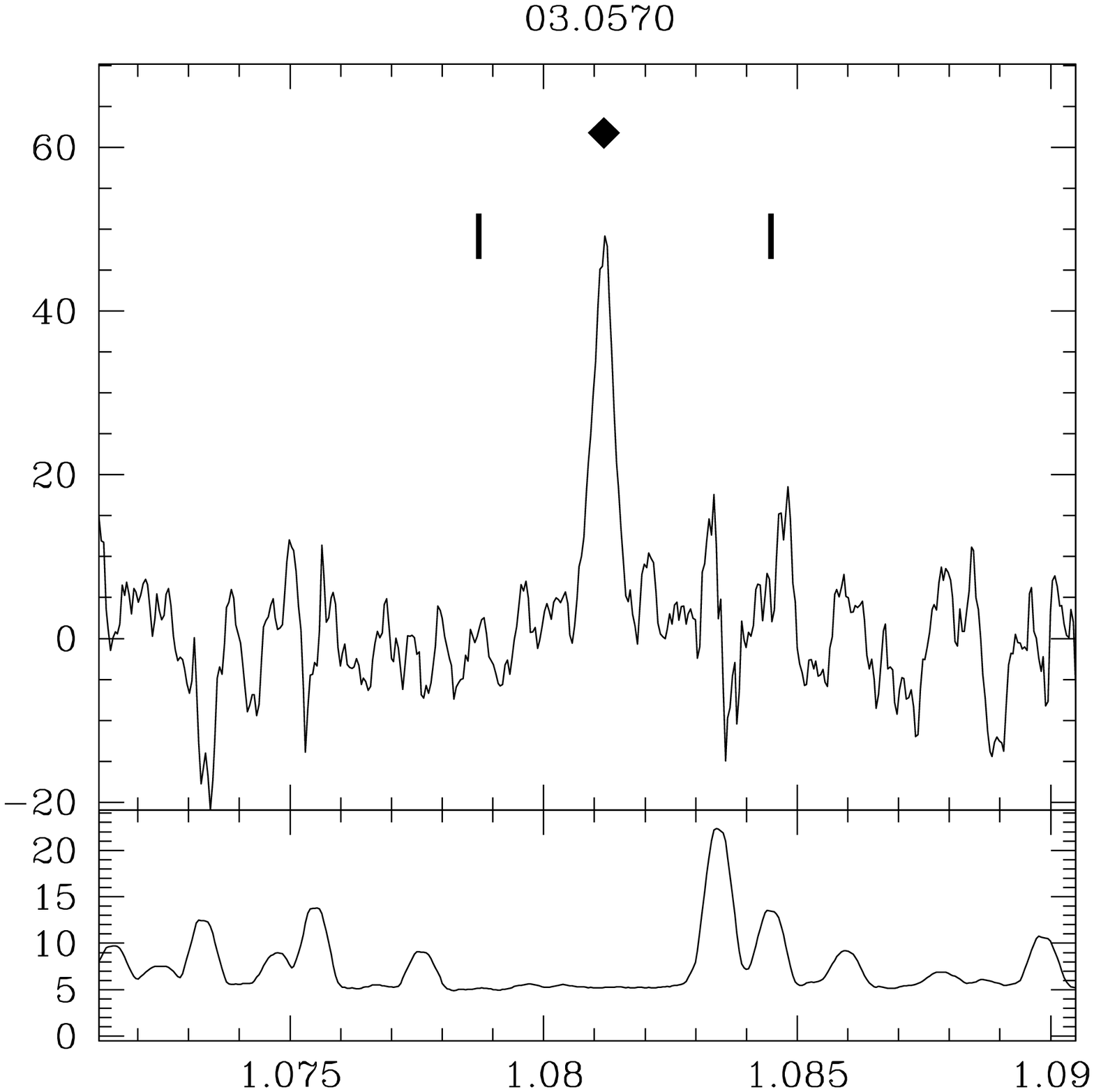} \\                                                           
 \subpage{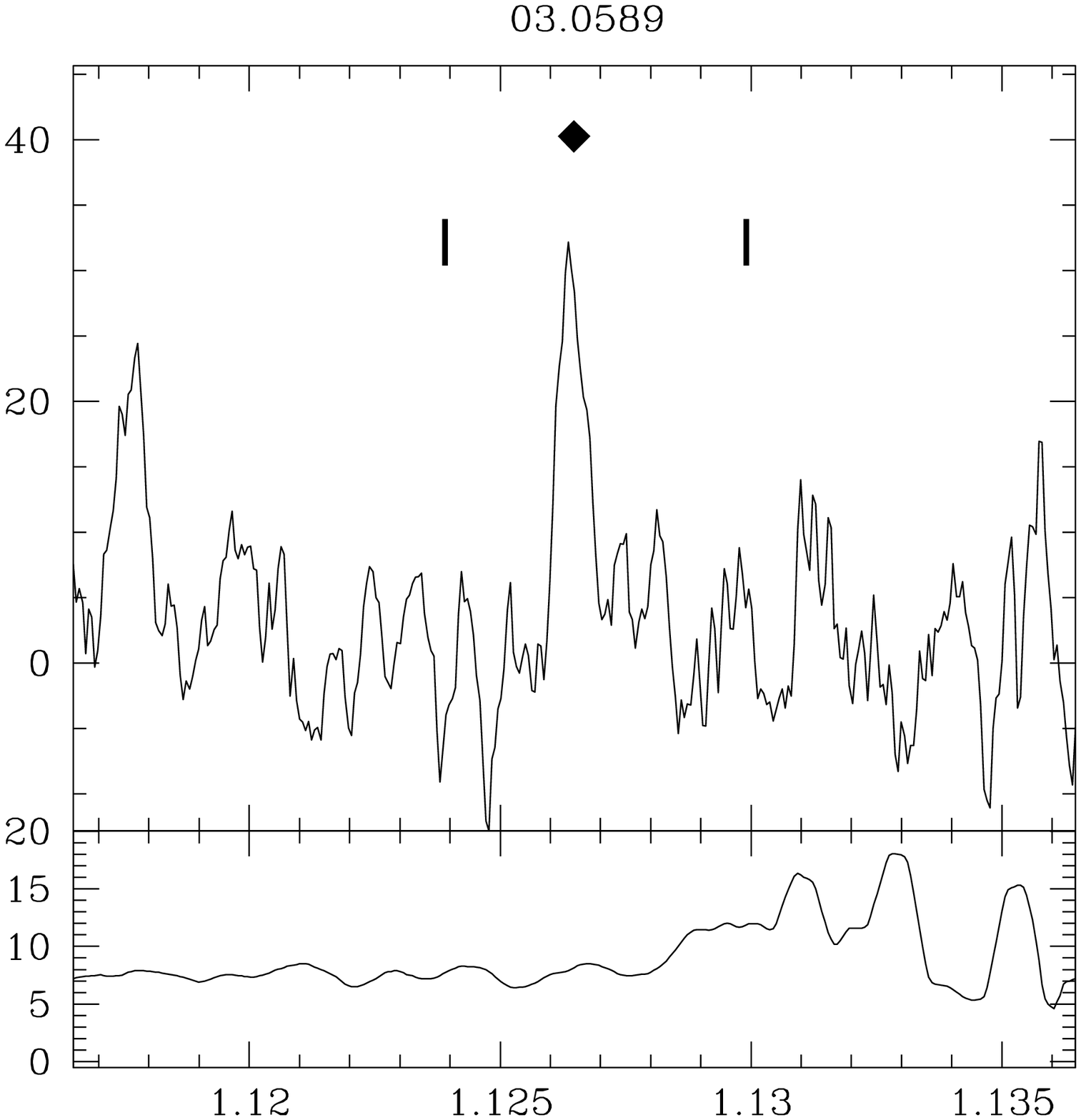}                                                              
 \subpage{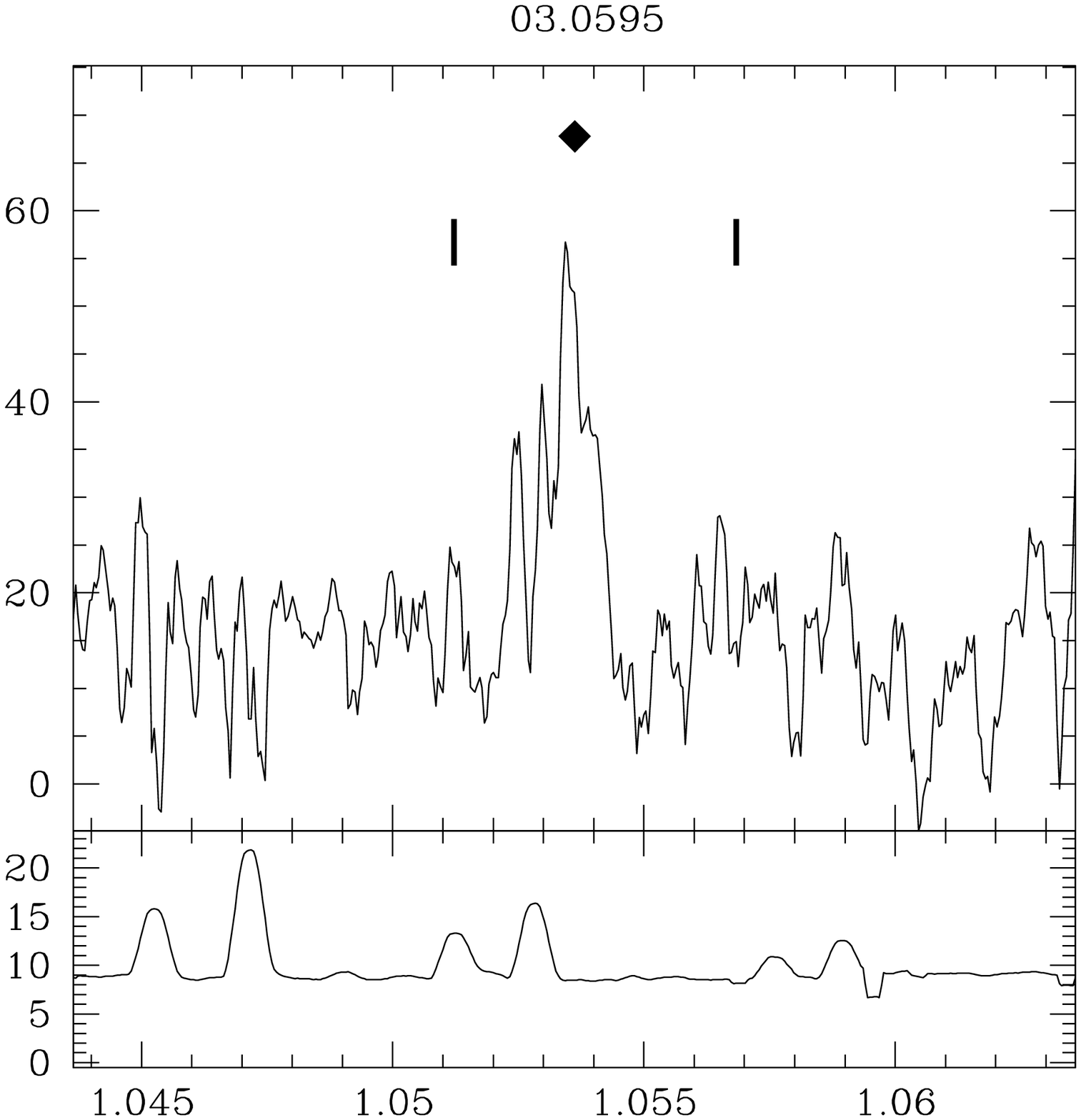}                                                              
 \subpage{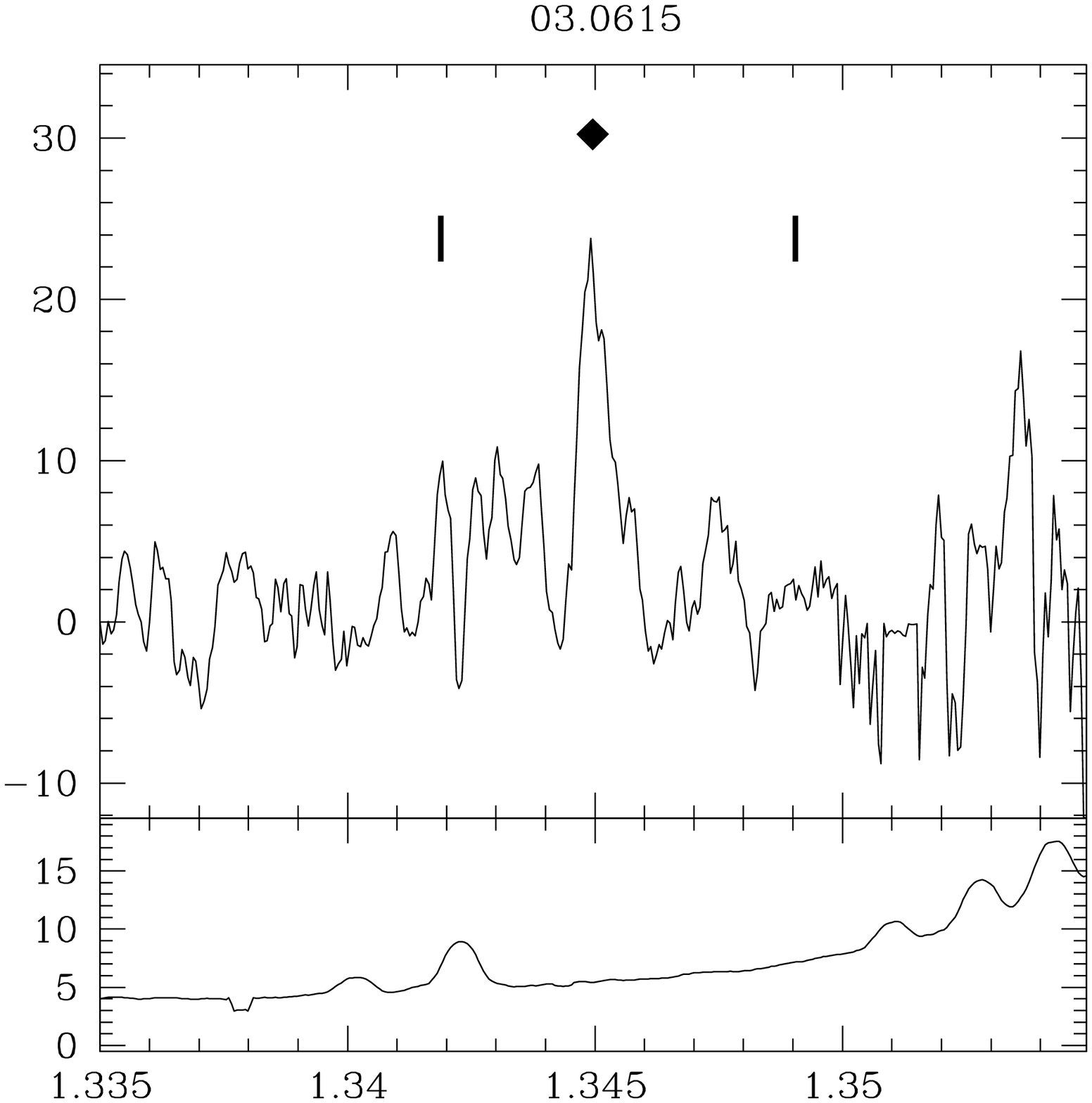} \\                                                           
 \subpage{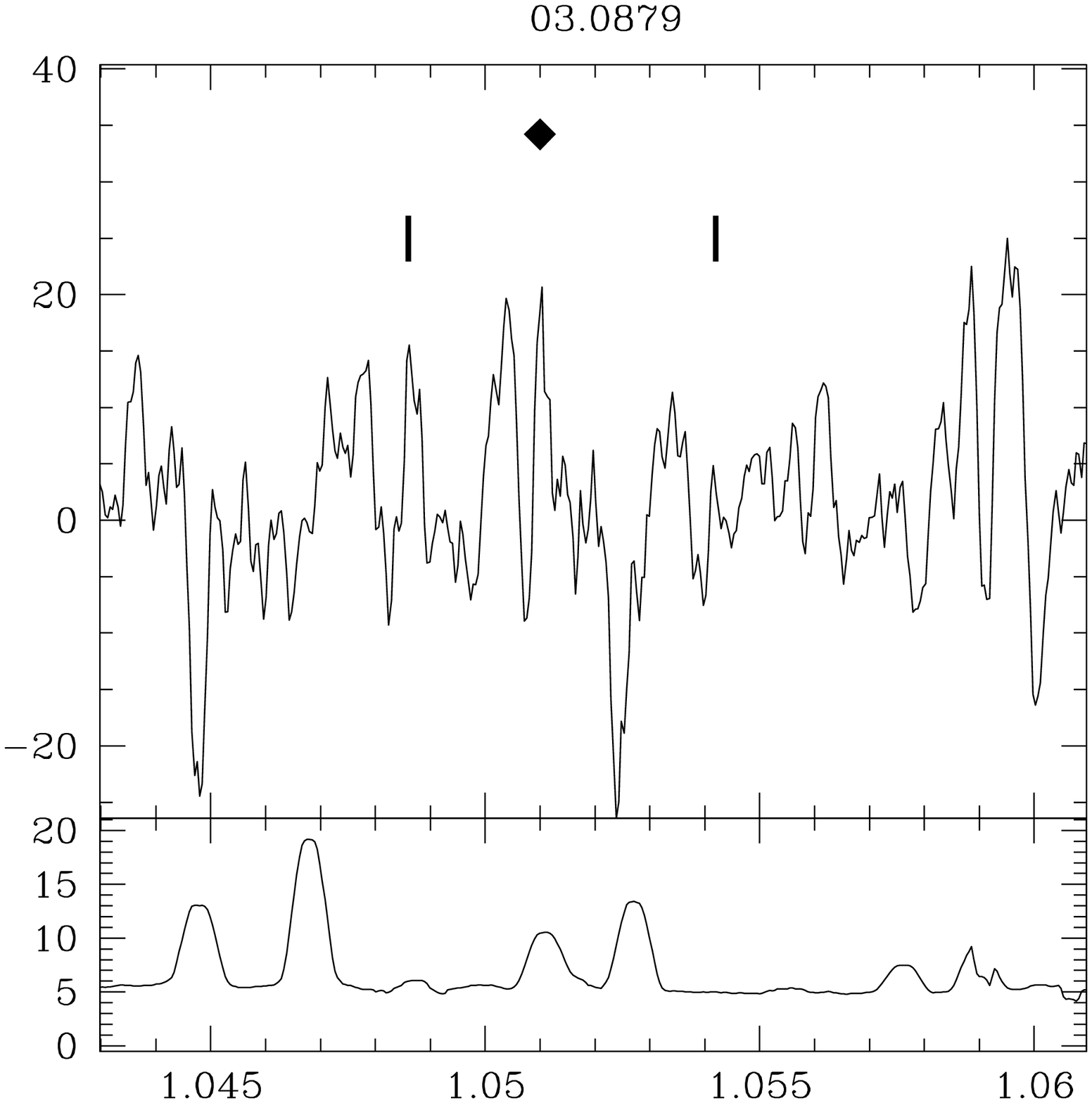}                                                              
 \subpage{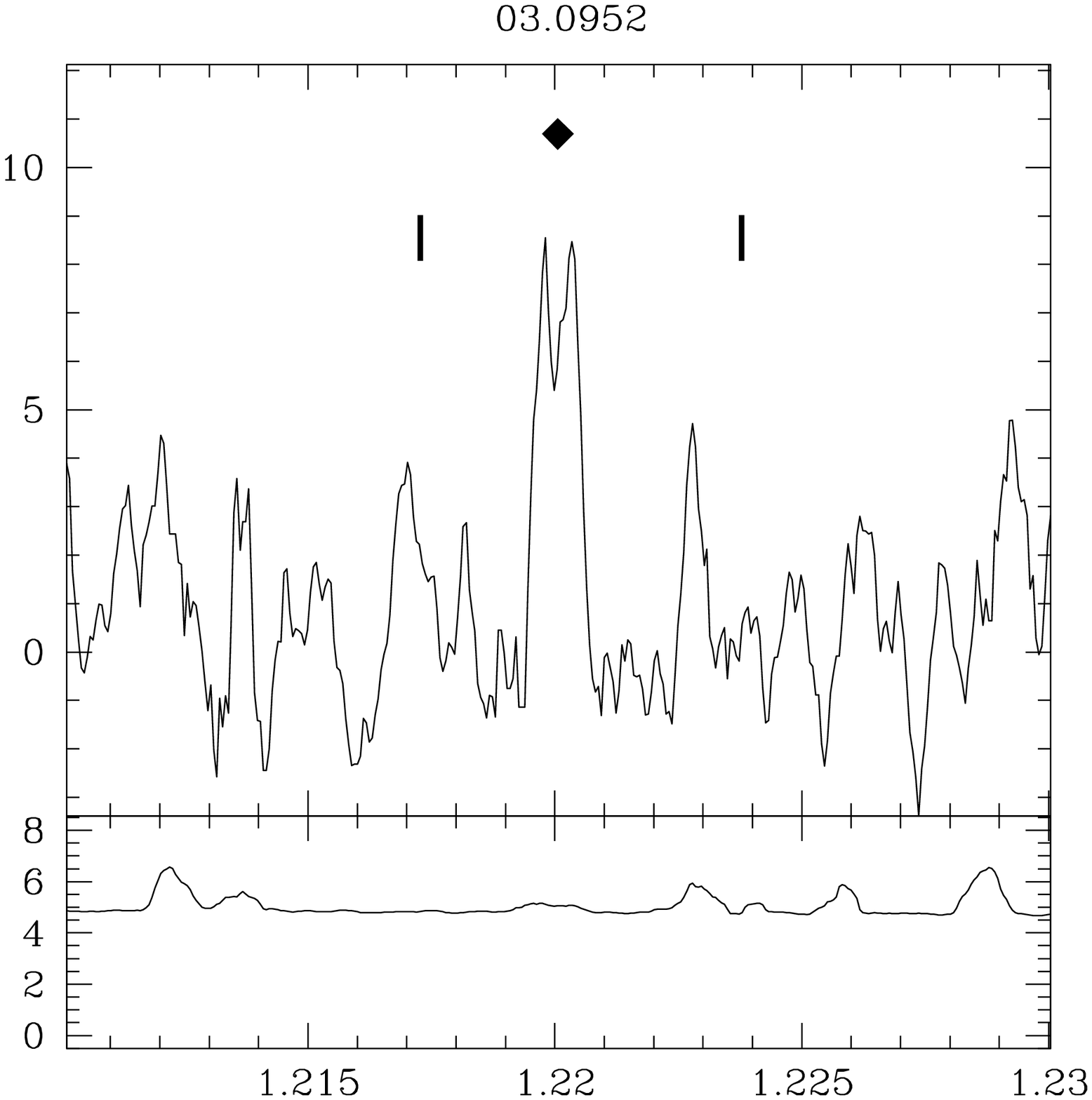}                                                              
 \subpage{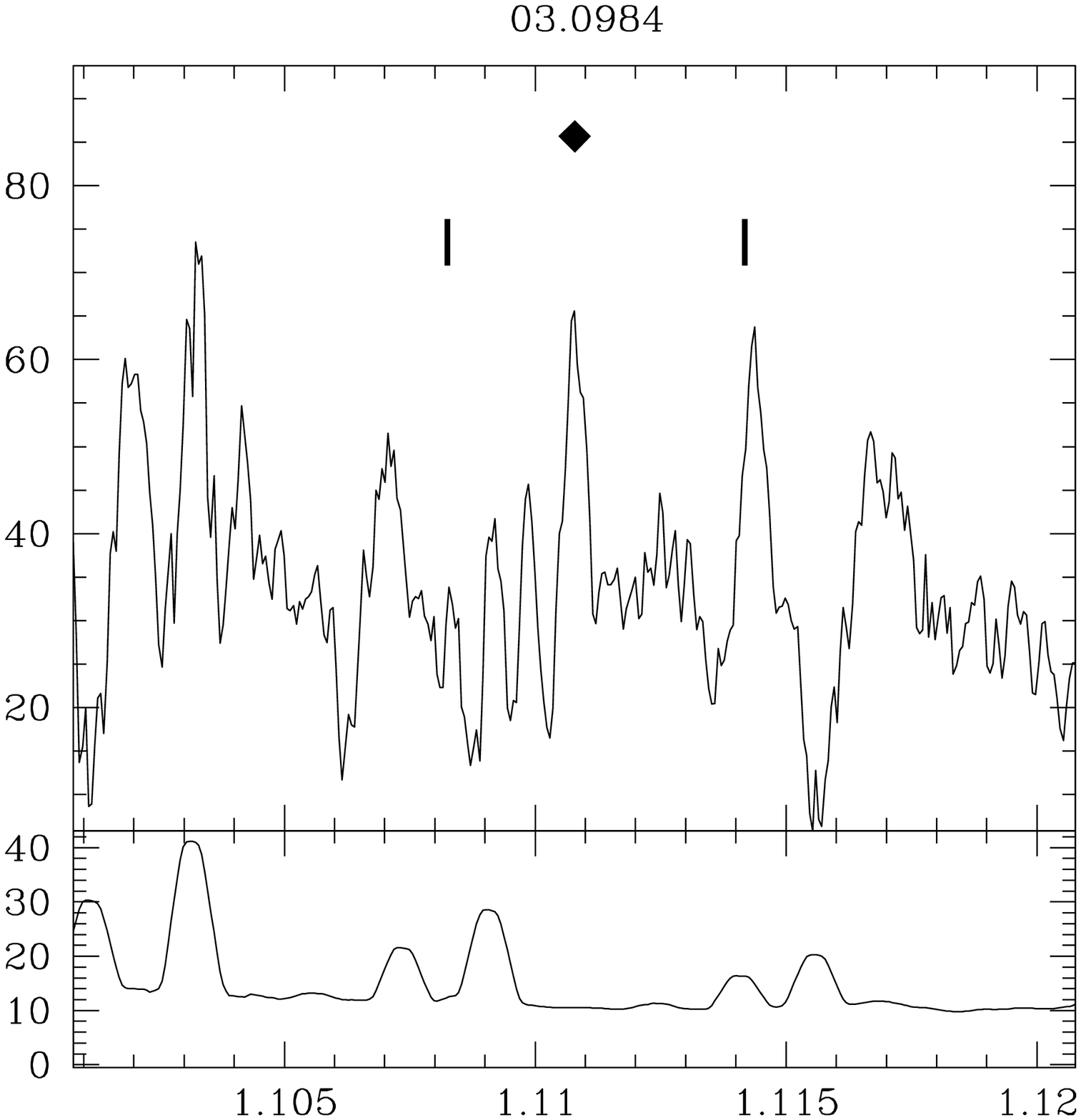}                                                              
}                                                                                      
\end{minipage}                          
\caption{
  The 30 H$\alpha$ emission-line observed-frame spectra between
  $z=0.5$ and $1.1$, observed with the ISAAC spectrograph at the
  ESO-VLT (top panel). In the plot, the spectra have been smoothed
  using a 7 pixel boxcar filter, the X-axis is the observed wavelength
  in $\mu$m, the Y-axis is the flux in units of 10$^{-18}$ erg
  s$^{-1}$ cm$^{-2}$, the diamond indicates the position of the
  detected H$\alpha$~$\lambda$6563 line, the vertical bars indicate
  the positions of the [\ion{N}{2}]$\lambda$6548 and
  [\ion{N}{2}]$\lambda$6583 lines. The the sky noise level is shown in
  the bottom panel. In each case, both the object and the sky have
  been extracted with the same aperture.  The CFRS identification is
  written at the top of each window (see also Table~\ref{tab1}).
\label{fig2}}
\end{figure*}
\setcounter{figure}{1}
\begin{figure*}   
\begin{minipage}{170mm}{                                                               
 \subpage{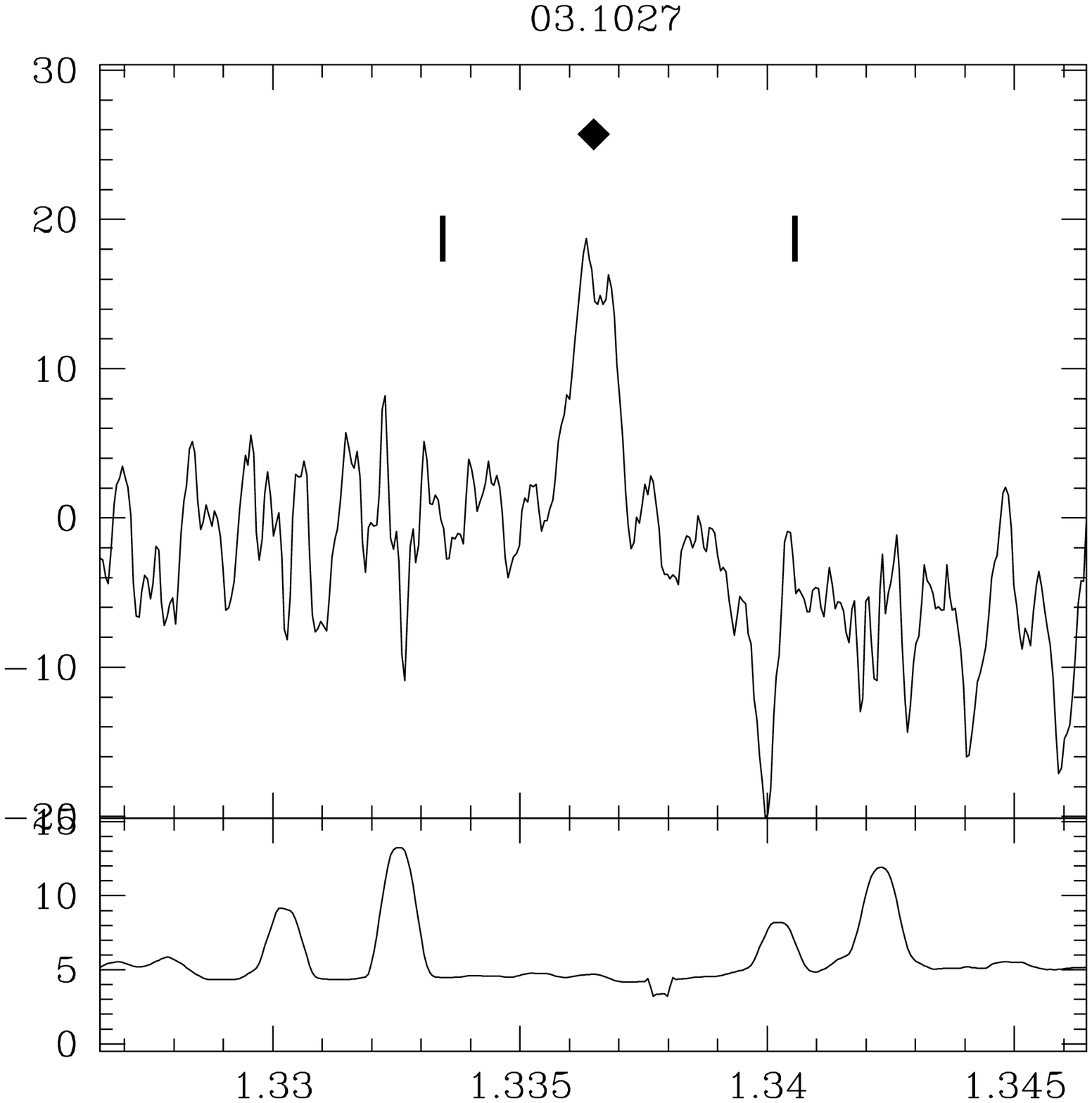}                                                              
 \subpage{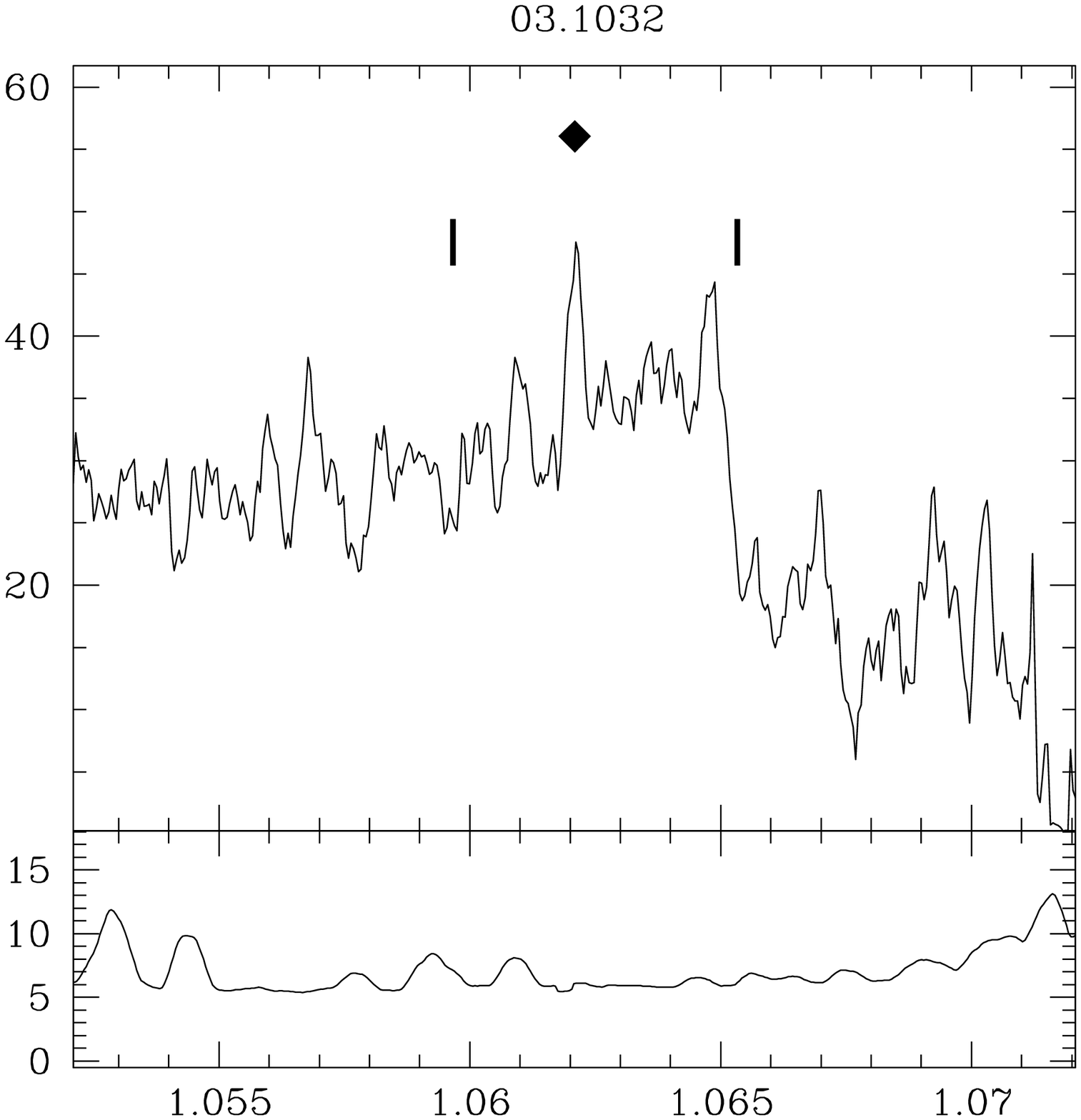}                                                              
 \subpage{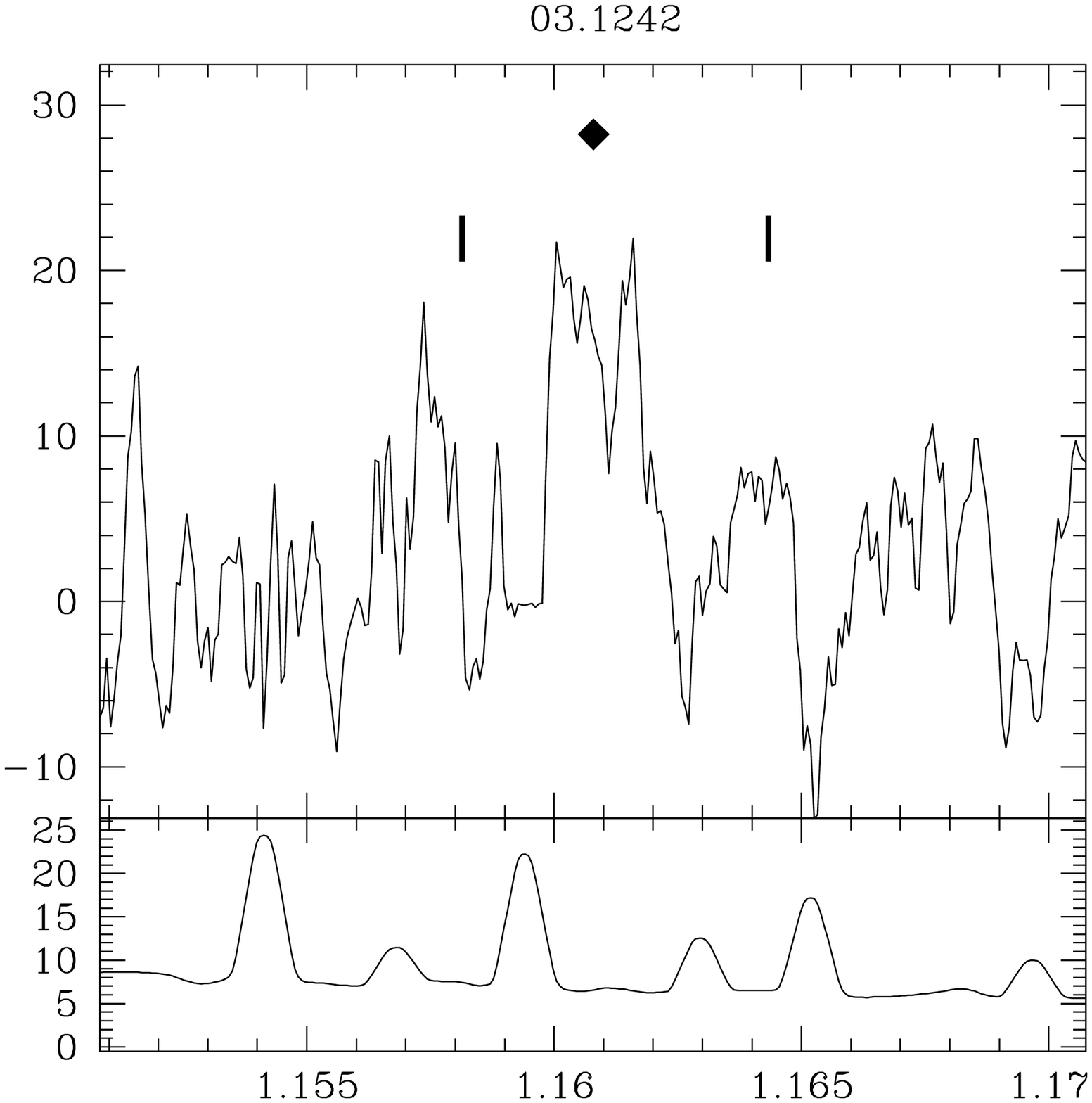} \\                                                           
 \subpage{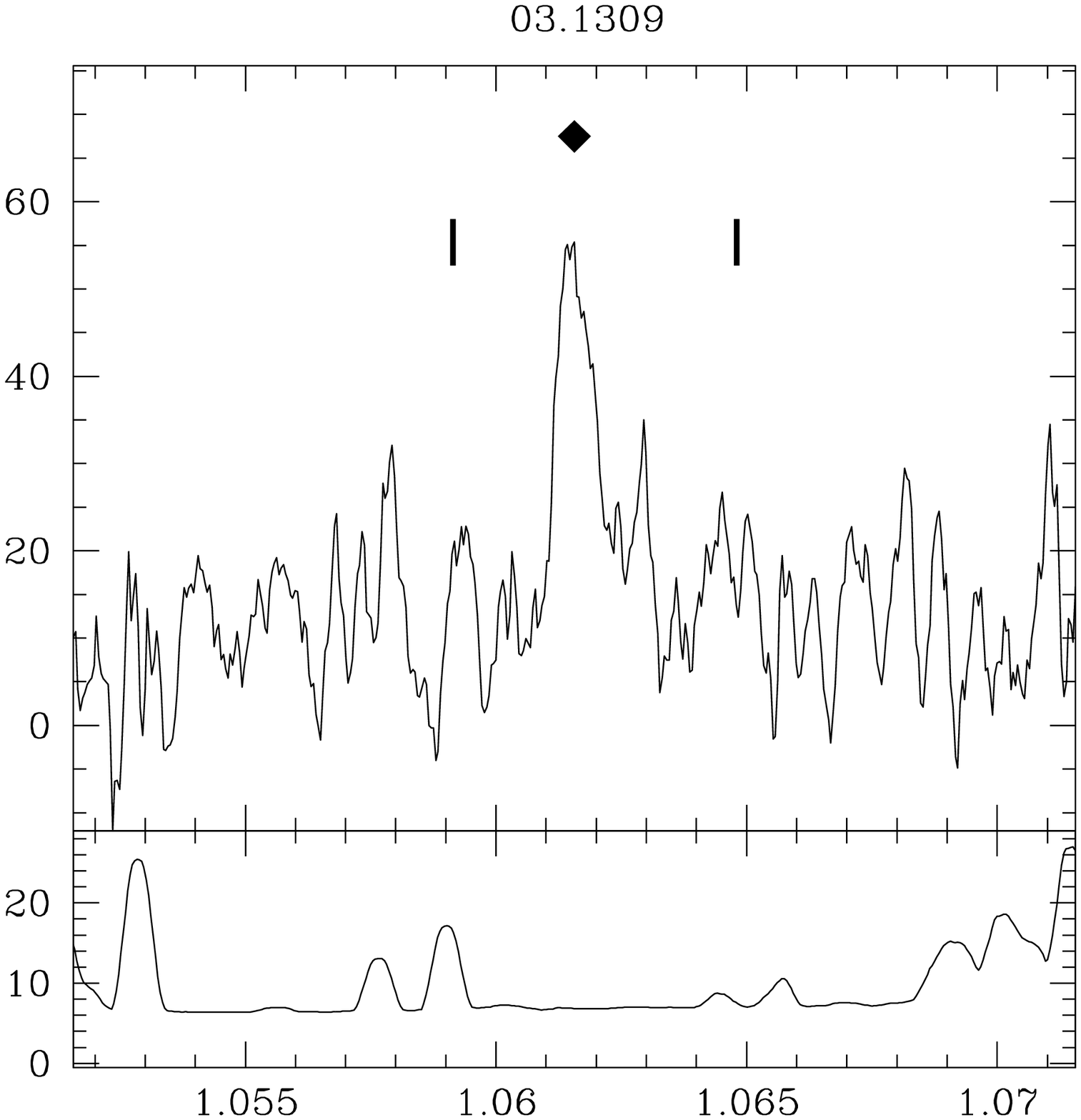}                                                              
 \subpage{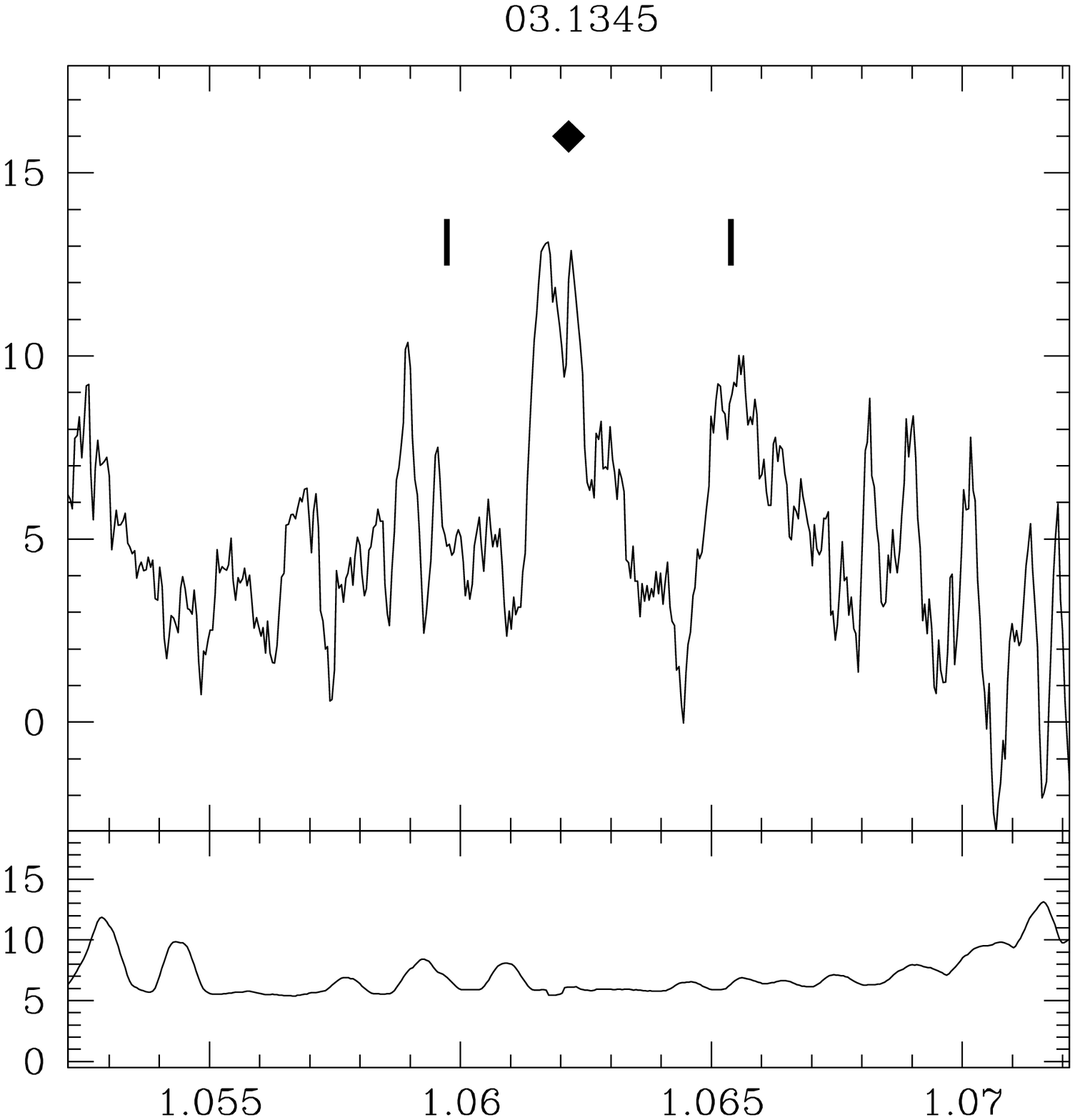}                                                              
 \subpage{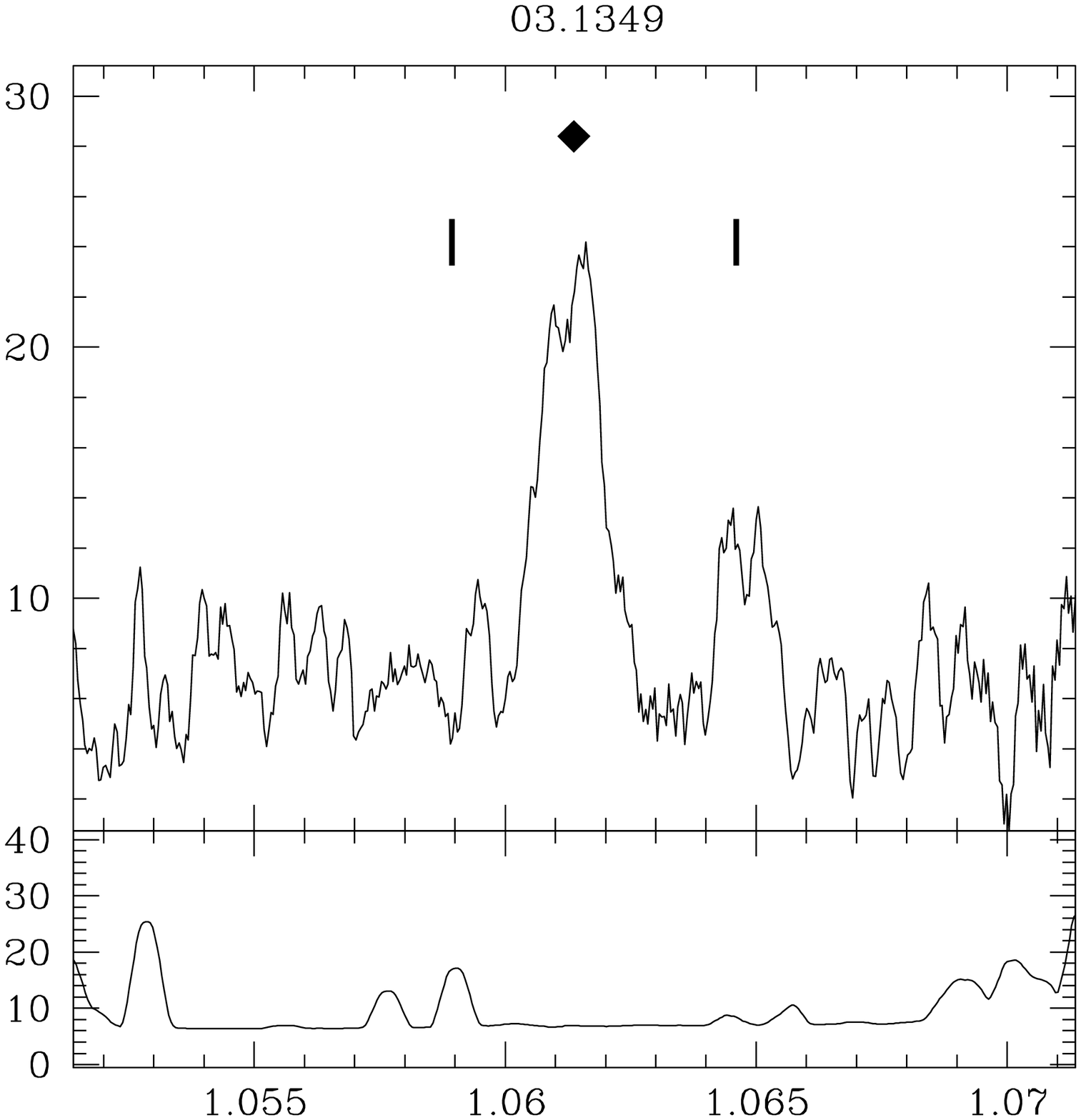} \\                                                           
 \subpage{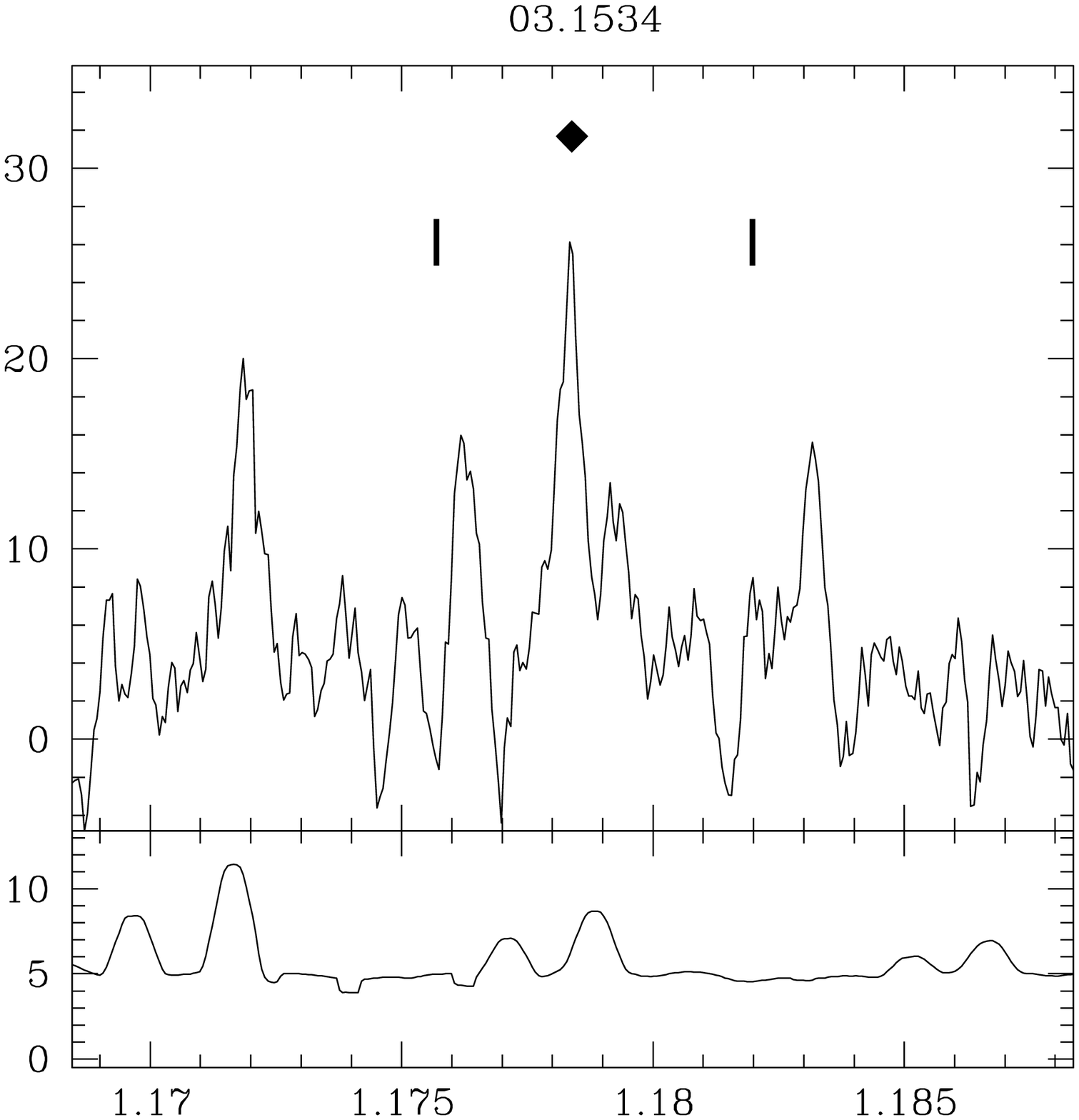}                                                              
 \subpage{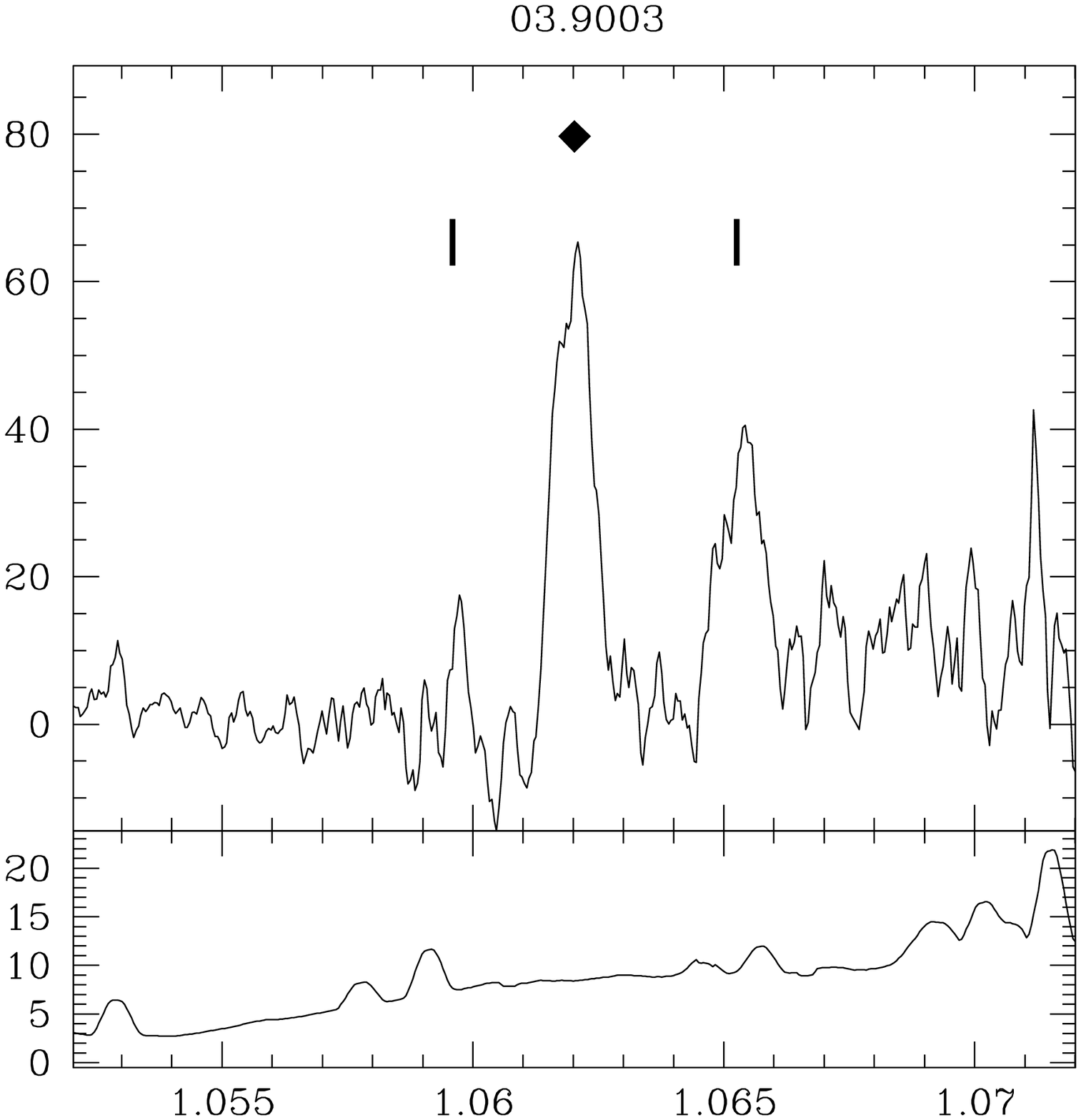}                                                              
 \subpage{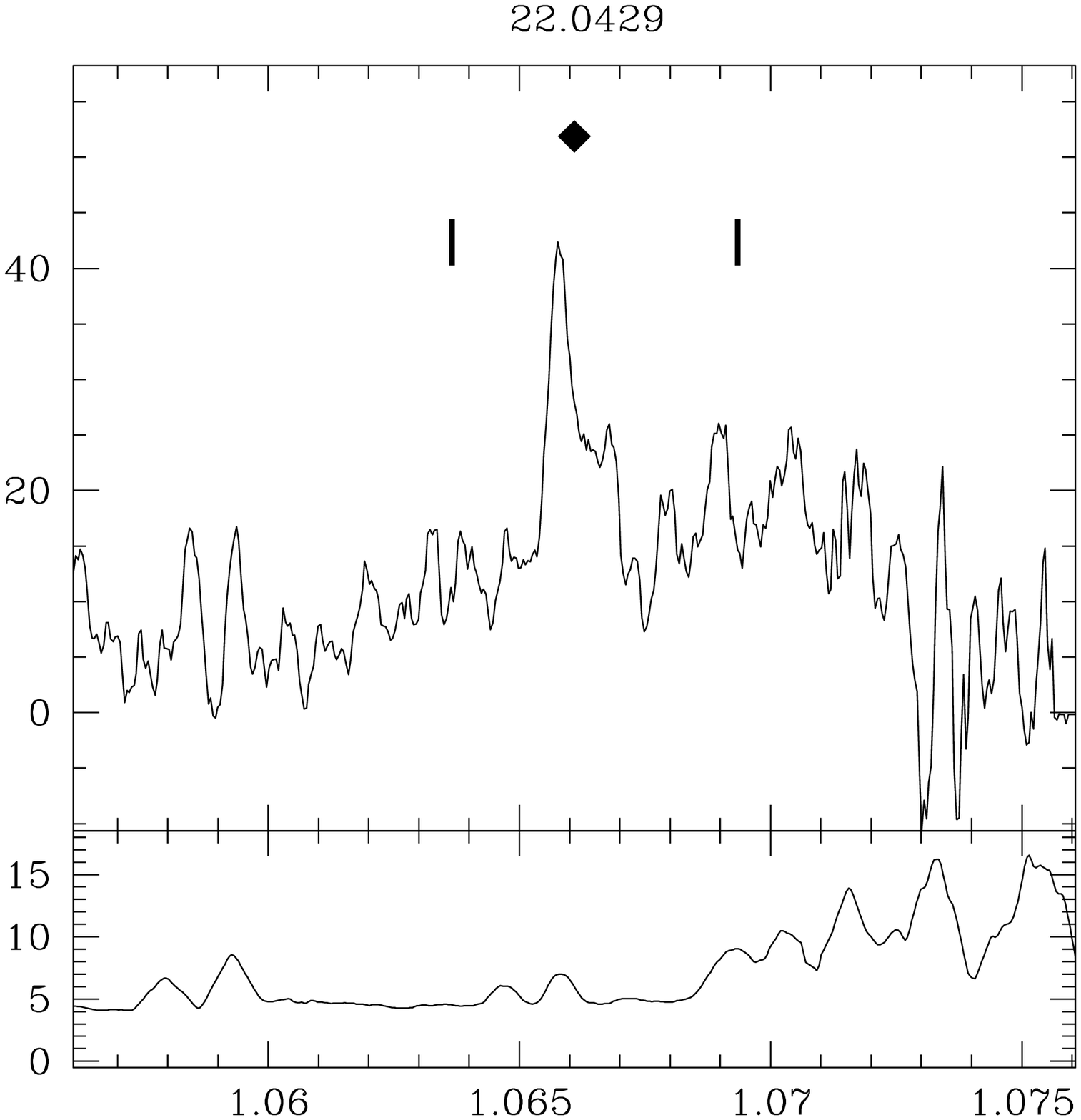} \\                                                           
 \subpage{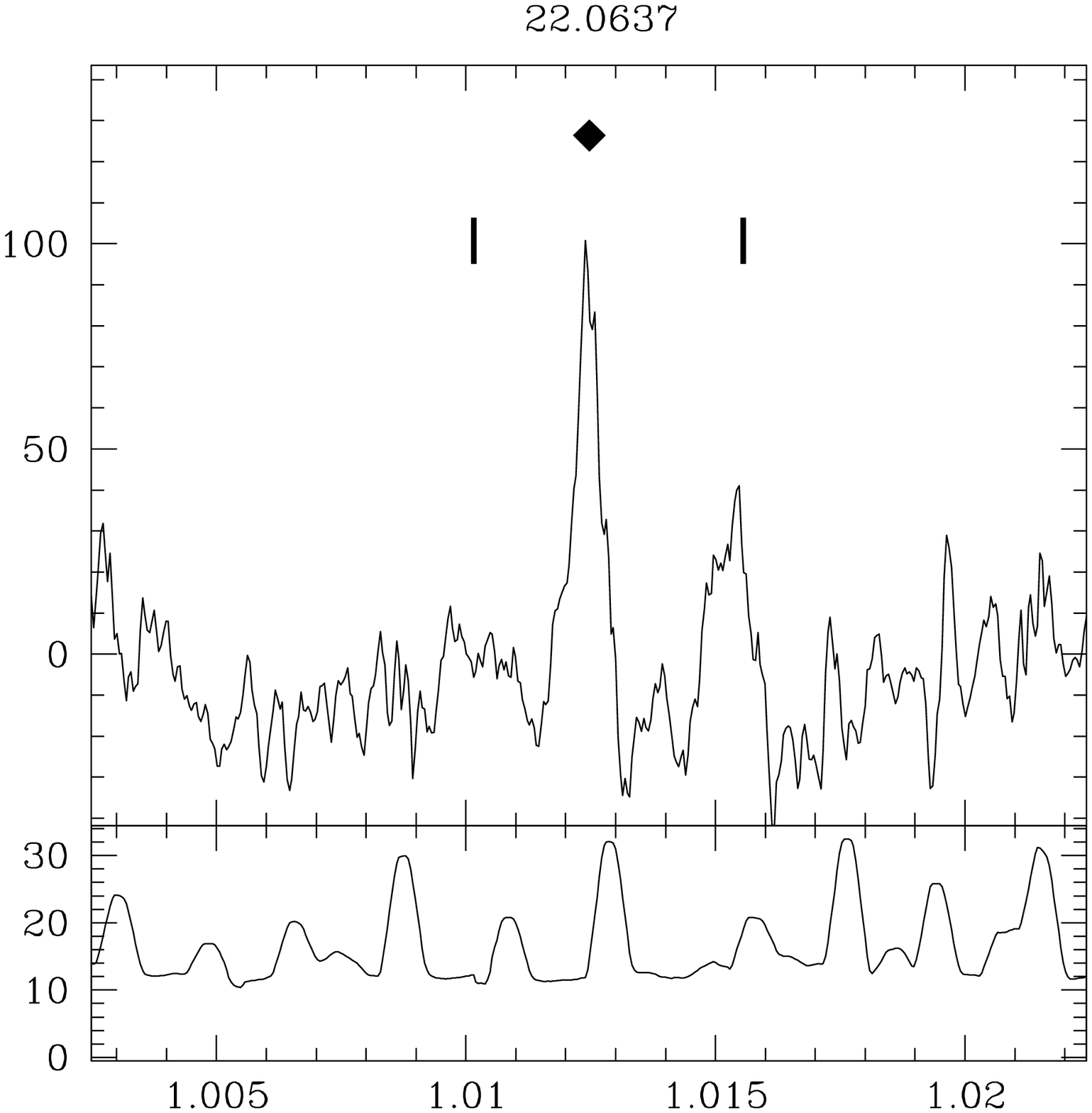}                                                              
 \subpage{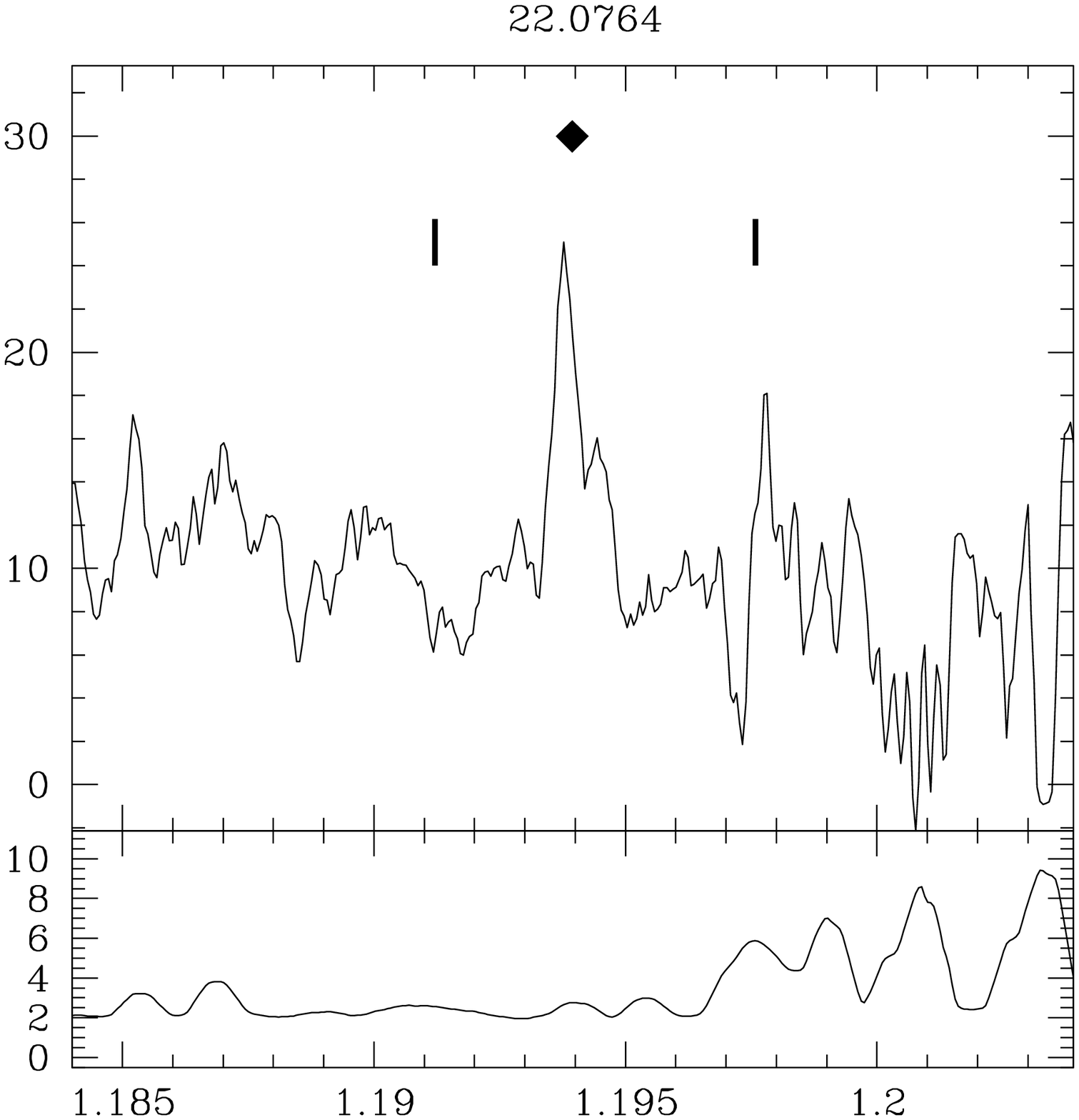}                                                              
 \subpage{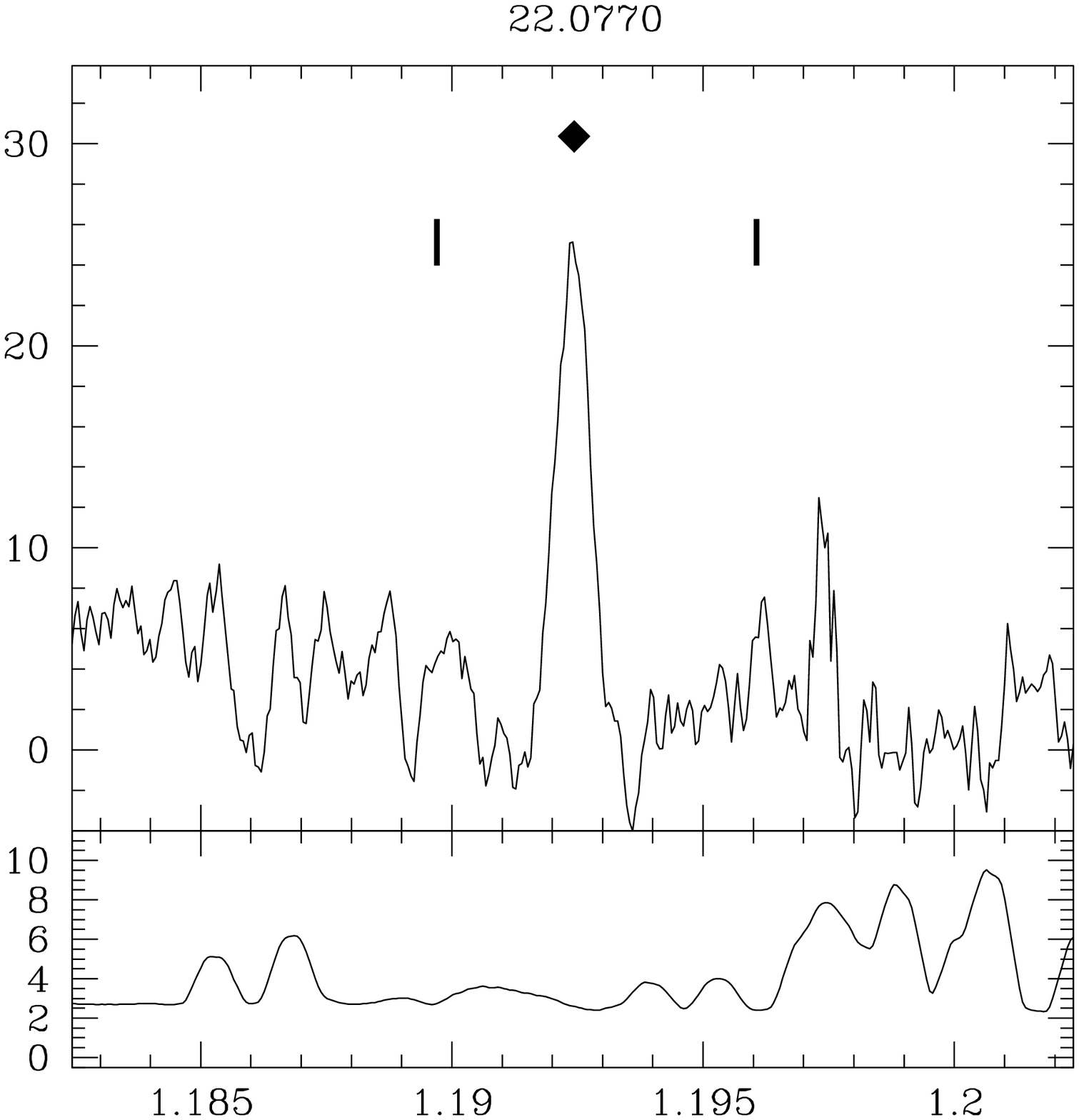} \\                                                           
 \subpage{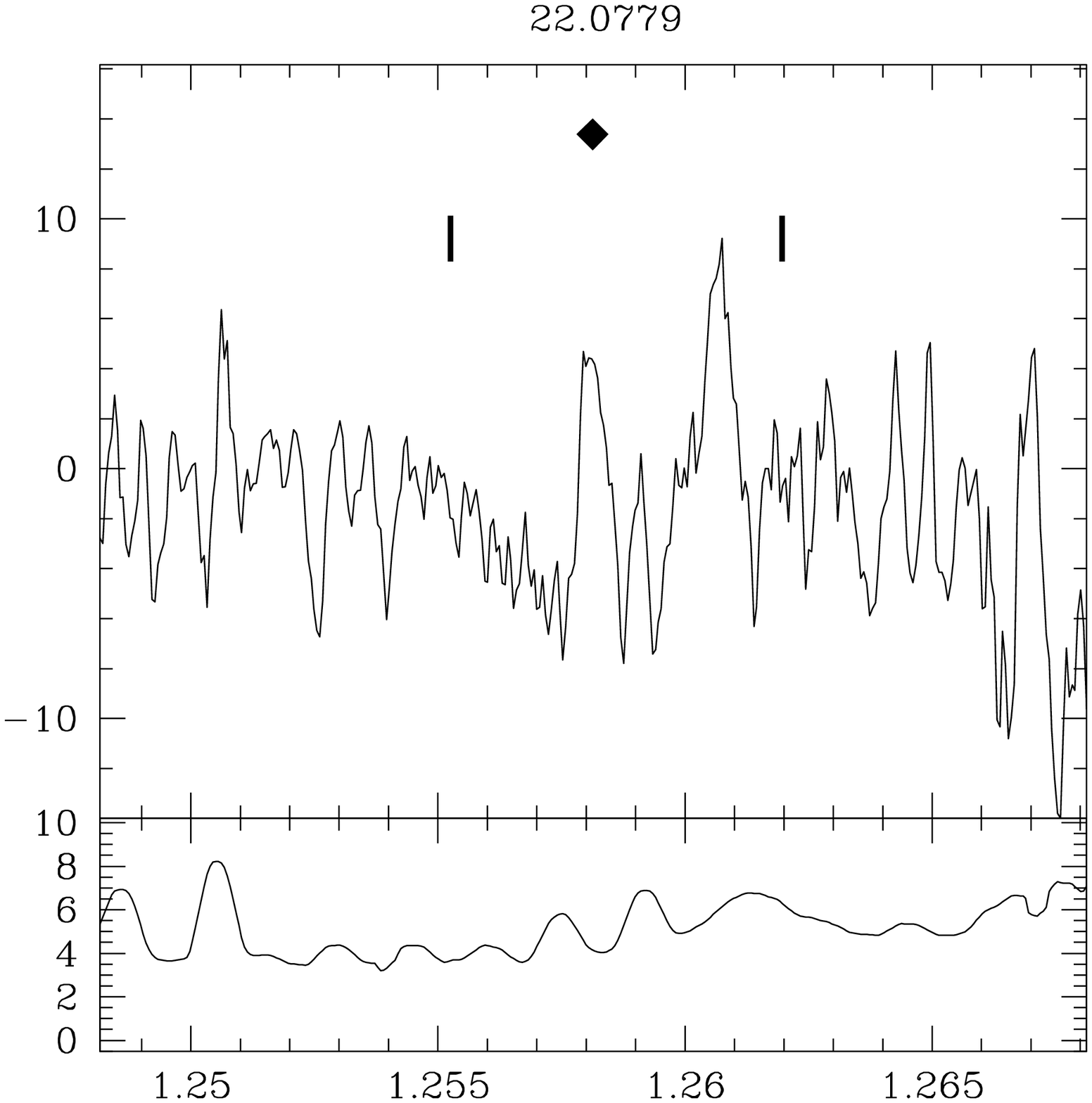}                                                              
 \subpage{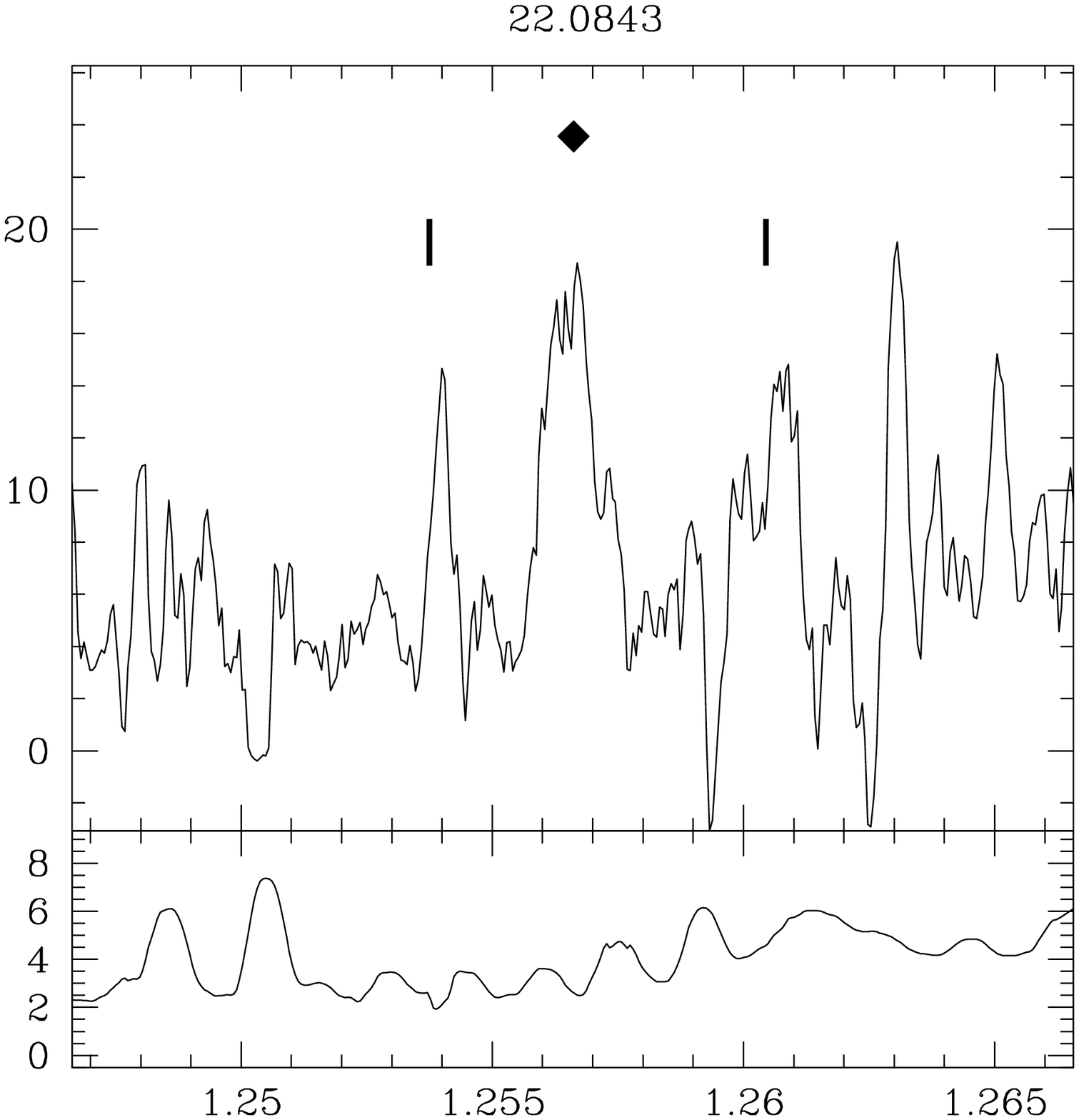}                                                              
 \subpage{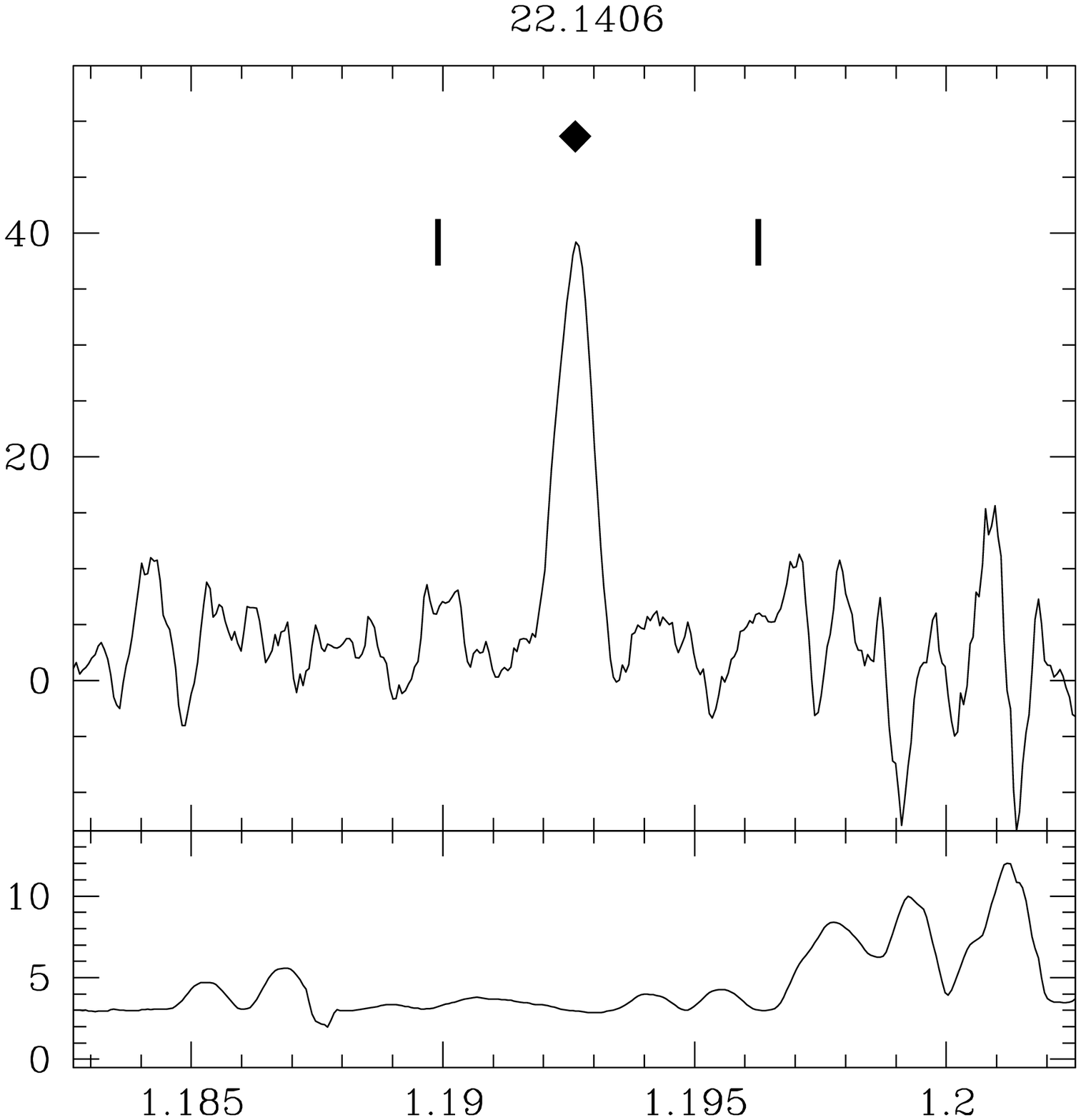}                                                              
}                                                                                      
\end{minipage}                          
\caption{Suite.}
\end{figure*}
\begin{figure}
\centerline{\psfig{figure=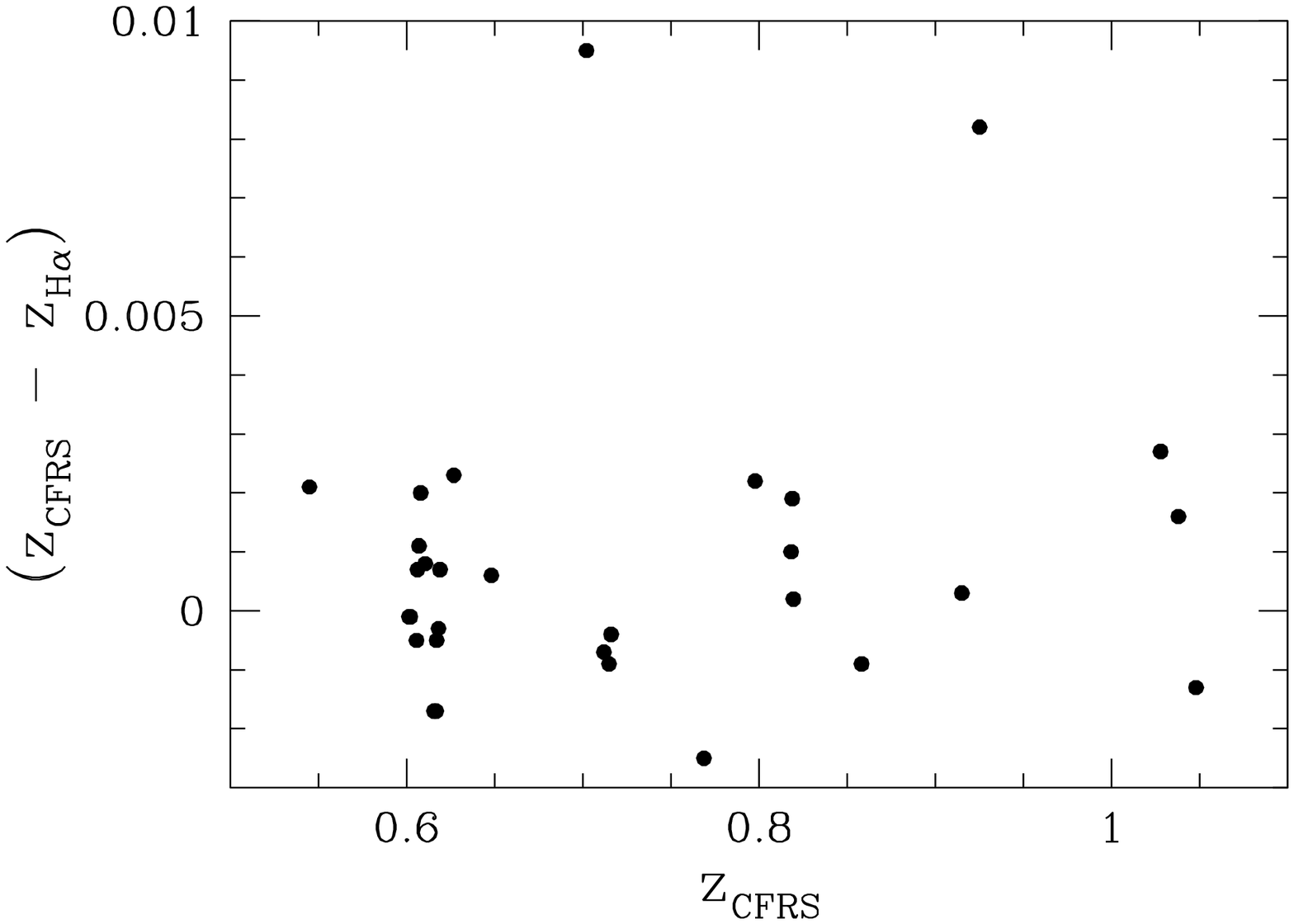,width=8cm}}
\vspace{-2.cm}
\caption{
  Difference in redshift between those measured from the original CFRS
  optical spectra, $z_{\rmn CFRS}$, and those from our H$\alpha$ line
  observed in the near infrared, $z_{\rmn H\alpha}$. The $rms$ is
  $0.0025$, or $0.0014$ when excluding the two most discrepant values.
\label{fig3}}
\end{figure}

\subsection{Emission-line measurements}

We measured the integrated fluxes and the corresponding 1$\sigma$
errors using the package {\scriptsize SPLOT} under {\scriptsize
  IRAF/CL}, interactively marking two endpoints around the line to be
measured.  This method allows the accurate measurement of lines which
are not well fit by simple Gaussian profiles, including lines with
asymmetric shapes. Out of 33 observation blocks, we obtained 30
H$\alpha$ flux measurements.  The spectra are shown in
Fig.~\ref{fig2}, and the fluxes are presented in Table~\ref{tab1}. We
detect no emission line in the remaining 3 observations; for these
either H$\alpha$ is below our detection limit, or it falls on an OH
line due to a large error in the redshift measurement.  Discrepancies
between the redshifts measured from the original CFRS spectra and from
ours are less than 0.003, except for two cases (0.0095 and 0.0082
respectively for \#03.0984 and \#22.0779, see Fig.~\ref{fig3}) for
which the CFRS calibration appears less accurate. Excluding these
cases, the $rms$ discrepancy is 0.0014, i.e. 420~km~s$^{-1}$,
consistent with the redshift accuracy of 350--550~km~s$^{-1}$
estimated in Le F\`evre et al. (1995).  The high-resolution of our
spectra means that H$\alpha$ is well separated from the
[\ion{N}{2}]$\lambda\lambda$6548, 6583 lines which are detected in
se\-veral spectra.  Since the slit is 2 arcsec wide, it covers most of
the apparent size of galaxies at $z > 0.5$ (see the 5x5 arcsec$^2$
postage stamps from HST data shown in Fig.~1 of Schade et al. 1995).
Thus we did not apply any aperture correction to our measurements.
Line luminosities are given in erg s$^{-1}$ by, $$L = 4 \pi (3.086\ 
10^{24} d_{\rmn{L}})^2 f,$$
where $f$ is the integrated line flux in
erg s$^{-1}$ cm$^{-2}$, and $d_{\rmn{L}}$ is the luminosity distance
in Mpc. In the case of H$\alpha$, the line fluxes vary from 5 to
82~10$^{-17}$ erg s$^{-1}$ cm$^{-2}$, with an average flux of
22~10$^{-17}$ erg s$^{-1}$ cm$^{-2}$.
\begin{figure}
\centerline{\psfig{figure=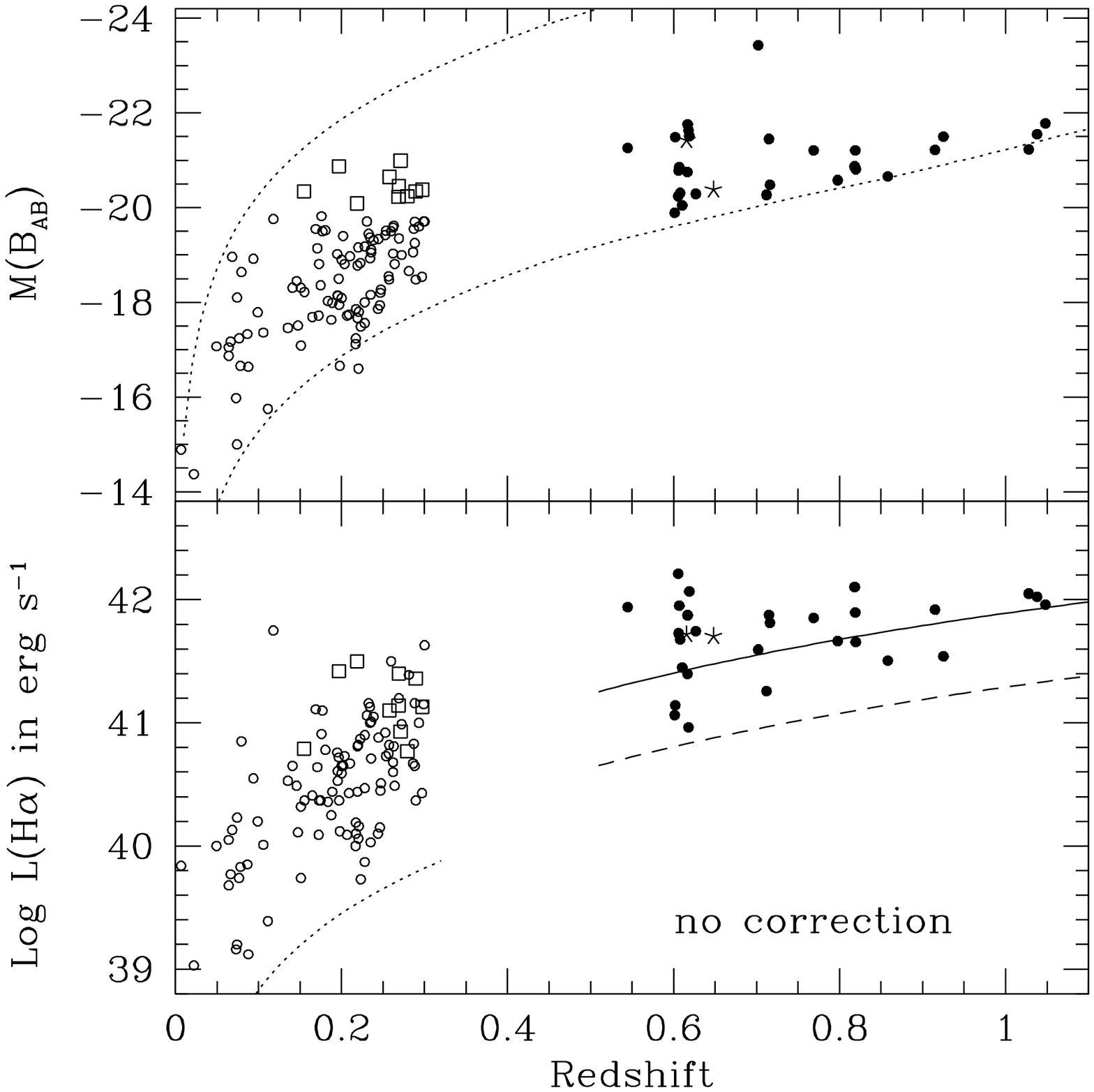,width=8cm}}
\caption{\MBAB\ and log L(H$\alpha$) versus redshift for gala\-xies at $z \le
  0.3$ (110 open symbols; Tresse \& Maddox 1998), and at $z > 0.5$ (30
  solid and starred symbols; this work). Low-$z$ galaxies with
  \MBAB~$<-20$ mag are plotted as open squares (10 out of 110).  In
  this figure, no reddening correction has been applied to the
  H$\alpha$ luminosities; they are derived directly from observed
  H$\alpha$ fluxes. In the following figures, all but two of the
  high-$z$ galaxies are corrected for reddening by assuming A$_{\rmn
    V}$~=~1 (solid dots). For the two galaxies shown by starred
  symbols we estimate the reddening from the Balmer decrement (see
  Section~4).  In the top panel, the dotted lines show the $I_{\rmn
    AB}$ CFRS apparent magnitude limits converted to \MBAB\ assuming
  the spectral energy distribution is like an Sab galaxy.  In the
  bottom panel, the dotted line shows the survey detection limit for
  an emission line observed in the $I$-band at the $I_{\rmn AB}=22.5$
  mag limit, and with observed EW~=~10~\AA.  The dashed line shows the
  limit for 22~\AA\ assuming $I_{\rmn AB} = J$, and the solid line
  shows the limit for 22~\AA\ assuming ($I_{\rmn AB}-J$)~=~1.5, as
  discussed in Section 5.1.
 \label{fig4}}
\end{figure}

\section{Reddening correction}
\label{sct_red}

Reddening is produced by interstellar extinction along the line of
sight to each observed galaxy. Interstellar extinction due to our
Galaxy is small (A$_{\rmn V} \la $ 0.1 mag), because the CFRS fields are
located at high galactic latitude ($\vert b_{\rmn{II}} \vert $ $\ge$
45$^\circ$); hence most of the reddening is intrinsic to the observed
galaxy.  For about two thirds of the galaxies, the H$\beta$ lines
observed in the CFRS spectra have too low signal-to-noise to be
accurately measured, so we cannot estimate A$_{\rmn V}$ for every galaxy.
Thus we present our results using two approaches: (a) with no
correction for reddening, (b) with a reddening correction derived from
two observed Balmer lines, and if not observed assuming an average
correction A$_{\rmn{V}}$ of 1 mag as found in several nearby samples
(e.g. for instance Kennicutt 1992, Tresse \& Maddox 1998). Note that
in all of these cases the estimates refer only to reddening that
occurs after the emission of the Balmer photons, and they do not
account for any obscuration of Lyman continuum photons before this.
Charlot et al. (2002) suggest that the absorption of Lyman continuum
may introduce a further attenuation of $\sim 20$\%.
\begin{figure}
\centerline{\psfig{figure=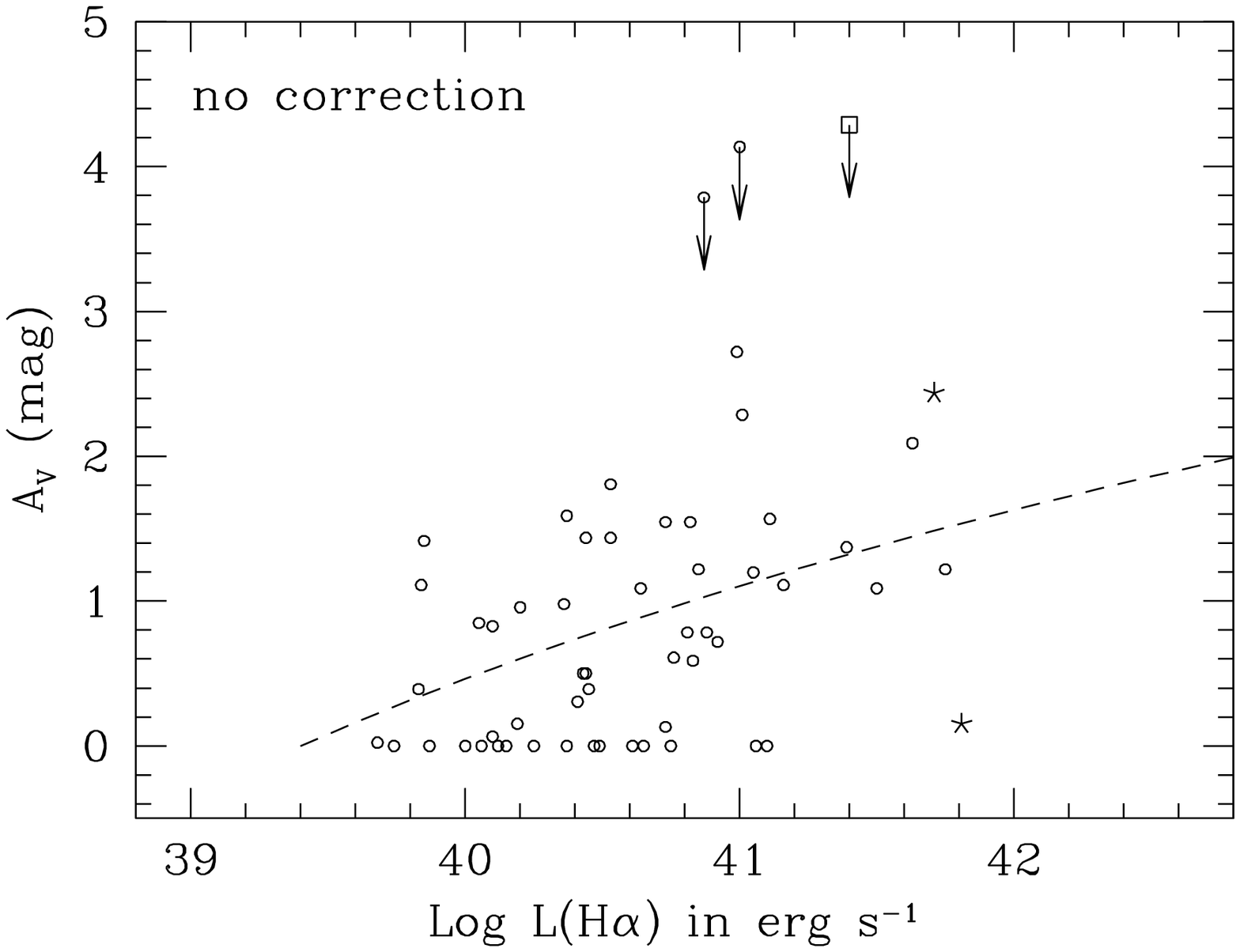,width=8cm}}
\vspace{-2.cm}
\caption{A$_{\rmn V}$ versus log L(H$\alpha$) for our galaxies 
  for which we could measure a Balmer decrement.  No reddening
  correction has been applied to the H$\alpha$ luminosities. The
  different symbols are defined in Fig.~\ref{fig4}. The dashed line
  shows the relation derived from the relation
  H$\alpha$/H$\beta$~=~0.82 log (L(H$\alpha$) 10$^{-41.09}$) + 4.24
  from Sullivan et al. (2001), using H$\alpha$/H$\beta$~=~2.86
  10$^{0.323C}$ and A$_{\rmn V}$~=~C 3.2/1.47. See Section~4.
\label{fig5}}
\end{figure}

For $z > 0.5$ galaxies we assume A$_{\rmn{V}} = CR/1.47 \simeq$ 1 mag
($R = 3.2$; Seaton 1979), which corresponds to an extinction
correction of $C=0.45$. Hence fluxes are dereddened as
$f$(H$\alpha$)10$^{0.677 C}$, and $f$([\ion{O}{2}])10$^{1.2553 C}$
using the galactic extinction law from Seaton~(1979).  We note that
Hammer et al.~(2000) estimated the $f($H$\beta)/f($H$\gamma)$ ratio
using high-resolution spectra of two of our $z > 0.5$ targets; the
ratios are 2.18 for \#03.0570 and 2.98 for \#03.1349. These ratios
corres\-pond to A$_{\rmn{V}}$ of 0.15 and 2.44 mag respectively, or
$C=0.07$ and 1.12 respectively, derived using
$I_{0}$(H$\gamma$/H$\beta$)~=~0.468 for case B recombination with a
density of 100 cm$^{-3}$ and a temperature of 10,000 K (Osterbrock
1989), and ap\-plying Seaton's law,
$f$(H$\gamma$/H$\beta$)~=~$I_{0}$(H$\gamma$/H$\beta$)10$^{0.129C}$.
These two galaxies are plotted as starred symbols in the figures. For
the $z\le 0.3$ galaxies, as described in Tresse \& Maddox (1998), for
half of the sample, we derived $C$ using measured
$f$(H$\alpha$)/$f$(H$\beta$) ratios and the same pres\-cription as
earlier with
$f$(H$\alpha$/H$\beta$)~=~$I_{0}$(H$\alpha$/H$\beta$)10$^{0.323C}$,
($I_{0}$(H$\alpha$/H$\beta$)~=~2.86; Osterbrock 1989). For the
remaining half we used A$_{\rmn{V}}$~=~1 mag.

Although we use an average reddening correction, the true
A$_{\rmn{V}}$ varies significantly from galaxy to galaxy. For
instance, in our CFRS low-$z$ spectra where the H$\beta$ lines have
sufficient signal-to-noise, we find values ranging from about 0 to 2
mag as shown in Fig.~\ref{fig5}. Variations in stellar absorption
introduce some scatter in apparent redde\-ning, but most of the
scatter represents genuine differences in reddening.  Ideally
H$\alpha$ and H$\beta$ should be corrected for stellar absorption, but
this is not possible with the CFRS spectra.  For galaxies without high
signal-to-noise spectra covering both H$\alpha$ and H$\beta$ we can do
no better than assume a constant A$_{\rmn{V}} = $ 1 mag as found in
seve\-ral nearby samples.  This applies to all, except two, of our
high-$z$ galaxies, and half of our low-$z$ sample. Although not ideal,
this simple approximation should be correct on average, and allows us
to examine some of the effects of reddening on our results.  We note
that Fig.~\ref{fig5} shows that our measured A$_{\rmn{V}}$ increases
towards high H$\alpha$ luminosities, and thus towards bright $B$-band
luminosities (see Fig.~\ref{fig6}).  $UV$-selected galaxy samples show
a similar trend (Sullivan et al.  2001), as do samples using FIR data
(for instance e.g. Wang \& Heckman 1996, Hopkins et al.  2001). Thus
it is possible that our high-$z$ H$\alpha$ sample would require a
correction larger than 1 mag. The fact that high-$z$ CFRS galaxies
exhibit more irregular/starbursting morphologies (Brinchmann et al.
1998) may also argue for a larger correction. However comparison of
our H$\alpha$ results with our 2800-\AA\ data (Section 6.4) suggests
that the average reddening correction cannot be much larger.  This
strengthens our 1~mag assumption for $I$-band selected CFRS sample.
We note also that H$\alpha$ attenuations based on radio data are
larger by $0.2\pm0.2$ mag on average than using Balmer decrements (see
e.g.  Bell \& Kennicutt 2001, Caplan \& Deharveng 1986).  All of these
uncertainties in reddening correction are small compared to the
statistical uncertainty in our final luminosity density estimate.

\section{Sample properties}

\subsection{Survey detection limits}

The CFRS $I$-band selection implies that intrinsically bright
galaxies, (M$^{\ast}\pm$1), are targeted at high redshift, i.e. with
\MBAB~$\simeq -21\pm1$ mag at $z > 0.5$. This bias is shown in the top
panel of Fig.~\ref{fig4}, which displays both our low- and high-$z$
H$\alpha$ samples. The ($z \le 0.3$) and ($z>0.5$) samples contain
respectively $\sim10$ and 100 per cent of galaxies brighter than
\MBAB~$=-20$ mag.  The low panel of Fig.~\ref{fig4} shows the
corresponding H$\alpha$ luminosities for the galaxies as a function of
redshift.  The detection limit of emission lines in the CFRS spectra
is about 10~\AA\ in terms of observed equivalent width.  The dotted
line represents a 10~\AA\ lower limit in terms of line luminosity
assuming the emission observed in the $I$-band ($\lambda_c =
8320$~\AA) at the survey magnitude limit, $I_{\rmn AB}=22.5$, i.e. a
line flux equivalent to $10\, {\rmn \AA} \times (1.5\times 10^{-9})
10^{-0.4 \times 22.5}$ in erg s$^{-1}$ cm$^{-2}$.  This limit is
correct for our low-$z$ sample since at $z\simeq 0.26$ H$\alpha$ falls
at the centre of the $I$-band.  There is no observational selection to
introduce an upper limit at large equivalent widths.  For our high-$z$
sample, we preselected the galaxies to have observed
EW([\ion{O}{2}])~$\sim 10$~\AA\, and this introduces an effective
detection limit of observed EW(H$\alpha$)~$\sim 22$~\AA\ (see
Table~\ref{tab2}).  The H$\alpha$ high-$z$ detection limit is shown
with EW~=~22~\AA\ assuming either a flat spectrum with $I_{AB}=J$, or
a colour term ($I_{\rmn AB}-J$)~$\simeq 1.5$.  In the latter case, we
see that H$\alpha$ can be detected below the `limit' because of the
spread in the observed [\ion{O}{2}]--H$\alpha$ relation, and also in
the colour term.  The rest-frame EW(H$\alpha$) varies from 1 to
140~\AA, with a median of 33~\AA. This range is similar for instance
to H$\alpha$ EW observed in the SAPM $b_J$-selected sample at $\langle
z \rangle = 0.05$ (Tresse et al.  1999).

\subsection{H$\alpha$ and $B$-band luminosities }

\begin{figure}
\centerline{\psfig{figure=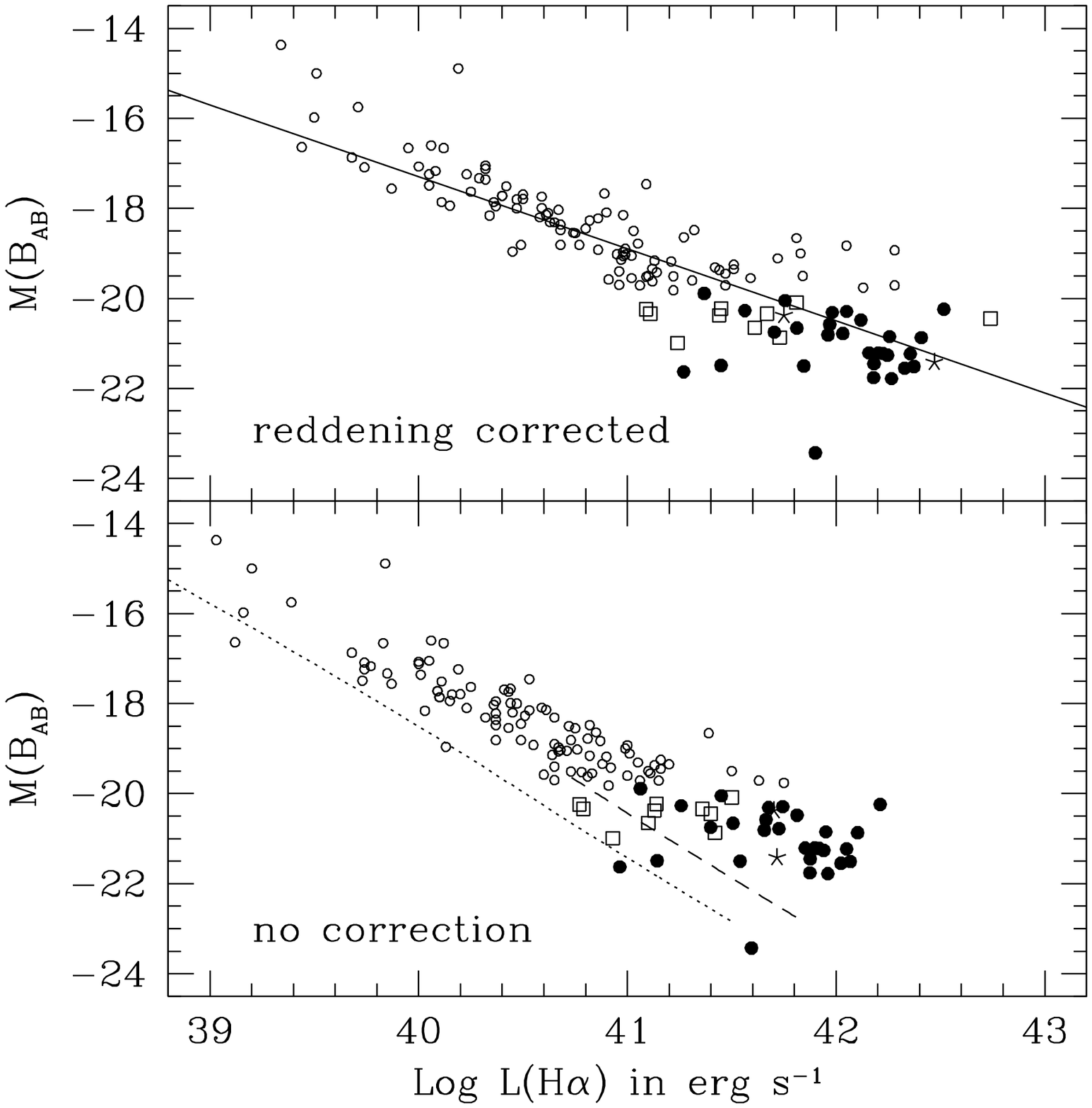,width=8cm}}
\caption{Log~L(H$\alpha$) versus \MBAB\ with reddening correction (top), with 
  no correction (bottom). The different symbols are defined in
  Fig.~\ref{fig4}.  The solid line is \MBAB~$= 46.7 -
  1.6$~log~$L$(H$\alpha$), as first noted by Tresse \& Maddox (1998).
  The dotted and dashed lines correspond to an Sab galaxy with $0<z<1$, 
  $I_{AB}=22.5$, and observed EW(H$\alpha$)~=~10 and 22~\AA\ 
  respectively.
\label{fig6}
}
\end{figure}

Fig.~\ref{fig6} shows the relation between L(H$\alpha$) and \MBAB\ of
the CFRS galaxies at $z \le 0.3$, and at $z > 0.5$.  The high-$z$
galaxies continue to follow the tight relation seen in the low-$z$
sample, first noted by Tresse \& Maddox (1998), \MBAB~$= 46.7 -
1.6$~log~$L$(H$\alpha$). The brighter in $B_{AB}$ a galaxy is, the
larger the H$\alpha$ luminosity.  The scatter in this relation is much
larger than the measurement uncertainties, and so reflects genuine
differences between the different physical conditions within each
galaxy, mainly the metallicity, the ionization parameter, the
intrinsic dust and the recent star-formation history.  The dotted line
in the bottom panel of Fig.~\ref{fig6} corresponds to an observed
10-\AA\ EW(H$\alpha$) for galaxies with an Sab spectral energy
distribution (SED) with $I_{AB}=22.5$ and $0<z<1$. The precise
location of the line depends on the assumed SED, but it corresponds
approximately to the detection limit for the low-$z$ sample. It is
clear that the galaxy distribution is almost all above this line.  As
discussed in Section~5.1, we preselected the high-$z$ galaxies on
their [\ion{O}{2}] emission, so although there is not a sharp limit on
EW(H$\alpha$), we are more likely to observe galaxies with
EW(H$\alpha)>22$~\AA. The dashed line in Fig.~\ref{fig6} shows the
22-\AA\ EW(H$\alpha$) limit.  Although there is no observational
constraint to exclude galaxies with high L(H$\alpha$) at a given
\MBAB, there clearly is an upper limit to the observed L(H$\alpha$)
which increases with \MBAB.  The highest L(H$\alpha$) for a given
\MBAB\ corresponds to the gala\-xies with the strongest and most
recent phases of star formation.

\begin{figure}
\centerline{\psfig{figure=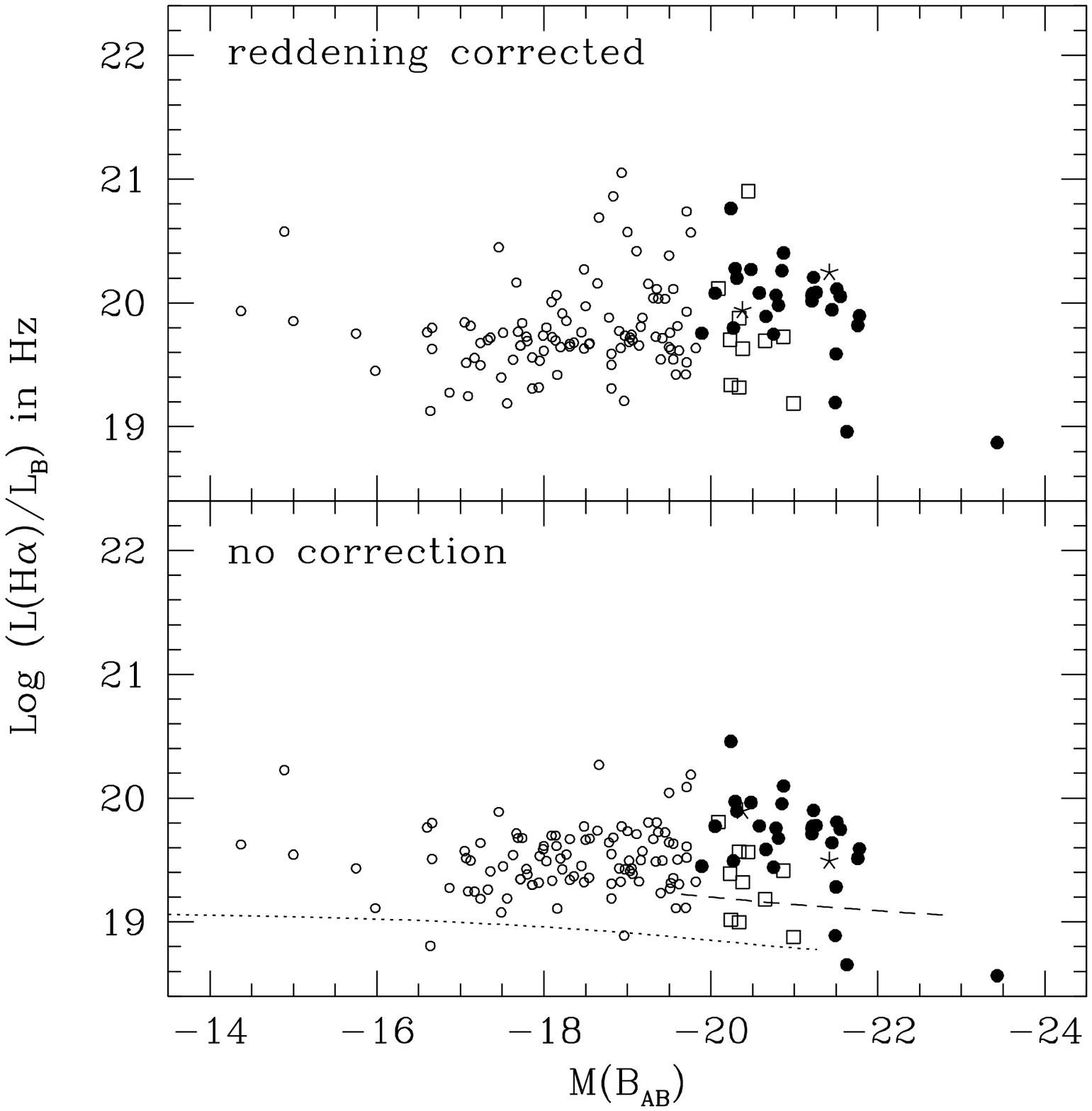,width=8cm}}
\caption{Logarithmic difference between L(H$\alpha$) and L(B$_{\rmn AB}$\hspace{-0.05cm}) versus \MBAB\
  with reddening correction (top), with no correction (bottom).  The
  different symbols are defined in Fig.~\ref{fig4}. The dotted and dashed 
  lines are the same as in Fig.~\ref{fig6}.
\label{fig7}
}
\end{figure}

Fig.~\ref{fig7} explicitly shows the ratio L(H$\alpha$)/L(B$_{\rmn
  AB}$\hspace{-0.05cm}) as a function of \MBAB.  The
L(H$\alpha$)/L(B$_{\rmn AB}$\hspace{-0.05cm}) ratio increases for
brighter galaxies, reflecting the non-linear relation between
L(H$\alpha$) and L(B$_{\rmn AB}$\hspace{-0.05cm}) inferred from
Fig.~\ref{fig6}. Within the low-$z$ sample, galaxies with \MBAB~$<-20$
appear to have a slightly lower ratio, but after reddening correction,
this is less apparent, probably because of the luminosity dependence
of the reddening. Also, after reddening correction the data appear
more scattered.

The high-$z$ galaxies all have \MBAB~$<-20$ and show a higher
L(H$\alpha$)/L(B$_{\rmn AB}$\hspace{-0.05cm}) ratio than the
equivalent low-$z$ galaxies.  This difference is mostly due to the
higher EW limit of our high-$z$ sample, as demonstrated by the dotted
and dashed lines, which show the expected ratio for EW(H$\alpha)=10$
\AA\ and EW(H$\alpha)=22$ \AA\ galaxies respectively, as in
Fig.~\ref{fig6}.  A small part of the difference is likely to be real
evolution in the L(H$\alpha$)/L(B$_{\rmn AB}$\hspace{-0.05cm}) ratio.
In Section 6.3 we find that the H$\alpha$ luminosity density increases
as $(1+z)^{4.1}$, whereas Lilly et al.  (1996) found that the $B$-band
luminosity density increases as $(1+z)^{2.7}$, so we expect
$\log$(L(H$\alpha$)/L(B$_{\rmn AB}$\hspace{-0.05cm})) to increase by
roughly 0.2 between the low- and high-$z$ samples, even if the
selection limits were the same.

Fig.~\ref{fig7} suggests that the upper limit observed in
Fig.~\ref{fig6} in terms of L(H$\alpha$)/L(B$_{\rmn
  AB}$\hspace{-0.05cm}) is independent of $B$-band absolute magnitude.
Since H$\alpha$ flux originates mainly from UV photoioni\-zing photons
($<$ 912 \AA) it traces only massive, young, short-lived stars (OB
stars, $t < {\rmn few} \times 10^6$ yr), while the rest-frame $B$-band
flux is dominated by longer-lived stars (type~A). The highest ratio
will be seen during a burst of star-formation, during which the ratio
directly reflects the Initial Mass Function. Thus a constant upper
limit is in agreement with the hypothesis of a universal Initial Mass
Function (IMF): for a certain amount of massive OB stars formed, there
is always the same amount of intermediate, type~A, stars.
\begin{figure}
\centerline{\psfig{figure=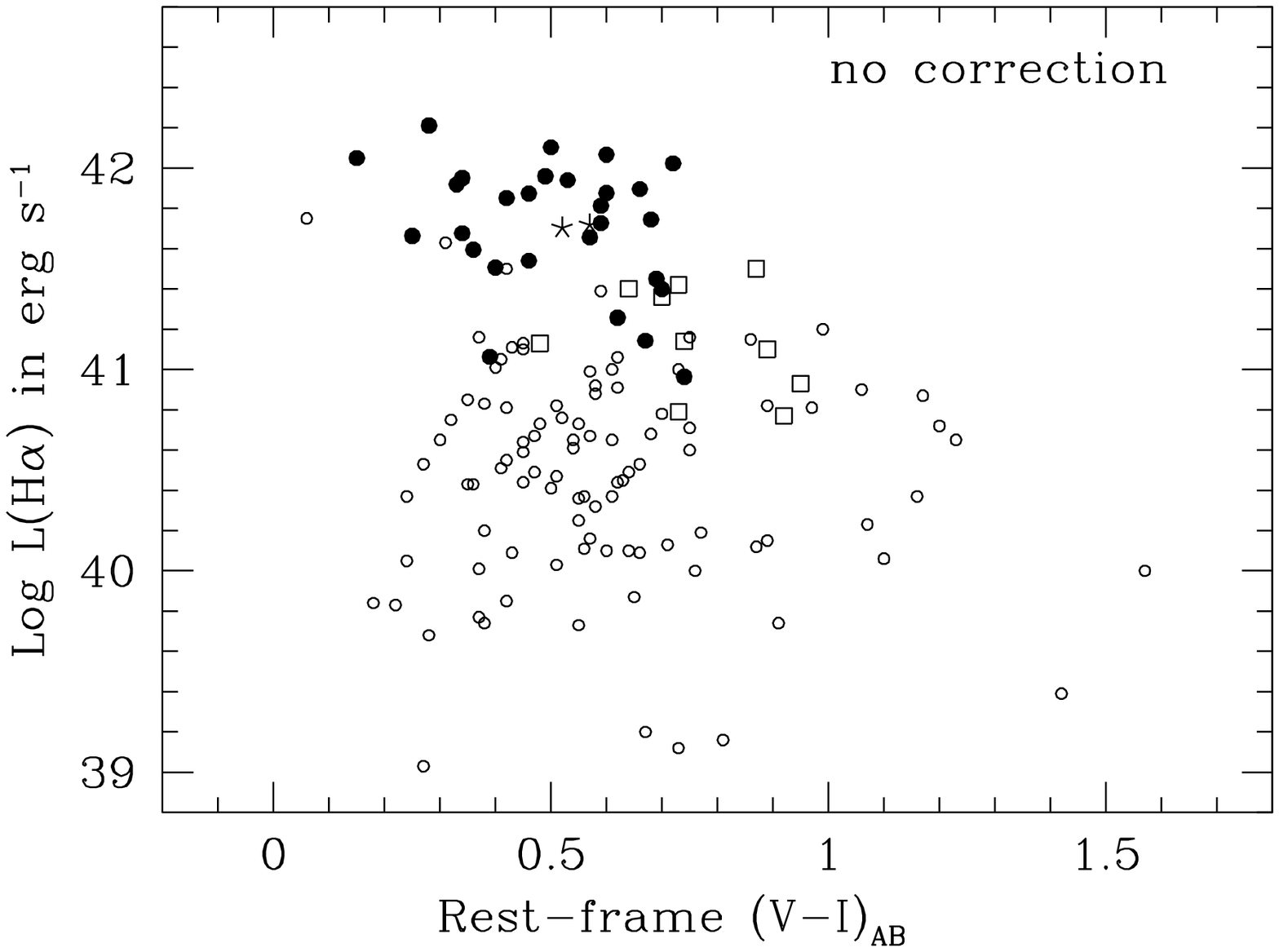,width=8cm}}
\vspace{-2.cm}
\caption{Log~L(H$\alpha$) versus rest-frame
  (V$-$I)$_{\rmn{AB}}$.  The different symbols are defined in
  Fig.~\ref{fig4}.
\label{fig8}}
\end{figure}

\subsection{H$\alpha$ luminosities and colours}

Fig.~\ref{fig8} shows (V$-$I)$_{\rmn AB}$ rest-frame colours of the
low- and high-$z$ galaxy samples.  The most striking feature of the
dia\-gram is that the high-$z$ galaxy sample has a much higher mean
L(H$\alpha$), as discussed in Section 5.2.

There is no significant correlation between H$\alpha$ luminosities and
the rest-frame (V$-$I) colours.  This is consistent with the whole
CFRS sample, which shows that no correlation between \MBAB\ and
rest-frame (U$-$V) colours (see Figure~5 in Lilly et al. 1995).  This
supports the idea that the H$\alpha$ flux depends on the instantaneous
star formation rate and on the time since the last burst.

The median colours of the low- and high-$z$ samples are the same
within the uncertainties: $0.58\pm0.15$ at low-$z$ and $0.53\pm0.11$
at high-$z$.  In the low-$z$ sample, there are a few strong H$\alpha$
emitters which have red colours. They show spectral features (Ca~H and
K, G~band, MgI) indicating the pre\-sence of a dominant old population
(see Tresse et al. 1996). In the high-$z$ sample we do not see
H$\alpha$ in gala\-xies redder than rest-frame (V$-$I)~=~0.8. This
could simply be due to the lower sampling rate in the observed
high-$z$ sample, but also the higher EW detection limit for the
high-$z$ sample (see Section 5.1) biases against detecting galaxies
with a red conti\-nuum and a weak H$\alpha$ flux.

\subsection{[\ion{O}{2}]--H$\alpha$ luminosity ratio}

\begin{figure}
\centerline{\psfig{figure=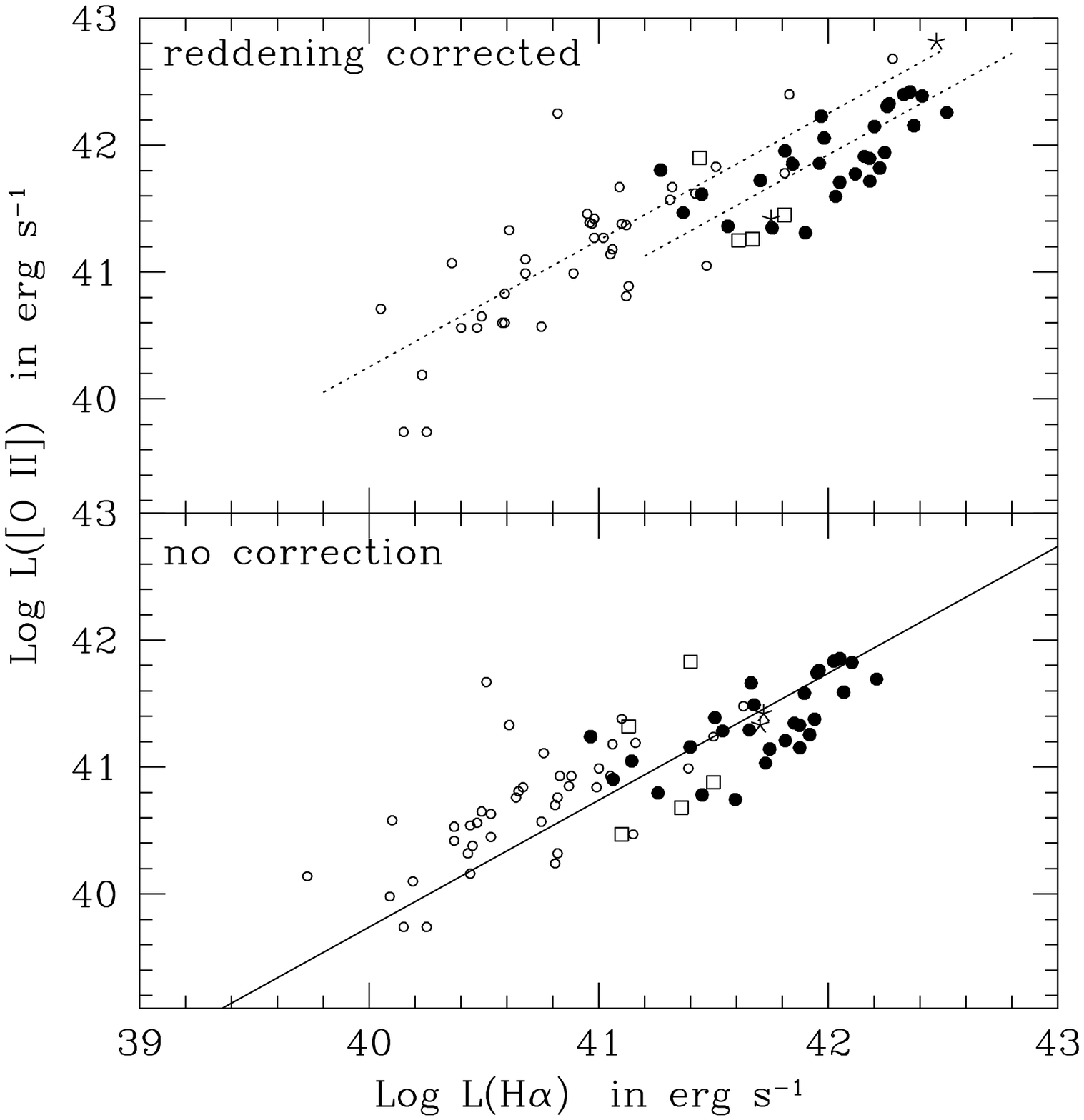,width=8cm}}
\caption{Log~L([\ion{O}{2}]) versus log~L(H$\alpha$) 
  with reddening correction applied to both emission lines (top), with
  no correction (bottom).  The dotted lines are the medians of the
  low- and high-$z$ ratios, and the plain line is the local ratio (see
  Table~\ref{tab1}).  The different symbols are defined in
  Fig.~\ref{fig4}.
\label{fig9}}
\end{figure}

In other surveys where H$\alpha$ could not be observed, the
[\ion{O}{2}] line has frequently been used to provide SFR estimates by
making the assumption that the [\ion{O}{2}]--H$\alpha$ ratio is
constant and equal to that derived from local samples. These two lines
are closely related since they both come from star formation activity
in \ion{H}{2} regions. However, [\ion{O}{2}] depends also on the metal
fraction present in the gas, and on the ionization parameter, which
means that the observed scatter about the mean relation is large. For
instance in the SAPM sample (Tresse et al. 1999) and the Kennicutt
(1992) data, the scatter in the EW ratio is as large as 60 per cent.
Va\-ria\-tions in the continuum colour may increase the scatter in the
EW measurements compared to line luminosities, but the scatter is very
large even in the luminosity ratio.  Fig.~\ref{fig9} shows the
correlation between [\ion{O}{2}] and H$\alpha$ luminosities for our
CFRS data.  The medians and {\it rms} dispersions are listed in
Table~\ref{tab2}.  The high-$z$ sample tends to have a lower
[\ion{O}{2}]--H$\alpha$ ratio than the overall mean.  Though it is
tempting to suggest that this is an evolutionary effect, we argue that
it is in fact evidence for a luminosity dependence in the
[\ion{O}{2}]--H$\alpha$ ratio.  \setcounter{table}{1}
\begin{table}
\caption{Medians and {\it rms} dispersions of the [\ion{O}{2}]--H$\alpha$ 
lumino\-sity ratio. 
\label{tab2}}
\begin{tabular}{lcccc}
\hline
Sample & N & $\langle z \rangle$  & Median & {\it rms} (\%)  \\
\hline
No correction                      &     &       &      &     \\
$0   < z <   0.1$, SAPM            & 859 & 0.05  & 0.55 &  60 \\
$0   < z \le 0.3$, CFRS$^\dagger$  & 40  & 0.24  & 0.97 & 164 \\
$0.5 < z <   1.1$, CFRS$^\ddagger$ & 30  & 0.73  & 0.46 &  68 \\
Reddening corrected                &     &       &      &     \\
$0   < z \le 0.3$, CFRS$^\dagger$  & 40  & 0.24  & 1.78 & 184 \\
$0.5 < z <   1.1$, CFRS$^\ddagger$ & 30  & 0.73  & 0.84 &  71 \\
\hline
\end{tabular}
{\footnotesize
$^\dagger$The $0 < z \le 0.3$ sample has been restricted to galaxies
with \MBAB~$\ge -20$.\\
$^\ddagger$The $0.5 < z < 1.1$ sample contains only galaxies
with \MBAB~$<-20$.}
\end{table}

\begin{figure}
\centerline{\psfig{figure=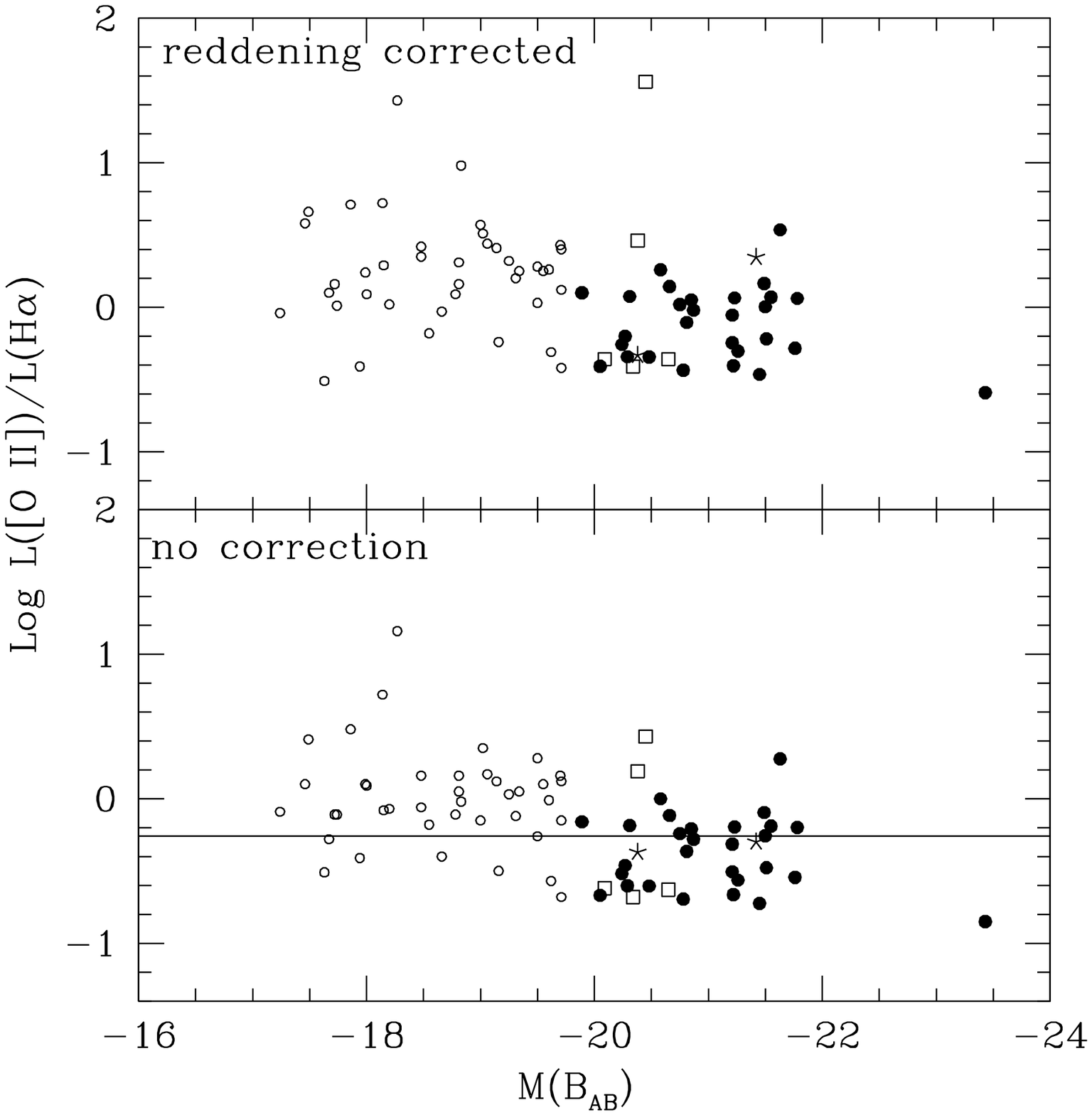,width=8cm}}
\caption{Logarithmic difference between  [\ion{O}{2}] and H$\alpha$ lumino\-sity  
  versus \MBAB\ with reddening correction applied to both emission
  lines (top), with no correction (bottom). The plain line is the
  local ratio (see Table~\ref{tab1}).  The different symbols are
  defined in Fig.~\ref{fig4}.
\label{fig10} }
\end{figure}

\begin{figure}
  \centerline{\psfig{figure=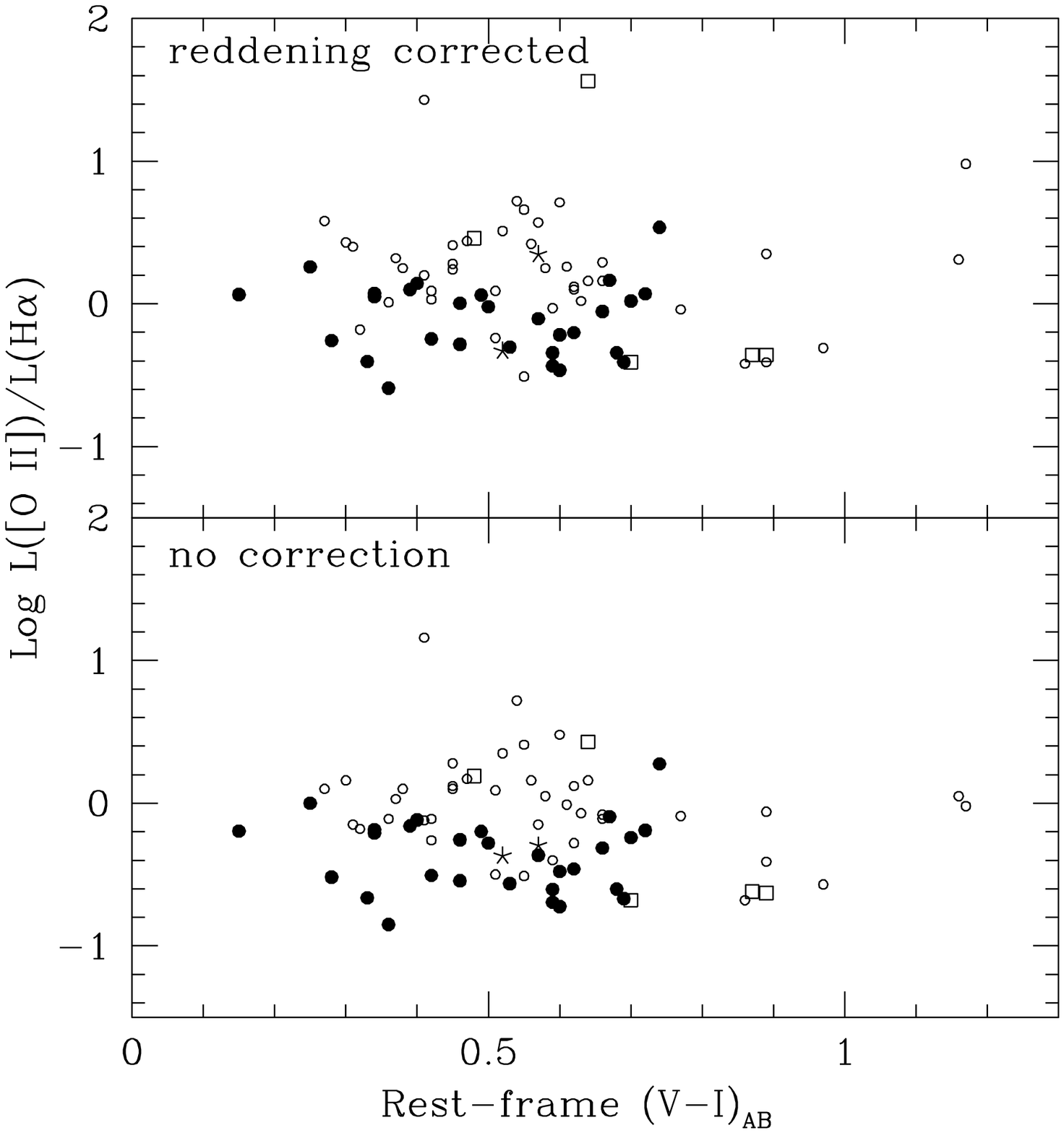,width=8cm}}
\caption{Logarithmic difference between [\ion{O}{2}] and H$\alpha$ luminosity 
  versus (V$-$I)$_{\rmn AB}$ with reddening correction applied to both
  emission lines (top), with no correction (bottom).  The different
  symbols are defined in Fig.~\ref{fig4}.
\label{fig11}}
\end{figure}

Fig.~\ref{fig10} shows the ratio of [\ion{O}{2}]/H$\alpha$ fluxes as a
function of \MBAB, and we see that there is a systematic decrease
towards brighter galaxies.  Most of the brighter galaxies are from the
high-$z$ sample, but there are a small number of bright galaxies at
low redshift (shown as open squares) which also follow the trend.
This argues against an evolutionary effect.  A similar luminosity
dependence in the [\ion{O}{2}]/H$\alpha$ ratio has been previously
noted in nearby surveys, in particular in the Nearby Field Galaxy
Survey (NGFS; Jansen, Franx \& Fabricant 2001), in the 15R survey
(Carter et al.  2001), and in the SAPM (Charlot et al. 2002).  Since
these surveys are both restricted to very low redshifts we can exclude
that the effect is due to evolution.  The correlation between
[\ion{O}{2}]/H$\alpha$ and \MBAB, and the scatter in the ratio is
related to changes in the effective gas parameters (ionization,
metallicity, dust content) as a function of luminosity, rather than
systema\-tic changes in SFR per unit \MBAB\ (see e.g. Jansen et al.
2001, Charlot et al. 2002).  This syste\-matic change in ratio means
that SFR estimates based on [\ion{O}{2}] line measurements may be
significantly biased if the local average ratio of
[\ion{O}{2}]/H$\alpha$ is assumed.  This effect will be particularly
important at higher redshifts where current samples tend to be
restricted to galaxies that are more luminous than the ave\-rage in
local samples. For instance, within the CFRS this ratio varies by a
factor~2 from the low- to high-$z$ samples.  Note that when we correct
for reddening we have assumed a constant A$_{\rm V}$ for our high-$z$
sample. If A$_{\rm V}$ is luminosity dependant as discussed in
Section~4, this constant correction will not remove the correlation
shown in Fig.~\ref{fig9}, upper panel.  We find no correlation between
[\ion{O}{2}]/H$\alpha$ and rest-frame (V$-$I) colour
(Fig.~\ref{fig11}).

\section{H$\alpha$ luminosity and SFR densities}

\subsection{Completeness correction}
\label{sct-halumfun}

In the following completeness correction, we account for the selection 
of our high-$z$ H$\alpha$ (see Section~2) from the CFRS survey.

For the original CFRS sample, objects were selected as spectroscopic
targets solely on the basis of their $I$-band photometry, $17.5 \le
I_{AB} \le 22.5$, with no star/galaxy classification (see Le F\`evre
et al. 1995).  Our low-$z$ sample includes all CFRS targets in the
relevant redshift range, and so the H$_\alpha$ luminosity function can
be estimated using the standard $V_{max}$ technique, as described by
Tresse \& Maddox (1998).  For our high-$z$ sample, the targets were
subject to a second selection based on their [\ion{O}{2}] emission,
and so estimating the H$_\alpha$ luminosity function requires a slight
modification to the $V_{max}$ technique.

For each galaxy we estimate the ob\-ser\-va\-ble volume by calculating
the volume between the ma\-xi\-mum and minimum redshift at which it
could have been in the sample given the apparent magnitude and
redshift limits and the solid angle of the CFRS.  Not all galaxies in
the CFRS photometric sample were targeted for spectroscopy in the
CFRS, thus we have effectively observed only a fraction of the total
observable galaxies in three CFRS fields, $f_{spec} =
N(CFRS|spec)/N(CFRS|phot)= 3 \times 434/2452$, where $N(CFRS|spec)$ is
the total number of galaxies with measured redshifts in the three
observed CFRS fields, and $N(CFRS|phot)$ is the corresponding number
of galaxies detected in the photometric sample.

Since we have also selected galaxies away from OH sky lines, we must
allow for the consequent gaps in redshift where we have excluded
galaxies.  For each galaxy we calculate the observable volume u\-sing 
($V_{max} - V_{min}$), and renormalize the resulting number
density by the ratio of the number of galaxies away from OH lines to
the total number CFRS galaxies within the allowed redshift range. This
cor\-res\-ponds to a fraction $f_{OH} = N(OH)/ N(CFRS|(spec \ \& \ 
0.50 \!\! < \!\! z \!\! < \!\! 1.05))=84/157$, where $N(OH)$ is the
number of CFRS galaxies with measured redshifts away from OH lines,
and $ N(CFRS|(spec \ \& \ 0.50 \!\! < \!\! z \!\! < \!\! 1.05))$ is
the total number of CFRS galaxies with measured redshifts in our
selected range.

We must make a further correction because we selected our target 
galaxies to have observed EW([\ion{O}{2}])~$> 12$~\AA.  This
introduces a bias to higher H$\alpha$ fluxes, as can be seen from
Fig.~\ref{fig9}.  We used the observed median [\ion{O}{2}]/H$\alpha$
ratio (see Table~\ref{tab2}) to estimate the average line flux limit
introduced by the preselection.  Since the scatter about the mean
relation is large ($\sim 50\%$), the [\ion{O}{2}] limit does not
introduce a sharp cut in H$\alpha$, but a smooth drop in completeness.
Assuming that the scatter is approximately Gaussian, the
incompleteness will be proportional to the corresponding error
function. Thus, when calculating the observable volume for each
galaxy, we weighted the volume integral according to $\frac{1}{2}({\rm
  erf}((z-z_{lim})/\sigma)+1)$, where $z_{lim}$ is the maximum
redshift that the median [\ion{O}{2}]--H$\alpha$ relation would have
introduced, and ($\sigma=0.2$) is determined by the scatter in the
relation.

This procedure corrects for the incompleteness near the observed
L(H$\alpha$) limit, but does not account for the fact that we have
observed only a fraction of the available gala\-xies.  We calculated
this correction factor assuming one of two extreme cases: either that
the high EW([\ion{O}{2}]) gala\-xies are representative of the complete
sample; or that they contribute all of the H$\alpha$ flux (equivalent
to assuming that H$\alpha$ is never seen in emission without
[\ion{O}{2}]).  In the first case there are two correction factors:
the fraction of galaxies with EW([\ion{O}{2}])~$>$~12~\AA, $f_{\rm [O
  II] } = 63/84$; and the fraction that we observed with ISAAC,
$f_{obs} = 33/63$.  The overall correction factor from observed
luminosity to estimated total lumino\-sity is then $f_{rep} = f_{obs}
f_{\rm [O II] } f_{OH} f_{spec} = 0.1116 $. In the second case, we
make the assumption that all of the H$\alpha$ luminosity would be seen
in the [\ion{O}{2}] selected sample, which is equivalent to setting
the factor $f_{\rm [O II] } = 1$.  This leads to an overall correction
factor $f_{all} = f_{obs} f_{OH} f_{spec} = 0.1488$.  The ratio 
between these two extremes is $f_{rep}/f_{all}=0.75$, and so a robust
limit on this source of uncertainty is 0.125 in log($\phi$).  

An intermediate estimate of the correction factor can be obtained from
the SAPM survey (Tresse et al. 1999) which has [\ion{O}{2}] and
H$\alpha$ measurements for all galaxies in a local sample. If we split
the SAPM into galaxies with observed EW([\ion{O}{2}])~$>$~12~\AA, and
galaxies with observed EW([\ion{O}{2}])~$\le$~12~\AA, including non
[\ion{O}{2}] emitters, we find that the mean H$\alpha$ luminosity for
high-[\ion{O}{2}] galaxies is about four times that for
low-[\ion{O}{2}] galaxies.  In the complete high-$z$ CFRS sample, we
have 251 galaxies with EW([\ion{O}{2}])~$>$~12~\AA, and 72 galaxies
EW([\ion{O}{2}])~$\le$~12~\AA.  Ignoring possible evolutionary effects
and complications due to weighting by different accessible volumes,
the ratio of mean H$\alpha$ luminosities for our high-$z$
EW([\ion{O}{2}]) samples will be the same. Thus the total H$\alpha$
luminosity would be 1.07 times our estimated total H$\alpha$
luminosity using the correction factor $f_{rep}$, which translates
into 0.03 in log H$\alpha$ luminosity density. This is much less than
the one sigma statistical uncertainty derived in the next sections,
and so for our best luminosity function estimate we use $f_{rep}$.

\subsection{The H$\alpha$ luminosity function} 

\begin{figure}
\centerline{\psfig{figure=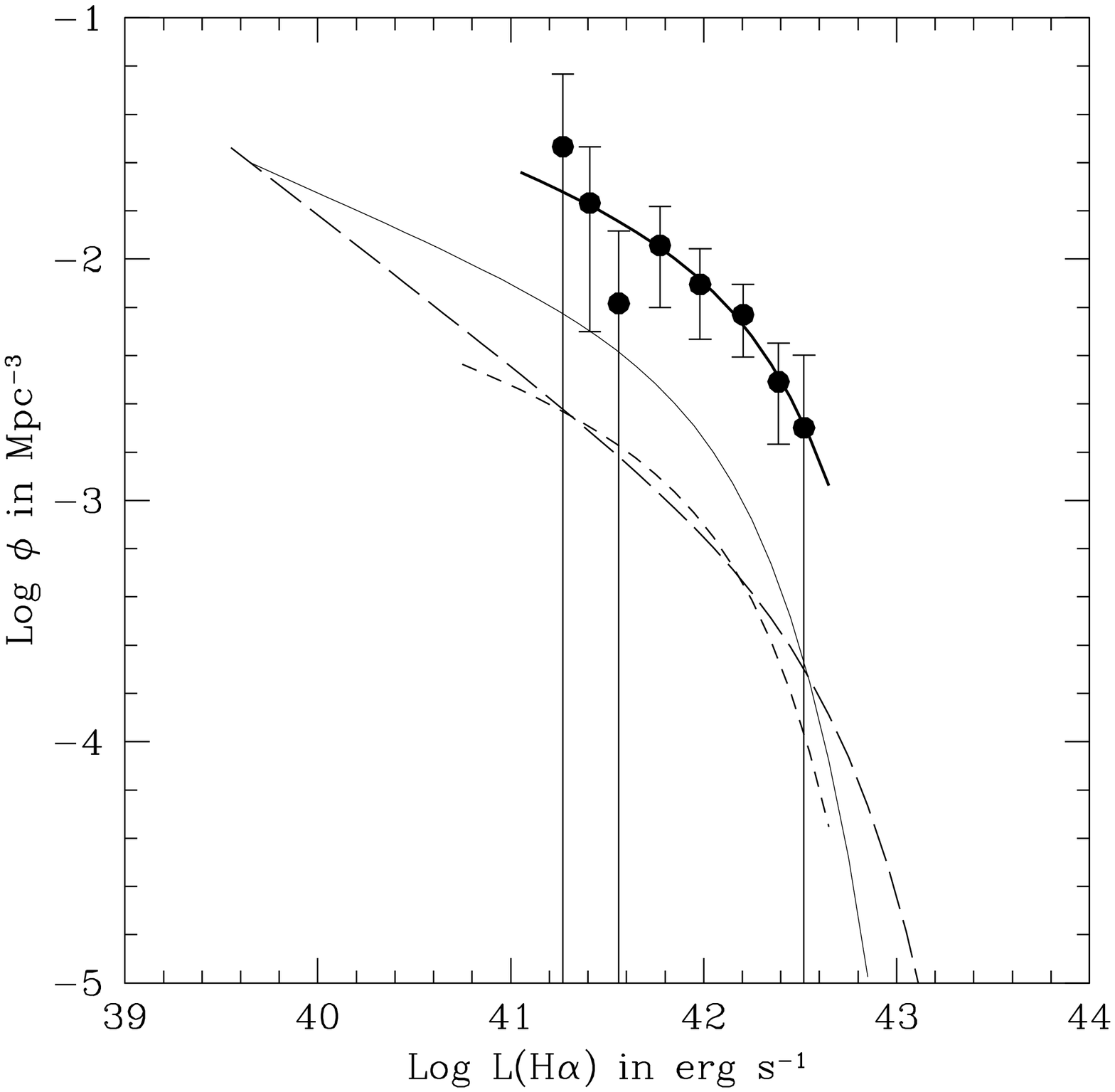,width=8cm}}
\caption{
  The reddening corrected H$\alpha$ luminosity function for four
  galaxy samples at different redshifts.  The filled circles are our
  data measurements at $0.5<z<1.1$, and the thick solid curve is the
  best Schechter fit to our data.  The density error bars assume
  Poisson statistics.  The thin solid curve is the Tresse \& Maddox
  (1998) H$\alpha$ LF at $\langle z \rangle = 0.2$ ($\alpha
  =-1.35\pm0.06$, $\phi^\ast= 10^{-2.83\pm0.09}$ Mpc$^{-3}$,
  $L^\ast=10^{42.13\pm0.13}$ erg~s$^{-1}$).  The short-dashed curve is
  the Gallego et al. (1995) H$\alpha$ LF at $z\simeq 0$ from an
  emission line-selected sample based on slitless spectra ($\alpha
  =-1.3\pm0.2$, $\phi^\ast= 10^{-3.2\pm0.2}$ Mpc$^{-3}$,
  $L^\ast=10^{42.15\pm0.08}$ erg~s$^{-1}$).  The long-dashed curve is
  the Sullivan et al. (2000) H$\alpha$ LF at $\langle z \rangle =
  0.15$ from a $UV$-selected sample ($\alpha =-1.62\pm0.10$,
  $\phi^\ast= 10^{-3.82\pm0.20}$ Mpc$^{-3}$,
  $L^\ast=10^{42.65\pm0.14}$ erg~s$^{-1}$).
\label{fig12}
}
\end{figure}

\begin{figure}
\centerline{\psfig{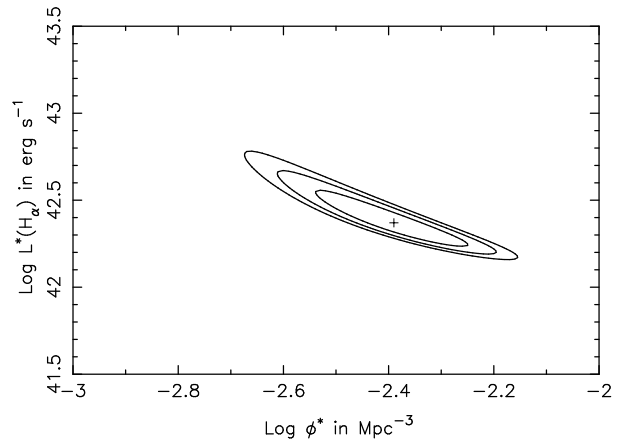}}
\caption{Contours of $\chi ^2 $ for Schechter function fit plotted 
  in Fig.~\ref{fig12}. The contours correspond to $1\sigma$ intervals
  in the parameters $\phi^\ast$ and $L^\ast$ with $\alpha=-1.31$.
\label{fig13}
}
\end{figure}

The resulting luminosity function (LF) and its Schechter (1976) fit
are shown in Fig.~\ref{fig12}. The best fit values are:
\begin{eqnarray*}
\alpha & =  & -1.31\pm0.11, \\ 
\phi^\ast & = & 10^{-2.39\pm0.06}\  {\rmn Mpc}^{-3}, \\
L^\ast & = & 10^{42.37\pm0.06}\  {\rmn erg s}^{-1}. 
\end{eqnarray*}
The uncertainties on these parameters are correlated as can be seen
from the $\chi ^2 $ contours shown in Fig.~\ref{fig13}.  The evolution
of the H$\alpha$ LF from $z\sim0$ to $z\sim 1$ is now clearly
demonstrated.  The Gallego et al. (1995) H$\alpha$ LF at $z \simeq 0$
is a factor $\sim 2$ lower than Tresse \& Maddox (1998) H$\alpha$ LF
at $\langle z \rangle =0.2$, which is in turn a factor $\sim 5$ lower
than our present data at $\langle z \rangle =0.7$.  The strong
evolution appears closely related to the ($\phi^{\ast}$, L$^{\ast}$)
parameters rather than to the faint-end slope, $\alpha$.

\begin{figure}
\centerline{\psfig{figure=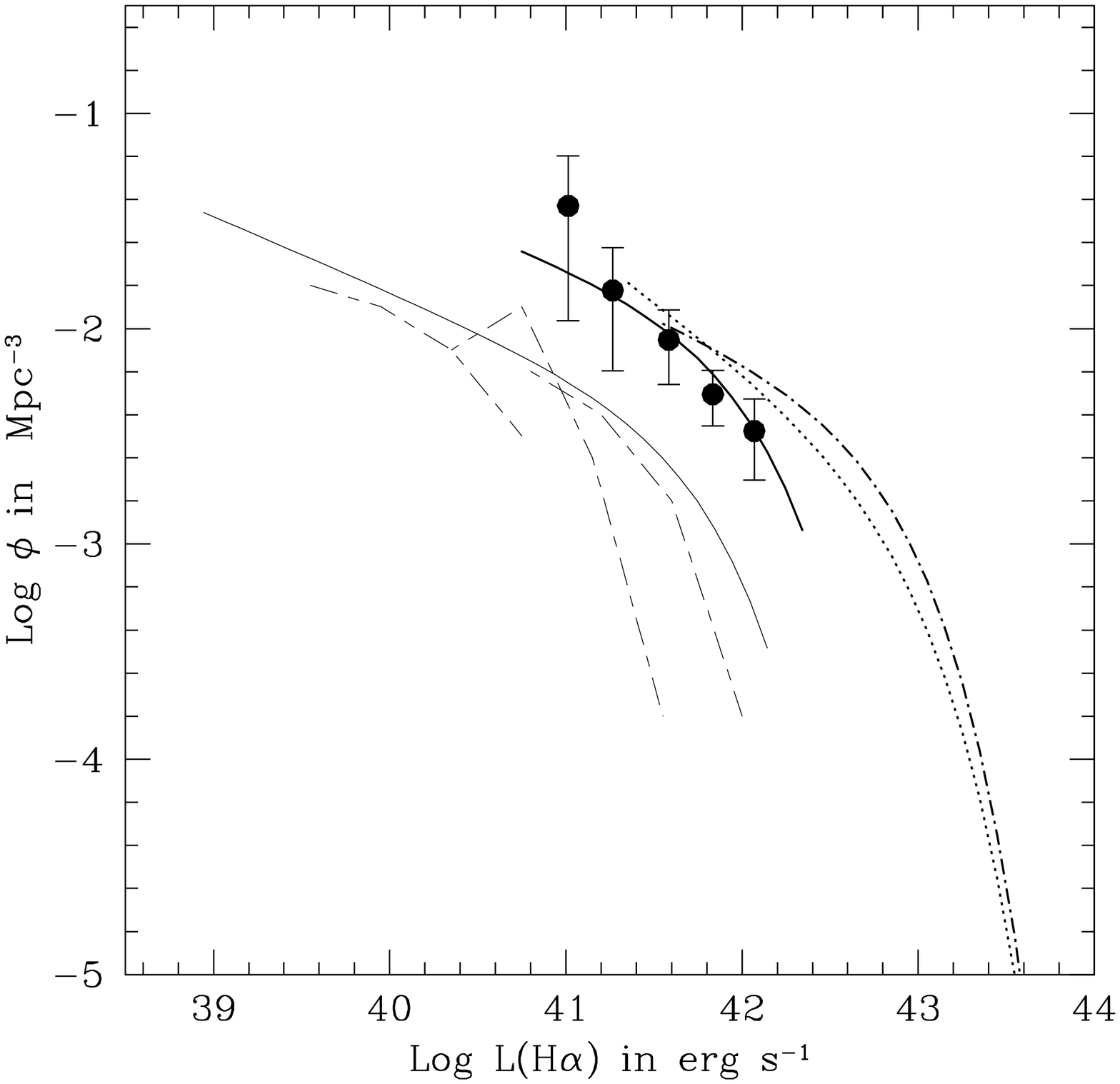,width=8cm}}
\caption{The H$\alpha$ luminosity function with no reddening 
 correction applied for five galaxy samples at different redshifts.  
  The filled circles are our data measurements not corrected for
 reddening at $0.5<z<1.1$, and the solid curve is the best fit to our
 reddening corrected data plotted in Fig.~\ref{fig12} but shifted in X-axis by
 $-$log~2.02. The thin solid curve is the Tresse \& Maddox (1998)
 H$\alpha$ LF but shifted in X-axis by $-$log~2.02, at $0<z\le 0.3$.
 The short-long dashed curves are the preliminary H$\alpha$ LF from
 Jones \& Bland-Hawthorn (2001) from left to right respectively at
 $z=0.08$, $z=0.24$, $z=0.40$.  The dot-dashed curve is the Yan et al.
 (1999) H$\alpha$ LF at $0.7<z<1.9$ with $\phi^\ast=
 10^{-2.77}$~Mpc$^{-3}$, $L^\ast=10^{42.85}$ erg~s$^{-1}$, and
 $\alpha$ assumed equal to $-1.35$, and the dotted curve is the
 Hopkins et al. (2000) H$\alpha$ LF at $0.7<z<1.8$ which includes the
 data of Yan et al. (1999).
\label{fig14}
}
\end{figure}

The LF from our non reddening corrected H$\alpha$ flux data is
presented in Fig.~\ref{fig14} where our previous best LF fit is
shifted by a factor 2.02 in log towards fainter luminosities (2.02
corresponds to A$_{\rmn V}=1$ mag, see Section~4). The uncorrected LFs
from Yan et al.  (1999) and Hopkins, Connolly \& Szalay (2000) are
also shown. The data of Yan et al. are composed of 33
H$\alpha$+[\ion{N}{2}] blends detected at $0.75<z<1.9$ with a similar
range of EW as our sample, from a few \AA\ to $\sim 130$ \AA.  We note
that their LF extends to brighter H$\alpha$ luminosities than ours;
this is certainly due to the fact that their galaxies have brighter
H$\alpha$ on average, but also due to their average
[\ion{N}{2}]$\lambda$6583/H$\alpha$ correction of 0.3.  In the local
SAPM data from Tresse et al. (1999), this ratio varies from 0.7 to 0.1
for a similar EW range, which implies that the correction to
H$\alpha$+[\ion{N}{2}] fluxes varies from 48\% to 12\%.  Hopkins et
al. (2000) combined their data with those of Yan et al. (1999), and
find a similar LF to Yan et al.  We also plotted the preliminary
narrow-band non reddening corrected H$\alpha$ LFs from Jones \& Bland-Hawthorn
(2001) at $z=0.08$, $z=0.24$ and $z=0.40$. Since they do not fit their
data with a Schechter function, we simply joined their data points in
Fig.~\ref{fig14}.  To compare to their data we plotted our ($0 < z \le
0.3$) H$\alpha$ LF shifted by a factor 2.02 in log towards fainter
luminosities. Their LFs are broadly consistent with our low-$z$ LF.
Their data at $z=0.24$ and $z=0.40$ show an evolution in H$\alpha$
luminosities. Overall the strong evolution of the H$\alpha$ LF is
clearly demonstrated from $z\sim 0$ to $z\sim 1.3$.

In Figs.~\ref{fig12}\&\ref{fig14}, the different shape of the LFs is
likely to be due to different sample selection. In particular the LF
for the $UV$-selected local sample of Sullivan et al. (2000) has a
shape quite different compared to the H$\alpha$-selected LF of Gallego
et al.  (1995), or with ours at low-$z$ (roughly corres\-ponding to a
rest-frame $R$ selection) and at high-$z$ (roughly corresponding to a
rest-frame $B$ selection).  A similar difference in shape has already
been noted in comparisons of broad-band LFs from $B$-selected samples
such as the Au\-tofib Redshift Survey (Ellis et al. 1996) to LFs from
$I$- or $R$-selected samples such as the CFRS (Lilly et al. 1995) and
the CNOC2 (Lin et al. 1999).  Samples which select galaxies on their
$UV$ continuum are more dominated by blue, late-type galaxies, which
steepen the slope $\alpha$, and brighten L$^{\ast}$, because of
$k$-correction effects.  The LF shape derived from samples with
narrow-band emission-line selection (Gallego et al.  1995), or with
rest-$B$ continuum selection (this paper) are similar. This is not
surprising if we account for the close relation between $B$ magnitudes
and H$\alpha$ luminosities (see Fig.~\ref{fig4}).  Since we observe
that H$\alpha$ is tightly correlated to absolute $B$ magnitude, we
expect H$\alpha$ LFs to be about same shape as the \MBAB\ LF of mid-
and late-type galaxies.

\subsection{H$\alpha$ luminosity densities}

A minimal estimate of the luminosity density can be obtained by simply
summing up $L$(H$\alpha$)/$V_{max}$ for each galaxy, weighted by
$f_{rep}$ (see Section~\ref{sct-halumfun}). This gives the directly
observed luminosity density; ${\cal L}$(H$\alpha$)~=~$10^{40.04}$ erg
s$^{-1}$ Mpc$^{-3}$ at $0.5<z<1.1$.  For a Schechter function, the
lumino\-sity density is dominated by $\sim$L$^{\ast}$ galaxies, so
galaxies outside the observed luminosity range will not introduce
large errors so long as galaxies near L$^{\ast}$ are included, and the
faint-end slope $\alpha$ is not as steep as $-2$.  Indeed the estimate
from analytically integrating our best Schechter function gives ${\cal
  L}$(H$\alpha$)~=~$10^{40.10 \pm 0.05}$ erg s$^{-1}$ Mpc$^{-3}$ at
$\langle z \rangle \simeq 0.7$.  If we account for the uncertainty in
log($\phi$) as described in Section~\ref{sct-halumfun}, we obtain
$10^{40.22\pm0.05}$ erg s$^{-1}$ Mpc$^{-3}$.  For the low-$z$ sample
of Tresse \& Maddox~(1998), using the sum of the individual densities
or the LF integration, we find the same total H$\alpha$ luminosity
density, respectively $10^{39.45}$ and $10^{39.44\pm0.04}$ erg
s$^{-1}$ Mpc$^{-3}$. The luminosity density contributed by galaxies
outside the observed range is very small, so the correction factor to
extrapolate to all luminosities is very close to unity.  In addition
to our high-$z$ estimate, we divided the high-$z$ sample at the median
redshift, $z=0.702$, into two bins. Using the sum of the individual
densities in each bin, we find $10^{40.05\pm0.05}$ erg s$^{-1}$
Mpc$^{-3}$ at $\langle z \rangle =0.6$ and $10^{40.21\pm0.17}$ erg
s$^{-1}$ Mpc$^{-3}$ at $\langle z \rangle =0.8$. We estimated the
uncertainties by slightly varying the redshift cut about the median
redshift and measuring the $rms$ dispersion. Even though the small
number of galaxies (15) in each bin makes the statistics poorer than
using the entire high-$z$ sample, there is a suggestion of an increase
in the luminosity density over the redshift range $0.5 <z <1.1$.
      
\begin{figure}
\centerline{\psfig{figure=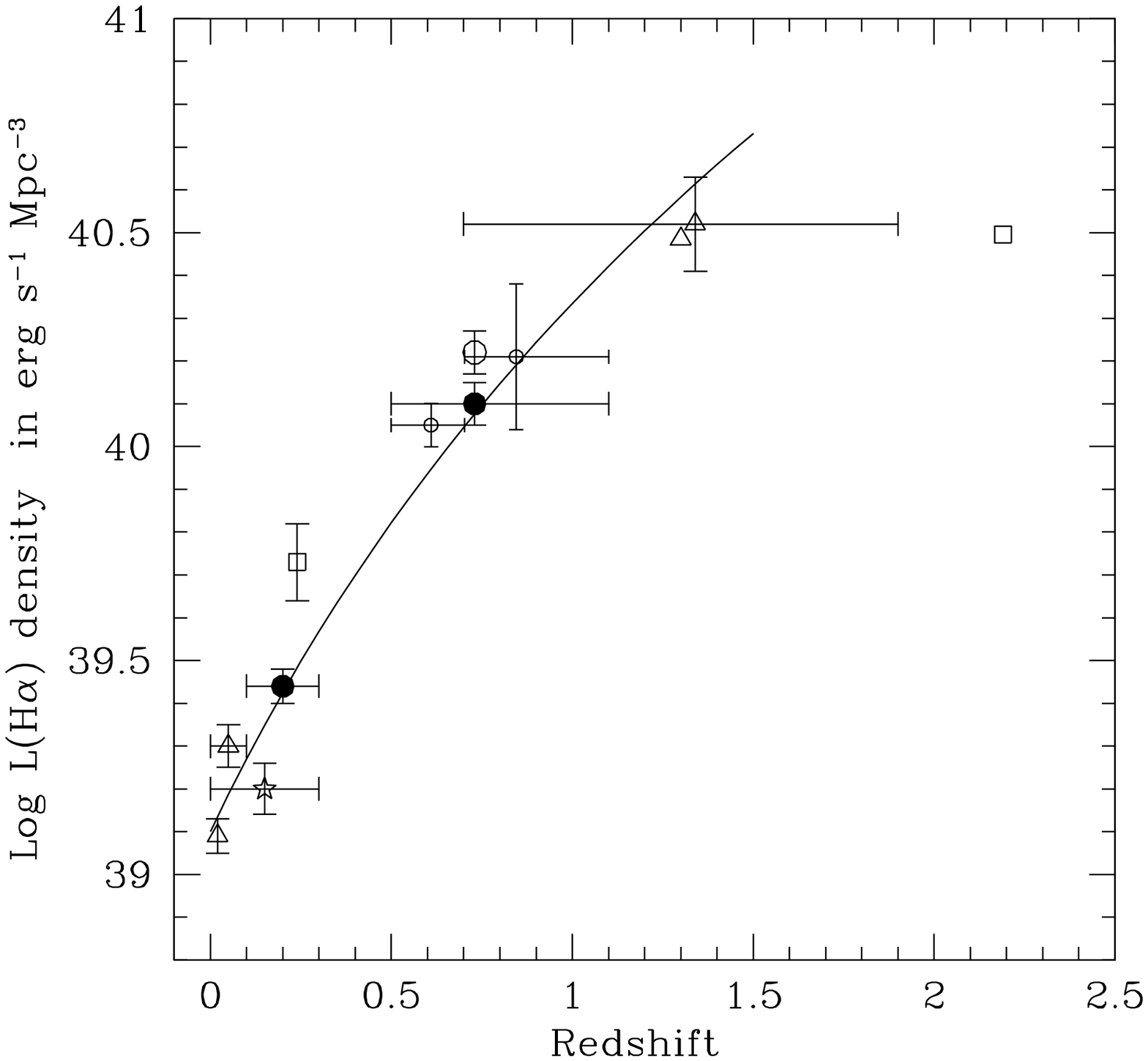,width=8cm}}
\caption{The evolution of H$\alpha$ luminosity density with redshift 
  with reddening-corrected H$\alpha$ data summarized  in Table~\ref{tab3}.
  The curve represents an evolution proportional to $(1+z)^{4.1}$.
  The filled circles represent our CFRS broad-band $I$-selected data at $\langle
  z \rangle = 0.3$ and $0.7$, and the large open  circle
  corresponds to our high-$z$ data but using the correction factor
  $f_{all}$ (see Section 6.1). The two small open circles represent
  our high-$z$ sample divided into two bins at $\langle z \rangle =
  0.6$ and $\langle z \rangle = 0.8$. The other points in the
  diagram are as follows.  
  Triangles represent slitless spectroscopic surveys: (UCM) Gallego et
  al. (1995) at $z\sim0$; (KISS) Gronwall et al. (1999) at $z=0.05$;
  Hopkins, et al. (2000) at $z = 1.3$ and Yan et al. (1999) at $z =
  1.34$. Squares represent narrow-band filter surveys: Pascual et al.
  (2001) at $z = 0.24$; Moorwood et al. (2000) at $z=2.2$.  The star
  represents the FOCA broad-band $UV$-selected data of Sullivan et al.
  (2000) at $\langle z \rangle = 0.15$.
\label{fig15}}
\end{figure}

In Fig.~\ref{fig15}, we present reddening corrected H$\alpha$ lumino\-sity
densities derived from H$\alpha$ LFs as a function of redshift (see
also Table~\ref{tab3}).  The increase of the H$\alpha$ luminosity
density is proportional to $(1+z)^{4.1\pm0.3}$ from $z\sim 0$ to 1.3.
We note that the $UV$-selected data from Sullivan et al.  (2000) at
$z\sim 0.15$ is slightly below this evolution.  This is because their
LF has a brighter L$^{\ast}$ (due to two extremely bright H$\alpha$
luminosities), and has a lower $\phi^\ast$. Their $\alpha$ is also
steep, but this does not contribute much to the overall luminosity
density. See also Section~6.2.
\begin{table*}
\caption{H$\alpha$ luminosity densities estimated from the H$\alpha$ luminosity function. 
\label{tab3}} 
\begin{tabular}{lcccll}
\hline
Redshift range & N &  $\langle z \rangle$ &  Log $\cal L$(H$\alpha$)$^{\dagger}$ & Selection$^{\ddagger}$  & Reference \\
\hline
$z  \le  0.045$      & 176   & 0.02  & 39.09$\pm$0.04  & SLS H$\alpha$-selected, UCM & Gallego et al.  (1995) \\
$0 < z < 0.1$        & 1100  & 0.05  & 39.30$\pm$0.05 & SLS H$\alpha$-selected, KISS  & Gronwall et al. (1999) \\
$0 < z < 0.4$        &  88   & 0.15  & 39.20$\pm$0.06   & $UV$-selected, FOCA               & Sullivan et al. (2000) \\ 
$0 < z \le 0.3$      & 138   & 0.21  & 39.44$\pm$0.04  & $I$-selected, CFRS                   & Tresse \& Maddox (1998) \\
$0.228 < z < 0.255$        &  37   & 0.24  & 39.73$\pm$0.09   & NBF H$\alpha$-selected          & Pascual et al. (2001) \\
$0.5 < z < 1.1$      &  33   & 0.73  & 40.10$\pm$0.05  & $I$-selected, CFRS                   & This paper\\
$0.7 < z < 1.8$      & 17/37 & 1.3   & 40.18+log(2.02)$^{1,2}$   & SLS H$\alpha$-selected   & Hopkins et al. (2000) \\
$0.7 < z < 1.9$      &  33   & 1.34  & 40.21+log(2.02)$^1$          & SLS H$\alpha$-selected   & Yan et al. (1999) \\
$2.175 < z < 2.225$  &  6    & 2.19  & 40.19+log(2.02)$^{1,2}$ & NBF H$\alpha$-selected  & Moorwood et al. (2000) \\
\hline \\
\end{tabular}
\begin{minipage}{170mm}{
$^{\dagger}$$\cal L$(H$\alpha$) in units of erg~s$^{-1}$~Mpc$^{-3}$. \\
$^{\ddagger}$The H$\alpha$-selected surveys use either narrow-band filters         
(NBF) or slitless spectroscopy (SLS).  These samples can be                       
contaminated by other emission-line features if the lines are not                 
fully spectroscopically confirmed, in particular at high redshifts.\\     
$^1$Assuming the canonical A$_{\rmn V}$=1 mag extinction.\\
$^2$Based on the Yan et al. (1999) H$\alpha$ luminosity function. } 
\end{minipage}
\end{table*}

Regarding possible narrow-line AGN contamination, we do not exclude
Seyfert 2 or LINER spectra in either our low- or high-$z$ data.  This
allows direct comparison with SFRs derived from continuum luminosity
densities where no such exclusion is done. Also it avoids possible
confusions from the fact that different criteria may be used to
identify active galaxies in different samples.  At $z\la 0.3$, the
fraction of AGN-like spectra can vary from few per cent to 40 per cent
depending on the selection criteria. For example, it is 8--17 per cent
in the CFRS low-$z$ sample (Tresse et al. 1996); 8 per cent in the
local UCM (Gallego et al. 1995); 17--28 per cent in the 15R survey
(Carter et al.  2001). Detailed stu\-dies as for instance done by Ho,
Filipenko \& Sargent (1997) find a fraction of 43 per cent in their
blue-selected survey.  $UV$-selected samples contain almost no
narrow-line AGN (Contini et al.  2002), this agrees with the common
picture that active galaxies are more dusty than normal galaxies.  If
an active nucleus is present in the central region of a galaxy, then
the H$\alpha$ flux is not correlated directly to forming stars, but is
a mixture from both the AGN and SFR.  This makes it difficult to
separate the stellar from the AGN contribution in the total luminosity
density from optically-selected galaxy surveys.  The SFR density
diagrams derived from such surveys must depend on the fraction of
AGN-like galaxies, and on any global cosmic evolution in the nuclear
activity of galaxies. In the UCM, the 8 per cent of AGN-like galaxies
contribute to 15 per cent to the overall luminosity density (Pascual
et al. 2001).  If this fraction stays about constant up to $\sim 1$,
then the H$\alpha$-derived SFRs are contaminated by no more than 15
per cent AGN light.  This is an upper limit to the contamination,
because excluding AGN-like galaxies will exclude the light from
present star formation as well as their AGN contribution.

\subsection{Star-formation rate densities}

\begin{figure}
\centerline{\psfig{figure=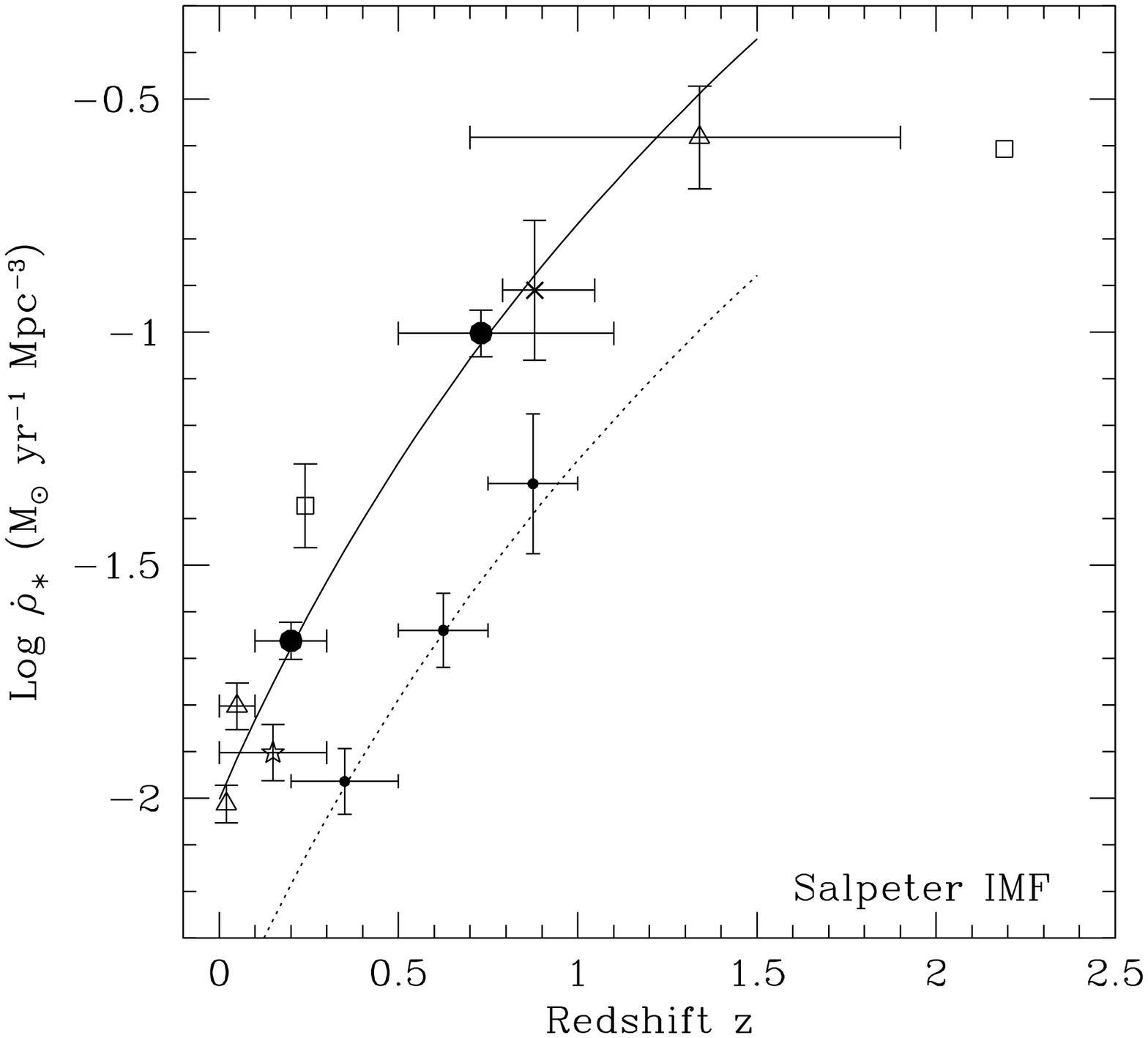,width=8cm}}
\caption{SFR densities based on reddening-corrected H$\alpha$ measurements. 
  We have applied a correction of A$_{\rmn V}=1$ mag for data which
  were are not reddening corrected (see Table~\ref{tab3}). SFR are
  derived from the K98 transformation,
  log~L(H$\alpha$)(erg~s$^{-1}$)~=~41.10 +
  log~$\dot{\rho_{\ast}}$(M$_{\odot}$ yr$^{-1}$).  Note that according
  the Charlot \& Longhetti (2001) model, the K98 H$\alpha$
  transformation gives a SFR ty\-pi\-cally two or three times too low,
  even after reddening correction (see Section~6.2).  The different
  symbols are defined in Fig.~\ref{fig15} with in addition the cross
  which shows the H$\alpha$ data on 6 CFRS gala\-xies at $\langle z
  \rangle = 0.9$ from which the SFR density has been derived using the
  broad-band CFRS luminosity function, and L(2800
  \AA)/L(H$\alpha$)~=~$3.1\ 10^{-14}$~Hz$^{-1}$ (Glazebrook et al.
  1999).  The solid curve shows the ($1+z$)$^{4.1}$ evolution as shown
  in Fig.~\ref{fig15}. For reference, we also show the 2800-\AA\ CFRS
  points from Lilly et al. (1996) u\-sing the K98 transformation log~
  L$_{UV}$(erg~s$^{-1}$~Hz$^{-1}$)~=~27.85 +
  log~$\dot{\rho_{\ast}}$(M$_{\odot}$ yr$^{-1}$) which includes no
  dust correction. The dotted curve is the same as the plain curve
  shifted by a factor~3.2.
\label{fig16}
}
\end{figure}

In Fig.~\ref{fig16} we plot SFR densities based on several recent
reddening-corrected H$\alpha$ measurements.  We use the Kennicutt
(1998, hereafter K98) transformation between H$\alpha$ luminosity
(units erg s$^{-1}$) and SFR (units M$_{\odot}$ yr$^{-1}$) for solar
abundances and a Salpeter IMF including stars in the
0.1--100~M$_\odot$ mass range: log~L(H$\alpha$)~=~41.10 +
log~$\dot{\rho_{\ast}}$.  For comparison, we also plot the 2800-\AA\ 
CFRS data from which Lilly et al. (1996) derived a growth proportional
to ($1+z$)$^{3.9\pm0.75}$. We use the K98 transformation, log~L$_{\rmn
  UV}$(erg~s$^{-1}$~Hz$^{-1}$)~=~27.85 +
log~$\dot{\rho_{\ast}}$(M$_{\odot}$ yr$^{-1}$) which includes no dust
correction. As shown in the figure, our H$\alpha$ SFR density growth,
$(1+z)^{4.1\pm0.3}$ is entirely consistent with the $UV$ SFR density
growth. To make compatible both the H$\alpha$ and $UV$ SFR densities,
a shift in SFR density of a factor of 3.20 is necessary.  This ratio
corresponds to the dust correction to the near $UV$ data, and implies
$A_{\rmn V} \simeq 1$ mag for a simple dust-screen model (Pei 1992).
This is in agreement with the L(2800\AA)/L(H$\alpha$) ratio of
3.1$\pm$0.4 10$^{-14}$ Hz$^{-1}$ found by Glazebrook et al. (1999)
with six CFRS H$\alpha$ data at $z\sim0.9$.  We note that a
correlation of the H$\alpha$/FUV ratio as a function of intrinsic
luminosity has been found by e.g. Bell \& Kennicutt (2001), Sullivan
et al.  (2001), Buat et al.  (1999). This will affect little our
result since the low-luminosity galaxies contribute little to the
total density, and since the FUV ($\le 2000$~\AA) is more sensitive to
dust attenua\-tion ($\sim 1.4$ mag) than the near UV at 2800~\AA\ 
($\sim 1$ mag).  In conclusion, our H$\alpha$ and 2800-\AA\ SFR
densities are broadly consistent assuming an attenuation of $A_{\rmn
  V}=1$ mag.

Charlot et al. (2002) have derived SFRs of the local SAPM galaxies,
using H$\alpha$, [\ion{O}{2}], [\ion{S}{2}] and [\ion{N}{2}].  They
conclude that the true SFRs are typically two or three times higher
than those derived from H$\alpha$ alone, using the K98 transformation,
even after applying the usual mean redde\-ning correction of A$_{\rmn
  V}$~=~1 mag. The Charlot \& Longhetti (2001) model accounts for the
absorption of ionizing photons by dust in \ion{H}{2} regions and the
contamination of H$\alpha$ emission by stellar absorption.  Including
these effects they find that SFR estimators derived from H$\alpha$,
[\ion{O}{2}], $UV$, and far-$IR$ fluxes give consistent results. This
would imply that the CFRS-$UV$ SFR densities should be increased by a
factor 5 or 6.

Our H$\alpha$ luminosity density estimates confirm the steep rise in
SFR density for $0.1<z<1.1$, as found by Lilly et al. (1996). Since
our results are based on independent H$\alpha$ measurements, the steep
rise is unlikely to be due to extrapolation of the CFRS data to the
2800-\AA\ luminosities as suggested by Cowie et al. (1999).  Our
measurements are inconsistent with the shallow rise of the 2500-\AA\ 
luminosity density, $\propto (1+z)^{1.5}$, found by Cowie et al.
(1999), and also the Sullivan et al. (2000) suggestion that the
2000-\AA\ luminosity density is $\propto (1+z)^{1.7}$.  At $z\sim0.2$
the Sullivan et al. (2000) luminosity density is consistent with both
our H$\alpha$ estimate and the Lilly et al. (1996) estimate, so the
suggestion of a shallow slope is mainly driven by the low luminosity
density at $z\sim 0.5-1$ found in the Cowie et al. (1999) survey.
This survey covers a very small area ($\sim 30$ arcmin$^2$) compared
to the CFRS ($\sim 500$ arcmin$^2$), and so must be subject to much
larger uncertainties due do cosmic variance. Thus we believe that the
steeper rise in SFR is more likely to be correct.

The exact slope of the rise also depends on which local value is
taken. For instance, using the H$\alpha$ SAPM data at $\langle z
\rangle =0.05$ Singleton (2001) found a lower value than Gallego et
al.  (1995), while Gronwall et al. (1998) finds a higher value using
the H$\alpha$ KISS survey. Soon, the 2dFGRS and the SDSS should
provide a more definitive local value.  Whatever the local value,
within our single survey the H$\alpha$ data demonstrate a rise in the
SFR density by a factor of $\sim 5$ from $z\sim0.2$ to $z\sim0.7$. Our
data and the Yan et al.  data (1999), require strong evolution in the
SFR density, which increases by a factor 12 from $z\sim0.2$ to
$z\sim1.3$.

\section{Conclusion}

We have measured H$\alpha$ in emission for 30 galaxies from the CFRS
at $0.5 < z < 1.1$ using the ISAAC ESO-VLT spectrograph. These data
combined with those at $z \le 0.3$ (Tresse \& Maddox 1998) enabled us
to derive measurements within a single survey with well-controled
selection criteria, and with secure redshift measurements. Our results
are the following: 
\begin{enumerate}
\item H$\alpha$ and $B$-band luminosities are  tightly correlated at
high-$z$ as seen at low-$z$ by Tresse \& Maddox (1998). This
strengthens the hypothesis of an universal IMF. 
\item We demonstrate that the [\ion{O}{2}]/H$\alpha$ ratio is luminosity
dependent. Bright galaxies have ratios larger by a factor~2 than
faint galaxies. This effect has already been observed in the local SAPM
data (Charlot et al. 2002), in the 15R survey (Carter et al. 2001), and 
in the NFGS (Jansen et al. 2001). 
\item Our best fit H$\alpha$ luminosity function at $\langle z \rangle
  = 0.7$ is $\alpha = -1.31\pm0.11$, $\phi^\ast = 10^{-2.39\pm0.06}$
  Mpc$^{-3}$, $L^\ast = 10^{42.37\pm0.06}$ erg s$^{-1}$.  Its shape is
  similar to the H$\alpha$-selected luminosity function at $z\sim0$ of
  Gallego et al. (1995), and also to the luminosity function at
  $\langle z \rangle=0.2$ of Tresse \& Maddox (1998).  The strong
  evolution of the H$\alpha$ LF appears closely related to the
  ($\phi^{\ast}$, L$^{\ast}$) parameters rather than to the faint-end
  slope $\alpha$ from $z \sim 0$ to $z \sim 1.3$.
\item We find an average SFR(2800\AA)/SFR(H$\alpha$) ratio of 3.2 using
the K98 SFR transformations. Since the $UV$ data are not dust corrected,
this corresponds to A$_{\rmn V}\sim 1$~mag assuming a simple dust-screen
model from Pei (1992). Accounting for Charlot et al. (2002) results of
H$\alpha$ SFRs being larger by a factor of $\sim 3$, it would imply
that the $UV$ CFRS densities need to be corrected by a factor of
$\sim 6$. 
\item The growth of the H$\alpha$ luminosity density is proportional to
$(1+z)^{4.1\pm0.3}$ from $z\sim 0$ to $z\sim 1.3$.  From
independent data, we confirm the steep rise of the  comoving SFR
density as found by the CFRS 2800-\AA\ continuum data.  
The SFR density derived from H$\alpha$ luminosities increases by a
factor~12 between $z\sim0.2$ and $z\sim 1.3$.  
\end{enumerate}

In the near future, the VIRMOS-VLT Deep Survey (VVDS, Le F\`evre et
al. 2001) will provide an wealth of data for measuring the
star-formation rate density from $z\sim0$ to $z\sim5$  within a single
galaxy survey.

\section*{Acknowledgments}
We thank the ESO staff at Garching and Paranal for their help 
in the acquisition of the data. We thank the referee for useful 
suggestions.

\bsp

\label{lastpage}

\end{document}